\documentclass[useAMS,usenatbib,twocolumn]{mn2e}
\usepackage{subfigure}
\usepackage{rotating}
\usepackage{times}
\usepackage{graphicx}
\usepackage{lscape}
\def\etal{{\it et al.}}

\title[Signatures of magnetar central engines in short GRB lightcurves]{Signatures of magnetar central engines in short GRB lightcurves}
\author[A. Rowlinson \etal]{A. Rowlinson$^{1}$\thanks{E-mail:b.a.rowlinson@uva.nl}, P.~T. O'Brien$^{2}$, B.~D. Metzger$^{3}$, N.~R. Tanvir$^{2}$, A.~J. Levan$^{4}$\\ 
$^{1}$Astronomical Institute ``Anton Pannekoek'', University of Amsterdam, Postbus 94249, 1090 GE Amsterdam, The Netherlands\\
$^{2}$Department of Physics \& Astronomy,University of Leicester, University Road, Leicester, LE1 7RH, UK\\
$^{3}$Department of Astrophysical Sciences, Peyton Hall, Princeton University, Princeton, NJ 08544, USA\\
$^{4}$Department of Physics, University of Warwick, Coventry CV4 7AL, UK}
\begin{document}

\pagerange{\pageref{firstpage}--\pageref{lastpage}} \pubyear{000}
\maketitle            

\label{firstpage}

\begin{abstract}
A significant fraction of the Long Gamma-ray Bursts (LGRBs) in the {\it Swift} sample have a plateau phase showing evidence of ongoing energy injection. We suggest that many Short Gamma-ray Bursts (SGRBs) detected by the {\it Swift} satellite also show evidence of energy injection. Explaining this observation within the typical SGRB progenitor model is challenging as late time accretion, often used to explain plateaus in LGRBs, is likely to be absent from the SGRB population. Alternatively, it is predicted that the remnant of NS-NS mergers may not collapse immediately to a BH (or even collapse at all), forming instead an unstable millisecond pulsar (magnetar) which powers a plateau phase in the X-ray lightcurve.

By fitting the magnetar model to all of the {\it Swift} SGRBs observed until May 2012, we find that about half can be clearly fitted with a magnetar plateau phase while the rest are consistent with forming a magnetar but the data are insufficient to prove a plateau phase. More data, both at early times and a larger sample, are required to confirm this. This model can be tested by detecting the gravitational wave emission from events using the next generation gravitational wave observatories.

\end{abstract}

\begin{keywords}

Gamma-Ray Bursts, Magnetars

\end{keywords}

\section{Introduction}

\begin{table*}
\begin{tabular}{cccccccccc}
\hline
GRB     & $\alpha_{1}$   & $\Gamma_{x,1}$ & T$_{1}$        & $\alpha_{2}$   & $\Gamma_{x,2}$ & T$_{2}$        & $\alpha_{3}$   & $\Gamma_{x,3}$ & Prob. Chance\\
        &                &                & (s)            &                &                & (s)            &                &               & Improvement \\
\hline 
\multicolumn{10}{l|}{2 or more breaks}\\
051221A & 1.43$^{+0.01}_{-0.01}$ & 2.03$^{+0.20}_{-0.19}$ & 2935$^{+714}_{-785}$ & 0.059$^{+0.22}_{-0.11}$ & 1.91$^{+0.23}_{-0.22}$ & 24370$^{+4631}_{-2823}$ & 1.41$^{+0.08}_{-0.07}$ & 2.06$^{+0.19}_{-0.18}$ & 1$\times10^{-4}$\% \\
060313  & 1.84$^{+0.85}_{-0.37}$ &   &  1.7$^{+1.2}_{-0.9}$    & 0.74$^{+0.08}_{-0.09}$ & 1.82$^{+0.16}_{-0.10}$ & 7467$^{+1511}_{-1491}$ & 1.65$^{+0.12}_{-0.11}$ & 2.50$^{+0.22}_{-0.28}$ & 6$\times10^{-20}$\% \\ 
061201  & 3.09$^{+0.66}_{-0.46}$ &     $^{     }_{     }$ & 1.85$^{+1.03}_{-0.53}$ & 0.54$^{+0.13}_{-0.14}$ & 1.44$^{+0.20}_{-0.19}$ & 2209$^{+802}_{-587}$ & 1.84$^{+0.17}_{-0.14}$ & 2.26$^{+0.38}_{-0.42}$ & 2$\times10^{-3}$\%\\
070724  & 0.97$^{+0.12}_{-0.05}$ & 1.45$^{+0.73}_{-0.64}$ &   79$^{+10}_{-35}$     & -1.13$^{+0.69}_{-1.01}$ & 1.66$^{+0.24}_{-0.23}$ & 110$^{+2}_{-3}$ & 10$^{+0.00}_{-4.33}$ & $^{}_{}$ & 1$\times10^{-5}$\% \\
090426  & 2.12$^{+0.36}_{-0.45}$ &                        & 33$^{+125}_{-3}$ & 0.21$^{+0.31}_{-0.34}$ & 1.85$^{+0.36}_{-0.24}$ & 260$^{+140}_{-127}$ & 1.04$^{+0.07}_{-0.06}$ & 2.14$^{+0.14}_{-0.14}$ & 2$\times10^{-14}$\% \\
090515  & 2.76$^{+0.55}_{-0.10}$ &                        & 0.30$^{+0.00}_{-0.30}$ & 0.28$^{+0.07}_{-0.03}$ & 1.85$^{+0.17}_{-0.16}$ & 156$^{+9}_{-27}$ & 2.51$^{+0.59}_{-0.87}$ & 2.12$^{+0.39}_{-0.33}$ & 3$\times10^{-32}$\% \\
100625A & 3.63$^{+0.01}_{-0.25}$ &                        & 1.90$^{+2.40}_{-1.10}$ & 0.36$^{+0.36}_{-0.63}$ & 2.09$^{+0.30}_{-0.29}$ & 222$^{+52}_{-50}$ & 3.15$^{+0.94}_{-0.85}$ & 2.66$^{+0.53}_{-0.83}$ & 0.11\% \\
100702A & 1.67$^{+0.15}_{-0.18}$ &     $^{     }_{     }$ & 0.59$^{+0.4}_{-0.4}$ & 0.74$^{+0.18}_{-0.18}$ & 2.05$^{+0.13}_{-0.13}$ & 194$^{+14}_{-6}$ & 4.86$^{+0.52}_{-0.26}$ & 2.41$^{+0.28}_{-0.26}$ & 2$\times10^{-43}$\% \\
101219A & & & & 0.79$^{+0.04}_{-0.04}$ & 1.33$^{+0.72}_{-0.75}$ & 195$^{+7}_{-12}$ & 10$^{+0.00}_{-2.40}$ & $^{}_{}$ &  9$\times10^{-3}$\% \\
120305A & 2.88$^{+0.30}_{-0.23}$ & & 2.2$^{+1.4}_{-0.9}$ & 0.18$^{+0.29}_{-0.29}$ & 1.94$^{+0.21}_{-0.20}$ & 156$^{+11}_{-10}$ & 5.11$^{+0.55}_{-0.52}$ & 2.51$^{0.72}_{0.44}$ & 0.17\% \\

\hline
\multicolumn{10}{l|}{1 break}\\
051210  & & & & 0.65$^{+0.04}_{-0.04}$ & 1.21$^{+0.25}_{-0.15}$ & 137$^{+8}_{-6}$ & 3.52$^{+0.25}_{-0.19}$ & 3.11$^{+0.44}_{-0.65}$ & 1$\times10^{-8}$\% \\
060801  & & & & 0.53$^{+0.05}_{-0.06}$ & 1.59$^{+0.23}_{-0.22}$ & 315$^{+21}_{-30}$ & 5.83$^{+0.86}_{-0.76}$ & 2.18$^{+0.63}_{-0.43}$ & 1$\times10^{-3}$\% \\
070714A & 2.23$^{+0.18}_{-0.04}$ &     $^{     }_{     }$ & 123$^{+4}_{-45}$ & 0.62$^{+0.06}_{-0.05}$ & 2.24$^{+0.33}_{-0.33}$ &  &  &  & 4$\times10^{-6}$\% \\
070809  & 1.42$^{+0.05}_{-0.04}$ & 1.65$^{+1.01}_{-0.40}$ & 233$^{+96}_{-68}$    & 0.52$^{+0.06}_{-0.06}$ & 1.35$^{+0.18}_{-0.13}$ &  &  &  & 3$\times10^{-3}$\% \\
080426  & 1.94$^{+0.15}_{-0.14}$ &     $^{     }_{     }$ & 15$^{+18}_{-7}$ & 1.18$^{+0.05}_{-0.05}$ & 2.03$^{+0.26}_{-0.24}$ &  &  &  & 0.018\% \\
080905A & & & & 0.44$^{+0.05}_{-0.05}$ & 0.89$^{+0.56}_{-0.41}$ & 126$^{+45}_{-55}$ & 2.51$^{+0.30}_{-0.25}$ & 1.53$^{+0.29}_{-0.27}$ & 0.03\% \\
080919  & & & & 0.86$^{+0.04}_{-0.03}$ & 2.31$^{+1.01}_{-0.83}$ & 351$^{+195}_{-55}$ & 4.83$^{+0.77}_{-0.84}$ & 2.35$^{+1.01}_{-0.83}$ & 0.02\% \\
090510  & & & & 0.80$^{+0.01}_{-0.01}$ & 1.78$^{+0.14}_{-0.14}$ & 1412$^{+136}_{-192}$ & 2.18$^{+0.17}_{-0.17}$ & 2.22$^{+0.20}_{-0.16}$ & 1$\times10^{-6}$\% \\
090621B & 4.06$^{+0.01}_{-0.49}$ &     $^{     }_{     }$ & 5$^{+5}_{-1}$ & 0.72$^{+0.18}_{-0.16}$ & 3.40$^{+1.40}_{-1.00}$ &  &  &  &  3$\times10^{-5}$\% \\
091109B & 4.02$^{+0.01}_{-0.32}$ &     $^{     }_{     }$ & 4$^{+1}_{-1}$ & 0.64$^{+0.08}_{-0.09}$ & 2.04$^{+0.55}_{-0.37}$ &  &  &  & 4$\times10^{-4}$\% \\
111020A & 1.63$^{+0.62}_{-0.05}$ &  & 124$^{+38}_{-123}$ & 0.76$^{+0.05}_{-0.04}$ & 2.18$^{+0.49}_{-0.43}$ &  &  &  & 0.02\% \\ 
120521A & & & & 1.20$^{+0.05}_{-0.05}$ & 1.81$^{+0.36}_{-0.29}$ & 283$^{+13}_{-17}$ & 9.98$^{+0.02}_{-2.25}$ &  & 0.12\% \\
\hline
\multicolumn{10}{l|}{No breaks}\\
050509B & 1.32$^{+0.06}_{-0.04}$ & 1.92$^{+1.13}_{-0.52}$ &  &  &  &  &  &  & \\ 
050813  & 1.27$^{+0.04}_{-0.03}$ & 2.70$^{+4.30}_{-1.20}$ &  &  &  &  &  &  &   \\
050906  & $>$1.28 &  & & & & & & & \\
051105  & $>$1.33 &  &  &  &  &  & & &  \\
060502B & 0.95$^{+0.04}_{-0.03}$ & 2.10$^{+2.77}_{-0.81}$ &  &  &  &  &  &  & \\
061217  & 1.29$^{+0.08}_{-0.05}$ & 1.40$^{+1.13}_{-0.86}$ & & &  &  &  &  &  \\
070209  & $>$1.23 &  &  &  &  &  &  &  & \\
070429B & 1.54$^{+0.05}_{-0.04}$ & 3.10$^{+1.00}_{-1.40}$ &  & &  &  &  &  &  \\
070729  & 1.29$^{+0.05}_{-0.04}$ & 1.62$^{+0.86}_{-0.43}$ &  & & &  &  &  &  \\
070810B & $>$1.36 &  &  &  &  &  &  & & \\
071112B & $>$0.87 &  &  &  &  &  &  & & \\
080702A & 1.13$^{+0.04}_{-0.04}$ & 1.99$^{+0.75}_{-0.67}$ &  &  &  &  &  &  &  \\
081024A & 0.99$^{+0.03}_{-0.02}$ & 1.82$^{+0.64}_{-0.55}$ &  &  &  &  &  &  & \\
081101  & $>$1.21 &  &  &  &  &  &  &  & \\
081226  & 1.45$^{+0.05}_{-0.04}$ & 3.84$^{+0.96}_{-1.93}$ & & &  &  &  &  &  \\
090305A & 1.42$^{+0.05}_{-0.04}$ &  &  &  &  &  &  &  & \\
100117A & 0.97$^{+0.01}_{-0.01}$ & 2.59$^{+0.48}_{-0.40}$ &  &  &  &  &  &  &  \\
100206A & 1.80$^{+0.05}_{-0.04}$ & 3.30$^{+3.30}_{-1.30}$ & & &  &  &  &  &  \\
100628A & 1.00$^{+0.01}_{-0.01}$ & &  &  &  &  &  & & \\
110112A & 1.00$^{+0.06}_{-0.05}$ & 2.15$^{+0.39}_{-0.31}$ &  &  &  &  &  &  &\\
111117A & 1.45$^{+0.05}_{-0.06}$ & 2.20$^{+0.40}_{-0.37}$ & & &  &  &  &  &  \\
\end{tabular}
\caption{The {\it Swift} SGRB sample and the results of broken powerlaw fits to the observed BAT-XRT data in the 0.3-10 keV band (as described in the text) and the X-ray spectral indicies for each regime ($\Gamma_{x}$). These are subdivided into those with 2 or more significant breaks in their lightcurves, those with 1 break and those with no significant breaks. Where values are left blank there was insufficient data available to constrain them. The last column shows the probability that this fit is a chance improvement on a simpler model.}
\label{lcfits}
\end{table*}

\begin{table*}
\begin{tabular}{cccccccc}
\hline
GRB     & z & T$_{90}$ & $\Gamma_{\gamma}$ & Fluence                  & Host & Host offset & Optical Afterglow\\
        &   & (s)      &          & (10$^{-7}$ erg cm$^{-2}$ s$^{-1}$) &          &(arcsec)    & \\
\hline
\multicolumn{8}{l|}{2 or more breaks}\\
051221A$^{(1)}$ & 0.55    & 1.4$\pm$0.2     & 1.39$\pm$0.06 & 11.6$\pm$0.4  & y & 0.12$\pm$0.04 & Y    \\
060313$^{(2)}$   & (0.72) & 0.7$\pm$0.1     & 0.71$\pm$0.07 & 11.3$\pm$0.5  & ? & 0.4$\pm$0.6   & Y    \\
061201$^{(3)}$  & 0.111   & 0.8$\pm$0.1     & 0.81$\pm$0.15 & 3.3$\pm$0.3   & ? & 17            & Y    \\
070724A$^{(4)}$ & 0.46    & 0.4$\pm$0.04   & 1.81$\pm$0.33 & 0.30$\pm$0.07 & y & 0.7$\pm$2.1   & N    \\
090426$^{(5)}$   & 2.6     & 1.2$\pm$0.3     & 1.93$\pm$0.22 & 1.8$\pm$0.3   & y & 18           & Y    \\
090515$^{(6)}$  & (0.72) & 0.04$\pm$0.02 & 1.60$\pm$0.20   & 0.21$\pm$0.04 & n & -              & Y    \\
100625A$^{(7)}$& (0.72)  & 0.33$\pm$0.03   & 0.90$\pm$0.10   & 2.3$\pm$0.2   & y & 0$\pm$1.8     & N    \\
100702A$^{(8)}$ & (0.72) & 0.16$\pm$0.03   & 1.54$\pm$0.15 & 1.2$\pm$0.1   & n & -             & N    \\
101219A$^{(9)}$ & 0.718   & 0.6$\pm$0.2     & 0.63$\pm$0.09 & 4.6$\pm$0.3   & y & -            & N   \\
120305A$^{(10)}$ & (0.72)   & 0.10$\pm$0.02     & 1.00$\pm$0.09 & 2.0$\pm$0.1   & n & -         & N   \\
120521A$^{(11)}$ & (0.72)   & 0.45$\pm$0.08     & 0.98$\pm$0.22 & 0.8$\pm$0.1   & n & -         & N   \\
\hline
\multicolumn{8}{l|}{1 break}\\
051210$^{(12)}$  & (0.72) & 1.4$\pm$0.2     & 1.10$\pm$0.30   & 0.8$\pm$0.1 & ? & 2.8$\pm$2.9   & N    \\
060801$^{(13)}$   & 1.13    & 0.5$\pm$0.1     & 0.47$\pm$0.24 & 0.8$\pm$0.1 & ? & 2.4$\pm$2.4  & N   \\
070714A$^{(14)}$  & (0.72) & 2.0$\pm$0.3     & 2.60$\pm$0.20   & 1.5$\pm$0.2   & n & -             & N    \\
070809$^{(15)}$   & 0.219   & 1.3$\pm$0.1     & 1.69$\pm$0.22 & 1.0$\pm$0.1 & y & 20           & Y   \\
080426$^{(16)}$   & (0.72) & 1.7$\pm$0.4     & 1.98$\pm$0.13 & 3.7$\pm$0.3   & n & -             & N    \\
080905A$^{(17)}$  & 0.122  & 1.0$\pm$0.1     & 0.85$\pm$0.24 & 1.4$\pm$0.2   & y & 9             & Y    \\
080919$^{(18)}$  & (0.72) & 0.6$\pm$0.1     & 1.10$\pm$0.26 & 0.7$\pm$0.1 & ? & -             & Y    \\
090510$^{(19)}$   & 0.9    & 0.3$\pm$0.1     & 0.98$\pm$0.20 & 3.4$\pm$0.4   & y & 1             & Y    \\
090621B$^{(20)}$ & (0.72) & 0.14$\pm$0.04   & 0.82$\pm$0.23 & 0.7$\pm$0.1   & n & -             & N    \\
091109B$^{(21)}$ & (0.72) & 0.30$\pm$0.03   & 0.71$\pm$0.13 & 1.9$\pm$0.2   & ? & 8             & Y    \\
111020A$^{(22)}$ & (0.72) & 0.40$\pm$0.09   & 1.37$\pm$0.26 & 0.7$\pm$0.1 & n &  & N   \\
\end{tabular}
\caption{Properties of the SGRB sample, including T$_{90}$, $\Gamma_{\gamma}$ and Fluence (15--150 keV). These observed quantities, including host galaxy associations, offsets and optical afterglow detections, are from published papers and GCNs (references listed below), host offsets are quoted with errors if published. When the redshift is not known, the average redshift 0.72 was used and this is shown using brackets.} 
\label{candidates}
$^{(1)}$\cite{cummings2005,soderberg2006}
$^{(2)}$\cite{markwardt2006a,roming2006}
$^{(3)}$\cite{markwardt2006b,stratta2007}
$^{(4)}$\cite{parsons2007,berger2009,kocevski2010}
$^{(5)}$\cite{sato2009,antonelli2009,xin2011}
$^{(6)}$\cite{barthelmy2009,rowlinson2010b}
$^{(7)}$\cite{barthelmy2010,tanvir2010}
$^{(8)}$\cite{baumgartner2010}
$^{(9)}$\cite{krimm2010, chornock2011}
$^{(10)}$\cite{palmer2012}
$^{(11)}$\cite{cummings2012}
$^{(12)}$\cite{sato2005,laparola2006}
$^{(13)}$\cite{sato2006,cucchiara2006}
$^{(14)}$\cite{barthelmy2007}
$^{(15)}$\cite{krimm2007,perley2008a}
$^{(16)}$\cite{cummings2008a}
$^{(17)}$\cite{cummings2008,rowlinson2010a}
$^{(18)}$\cite{baumgartner2008,immler2008,covino2008}
$^{(19)}$\cite{ukwatta2009,depasquale2010,mcbreen2010}
$^{(20)}$\cite{krimm2009}
$^{(21)}$\cite{markwardt2009, levan2009, malesani2009}
$^{(22)}$\cite{sakamoto2011b}
\end{table*}

\begin{table*}
\centering
\begin{tabular}{cccccccc}
\hline
GRB     & z & T$_{90}$ & $\Gamma_{\gamma}$ & Fluence                  & Host & Host offset & Optical Afterglow\\
        &   & (s)      &          & (10$^{-7}$ erg cm$^{-2}$ s$^{-1}$) &          &(arcsec)    & \\
\hline
\multicolumn{8}{l|}{No breaks}\\
050509B$^{(24)}$  & 0.23   & 0.024$\pm$0.009 & 1.50$\pm$0.40   & 0.2$\pm$0.1 & y & 17.9$\pm$3.4  & N    \\
050813$^{(25)}$ & (0.72)   & 0.6$\pm$0.1     & 1.19$\pm$0.33 & 1.2$\pm$0.5   & n & -         & N   \\
050906$^{(26)}$ & (0.72)   & 0.13$\pm$0.02     & 1.91$\pm$0.42 & 0.6$\pm$0.3   & ? & -         &  N  \\
051105$^{(27)}$ & (0.72)   & 0.028$\pm$0.004     & 1.38$\pm$0.35 & 0.2$\pm$0.05   & ? & -         & N   \\
060502B$^{(28)}$ & (0.72)   & 0.09$\pm$0.02     & 0.92$\pm$0.23 & 0.4$\pm$0.05   & n & -         & N   \\
061217$^{(29)}$ & (0.72)   & 0.3$\pm$0.05     & 0.96$\pm$0.28 & 0.46$\pm$0.08   & ? & -         & N   \\
070209$^{(30)}$ & (0.72)   & 0.1$\pm$0.02     & 1.55$\pm$0.39 & 0.11$\pm$0.03   & n &  -        & N   \\
070429B$^{(31)}$ & (0.72)   & 0.5$\pm$0.1     & 1.71$\pm$0.23 & 0.63$\pm$0.1   & ? & -         & ?   \\
070729$^{(32)}$ & (0.72)   & 0.9$\pm$0.1     & 0.96$\pm$0.27 & 1$\pm$0.2   & ? & -         & N   \\
070810B$^{(33)}$ & (0.72)   & 0.08$\pm$0.01     & 1.44$\pm$0.37 & 0.12$\pm$0.03   & ? & -         & N   \\
071112B$^{(34)}$ & (0.72)   & 0.3$\pm$0.05     & 0.69$\pm$0.34 & 0.5$\pm$0.1   & n & -         & N   \\
080702A$^{(35)}$  & (0.72) & 0.5$\pm$0.2     & 1.34$\pm$0.42 & 0.4$\pm$0.1 & n & -             & N    \\
081024A$^{(36)}$  & (0.72)  & 1.8$\pm$0.6     & 1.23$\pm$0.21 & 1.2$\pm$0.2   & n & -             & N    \\
081101$^{(37)}$ & (0.72)   & 0.2$\pm$0.02     & 1.25$\pm$0.20 & 0.62$\pm$0.1   & n & -         & N   \\
081226$^{(38)}$ & (0.72)   & 0.4$\pm$0.1     & 1.36$\pm$0.29 & 1.0$\pm$0.2   & n & -         & N   \\
090305A$^{(39)}$ & (0.72)   & 0.4$\pm$0.1     & 0.86$\pm$0.33 & 0.8$\pm$0.1   & n & -         & Y   \\
100117A$^{(40)}$ & (0.72) & 0.30$\pm$0.05   & 0.88$\pm$0.22 & 0.9$\pm$0.1 & y & 0.6           & Y    \\
100206A$^{(41)}$ & (0.72)   & 0.12$\pm$0.03     & 0.63$\pm$0.17 & 1.4$\pm$0.2   & ? & -         & N   \\
100628A$^{(42)}$ & (0.72)   & 0.36$\pm$0.009     & 1.26$\pm$0.25 & 0.3$\pm$0.1   & y & -         & N   \\
110112A$^{(43)}$ & (0.72) & 0.5$\pm$0.1     & 2.14$\pm$0.46 & 0.3$\pm$0.1   & y & -             & Y   \\
111117A$^{(44)}$ & (0.72) & 0.47$\pm$0.09   & 0.65$\pm$0.22 & 1.4$\pm$0.2   & y & 1.00$\pm$0.13 & N   \\
\end{tabular}
\contcaption{} 
$^{(24)}$\cite{barthelmy2005,gehrels2005}
$^{(25)}$\cite{sato2005b}
$^{(26)}$\cite{parsons2005, levan2008}, note this is a candidate extra-galactic magnetar giant flare
$^{(27)}$\cite{cummings2005b, barbier2005, klose2005}
$^{(28)}$\cite{sato2006b}
$^{(29)}$\cite{parsons2006,ziaeepour2006}
$^{(30)}$\cite{sakamoto2007}
$^{(31)}$\cite{tueller2007, antonelli2007, holland2007}
$^{(32)}$\cite{sato2007b, berger2007}
$^{(33)}$\cite{sakamoto2007b, thone2007}
$^{(34)}$\cite{fenimore2007}
$^{(35)}$\cite{krimm2008}
$^{(36)}$\cite{barthelmy2008}
$^{(37)}$\cite{barthelmy2008b}
$^{(38)}$Automated BAT analysis products
$^{(39)}$\cite{krimm2009b, cenko2009}
$^{(40)}$\cite{markwardt2010,levan2010,fong2010}
$^{(41)}$\cite{sakamoto2010, miller2010, levan2010b, berger2010}
$^{(42)}$\cite{barthelmy2010b, starling2010, berger2010b}
$^{(43)}$\cite{barthelmy2011,levan2011}
$^{(44)}$\cite{sakamoto2011,cenko2011,berger2011}
\end{table*}

\begin{figure*}
\centering
\includegraphics[width=7.5cm]{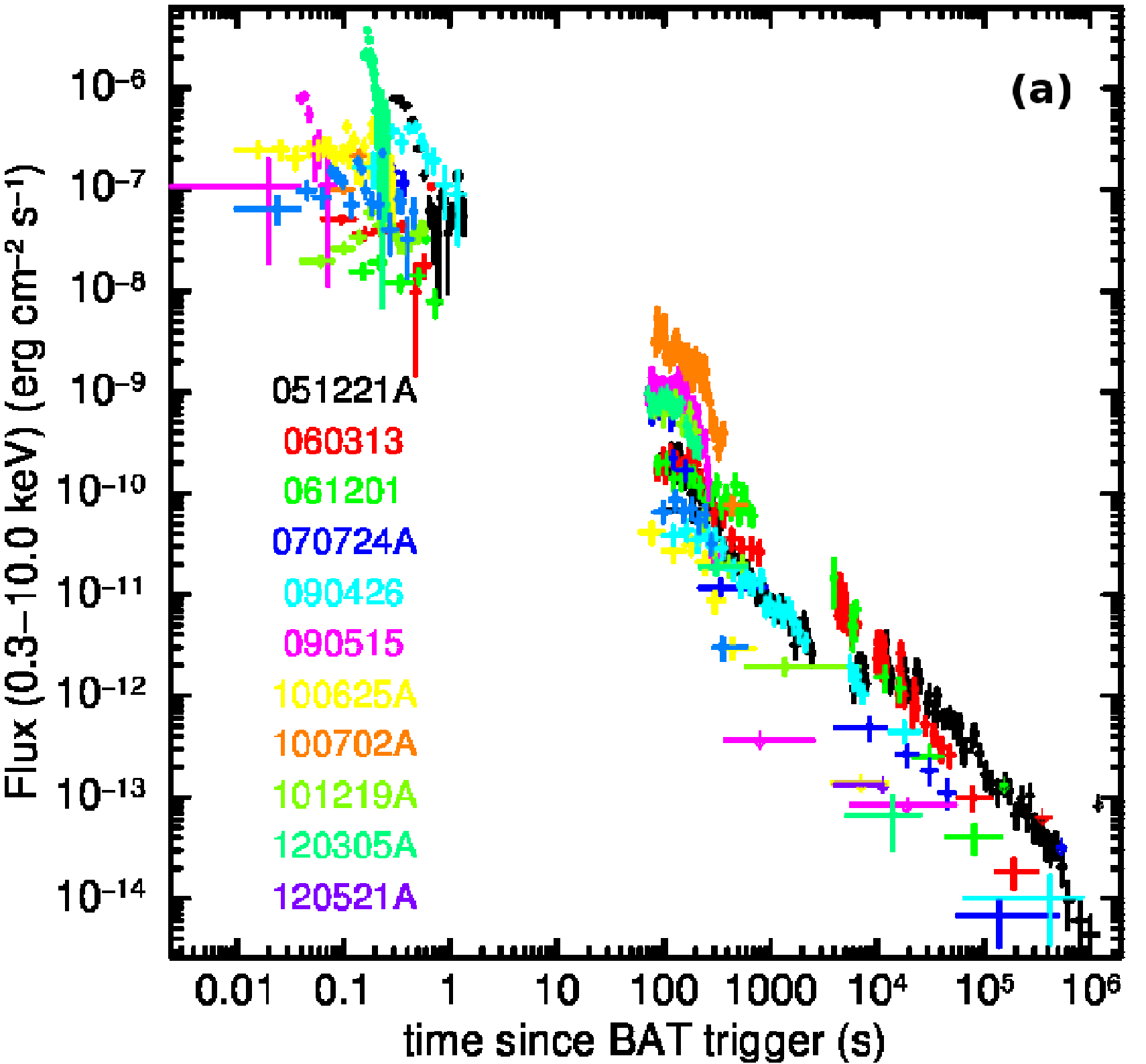}
\includegraphics[width=7.5cm]{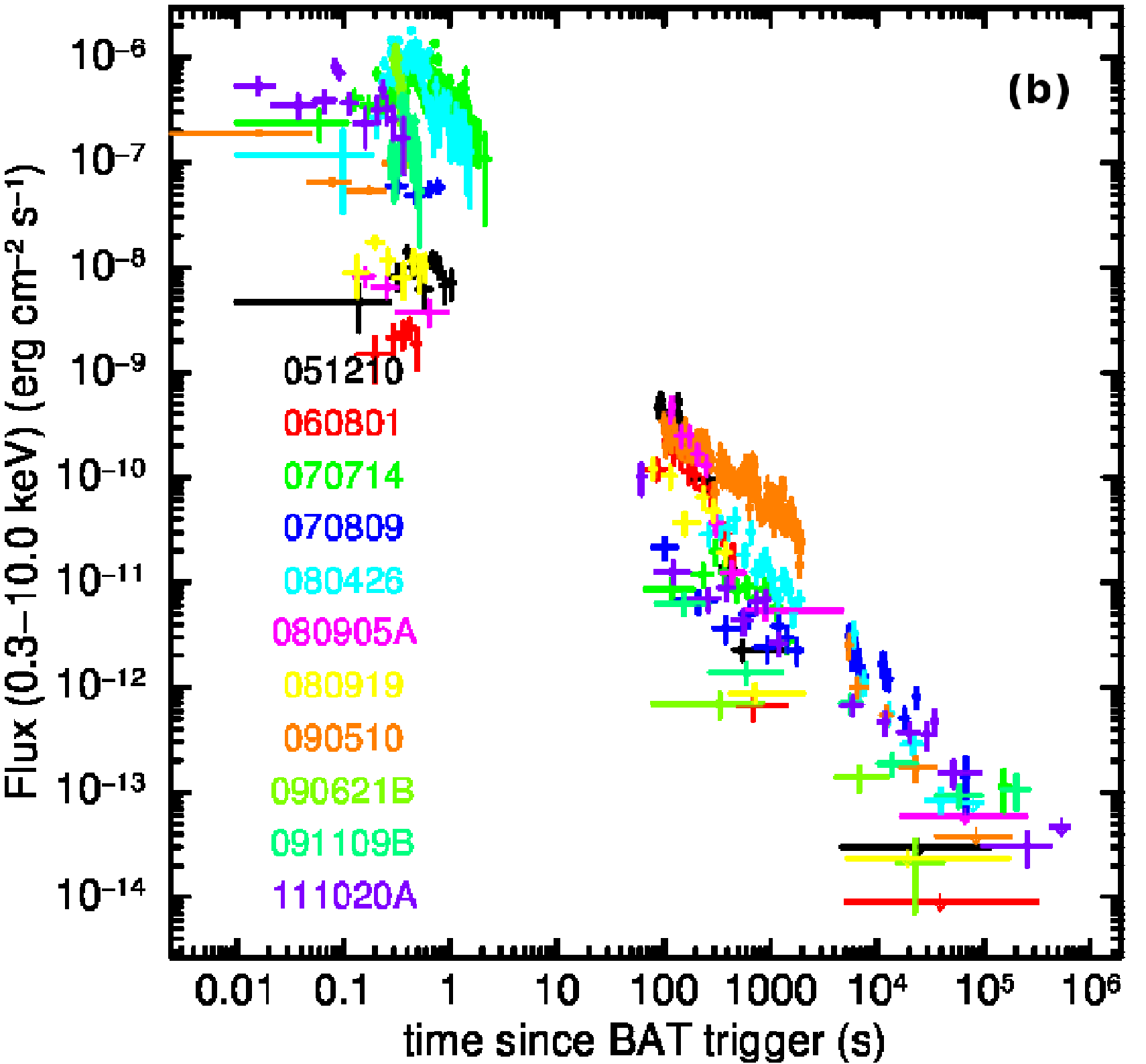}
\includegraphics[width=7.5cm]{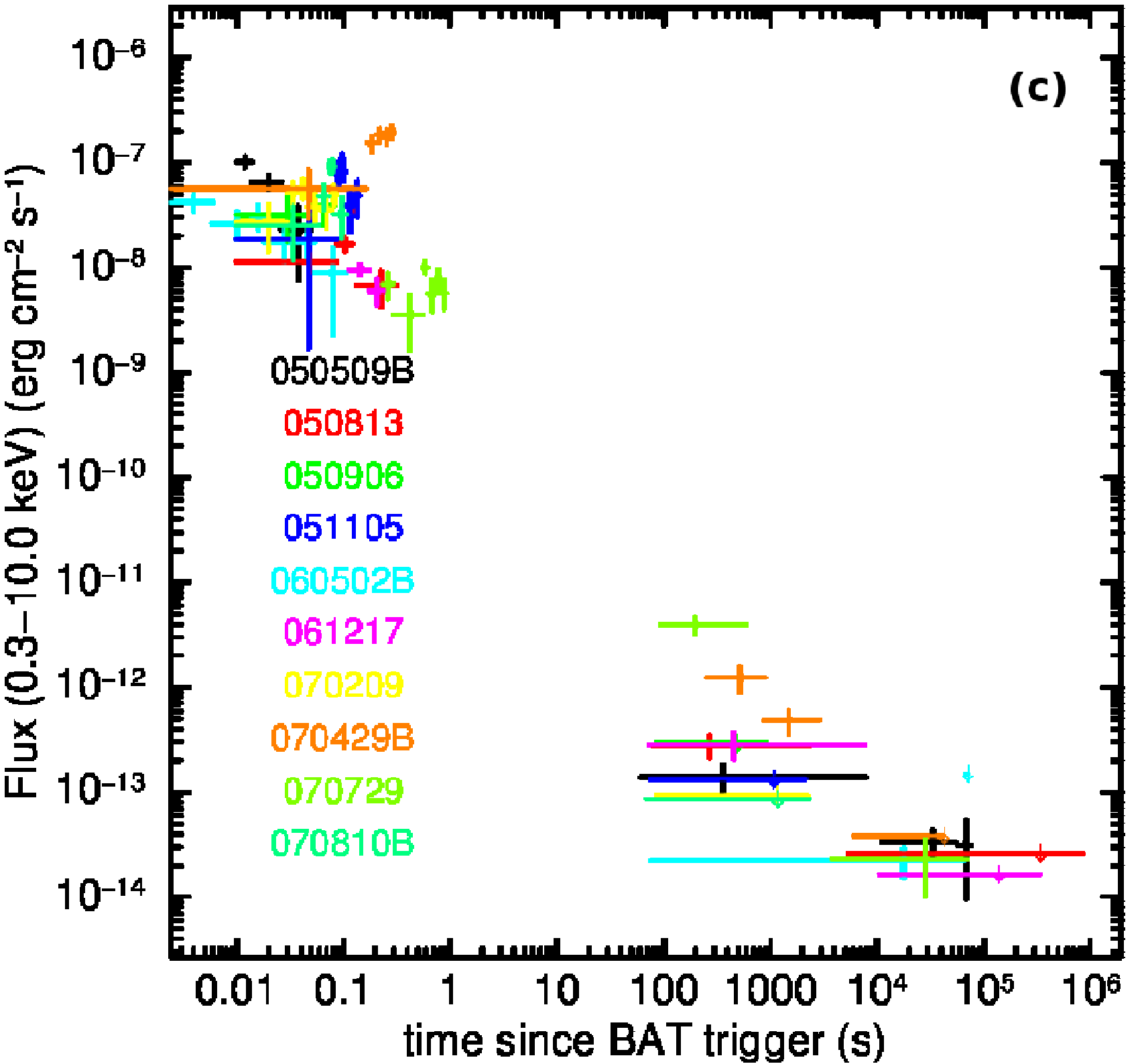}
\includegraphics[width=7.5cm]{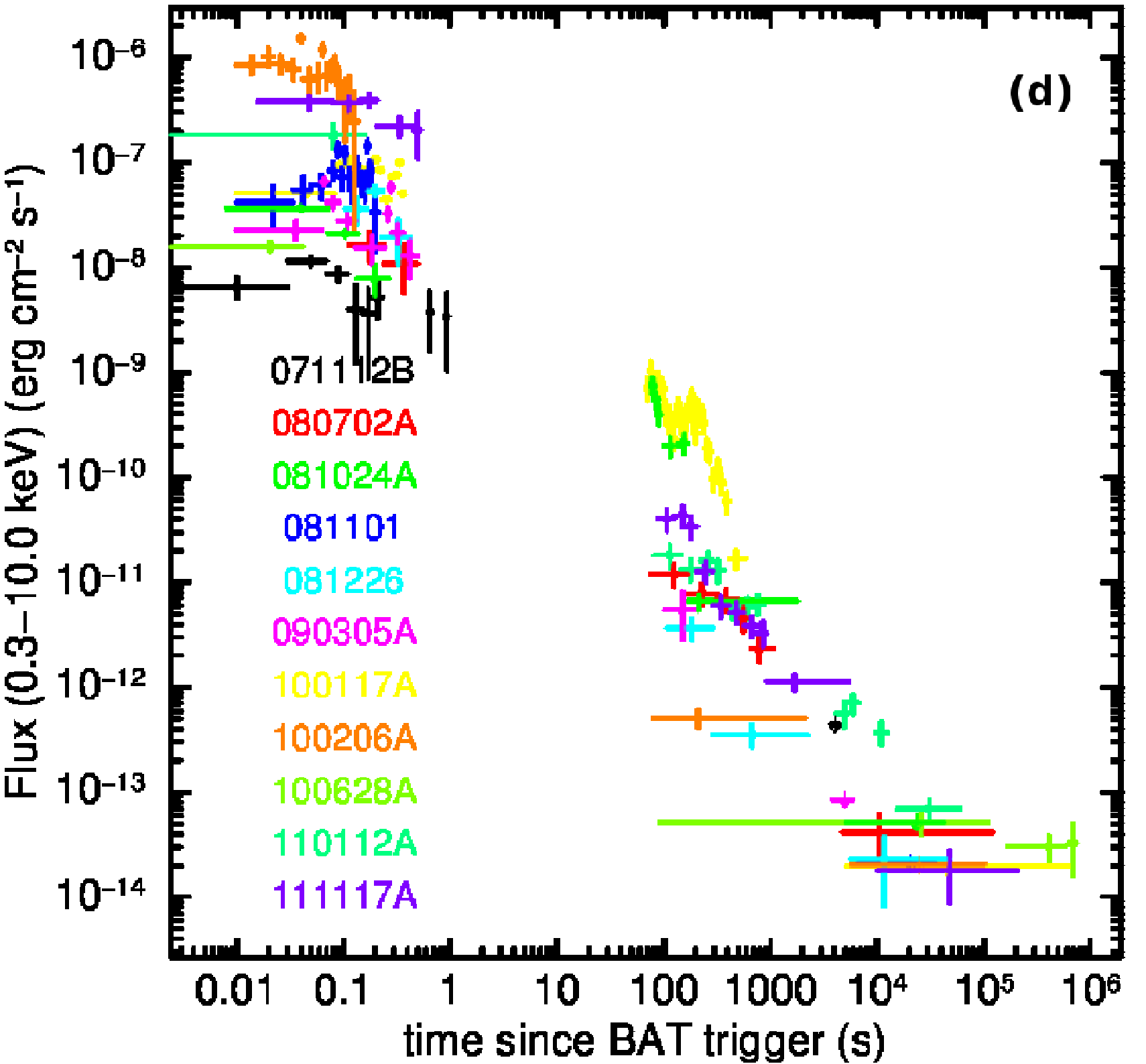}
\caption[BAT-XRT lightcurves for SGRB sample]{These are the BAT-XRT lightcurves (0.3 -- 10 keV, observed flux) sorted into 3 groups. a) These GRBs have 2 or more breaks in their lightcurve. b) GRBs with 1 break in their lightcurve. c) and d) GRBs with no significant breaks in their lightcurve.}
\label{fig0.1}
\end{figure*}

\begin{figure}
\centering
\includegraphics[width=7.5cm]{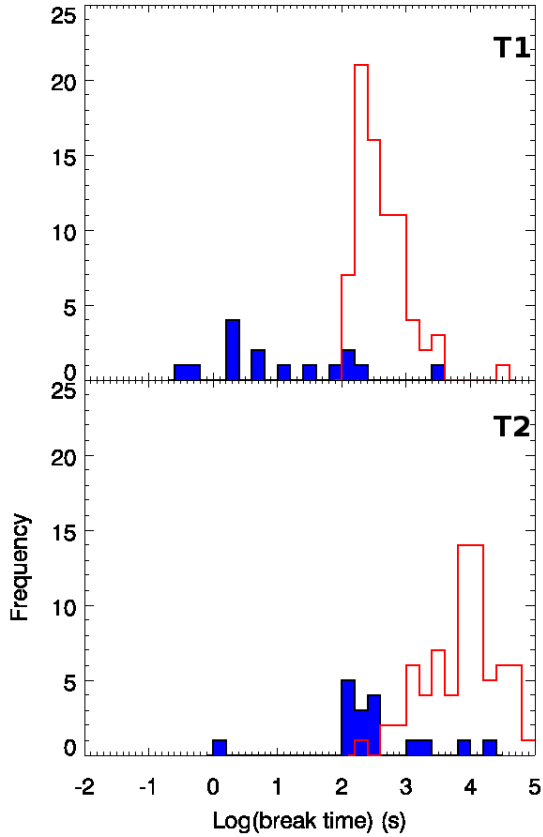}
\caption[Break times of ``canonical'' like SGRB lightcurves]{Histograms showing the break times for the SGRB lightcurves with a plateau phase. T$_{1}$ is the break from the steep decay phase to the plateau phase while T$_{2}$ marks the end of the plateau. The blue filled histograms correspond to the SGRB sample used in this Paper and overplotted in red are the LGRB values determined by \citet{evans2009}.}
\label{fig0.1c}
\end{figure}

\begin{figure}
\centering
\includegraphics[width=7.5cm]{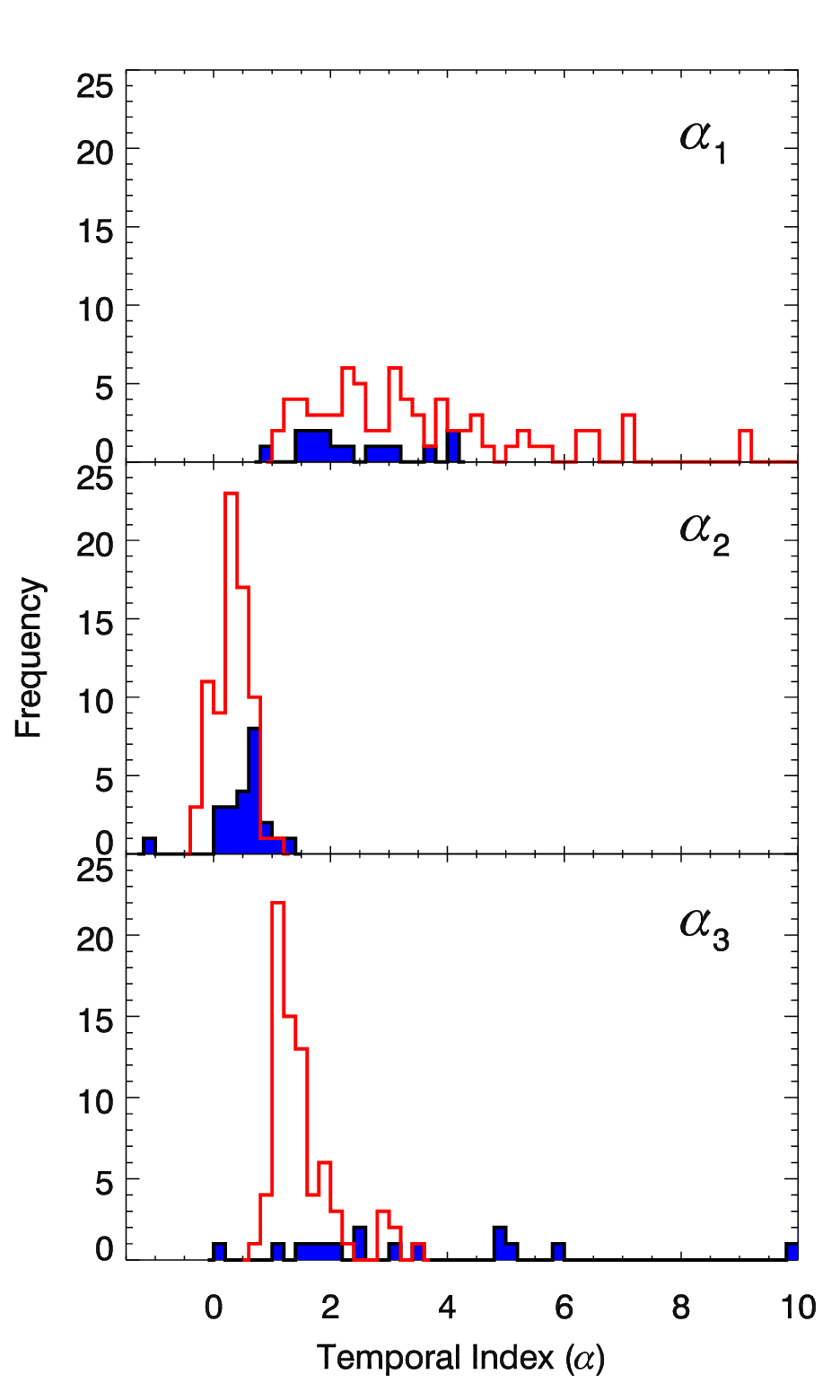}
\caption{Histograms showing the temporal indices of the SGRB lightcurves with a plateau phase. $\alpha_1$ is the initial steep decay phase from the last decay in the prompt emission. $\alpha_2$ are the plateau and shallow decay phase slopes. $\alpha_3$ is the final afterglow decay slope. The filled histograms correspond to the SGRB sample used in this paper and overplotted are the LGRB values determined by \citet{evans2009}.}
\label{fig0.1b}
\end{figure}

Following the launch of the {\it Swift} satellite \citep{gehrels2004}, it has been possible to place tighter constraints on the nature of short gamma-ray bursts (SGRBs). The detection of their faint and rapidly fading X-ray afterglows has led to the identification of optical afterglows and in many cases candidate host galaxies \citep[for example GRB 050509B, ][]{gehrels2005, hjorth2005}. These observations have provided significant support for the popular compact binary merger progenitor theories, i.e. the coalescence of two neutron stars (NS) or a NS and a black hole (BH) \citep{lattimer1976, eichler1989,narayan1992}. However, without the coincident observation of gravitational waves by observatories like LIGO (Laser Interferometry Gravitational-wave Observatory) we are missing the supporting ``smoking gun'' observation for this progenitor theory. 

Observed features in X-ray lightcurves suggest longevity of the central engine of GRBs, for example late time flares \citep[e.g. ][]{curran2008, margutti2010, bernardini2011} and plateaus \citep[e.g. ][]{nousek2006, zhang2006}. GRBs whose X-ray lightcurves have a steep decay and a plateau phase followed by a standard afterglow phase, have been identified as ``canonical'' lightcurves \citep{nousek2006, obrien2006, zhang2006, evans2009}. The steep decay phase is associated with high latitude emission from the prompt emission followed by a late emission plateau giving the plateau phase \citep{tagliaferri2005, goad2006}. The fluence of this plateau can be comparable to the fluence of the prompt emission \citep{obrien2006, margutti2012}, and typically they occur from $10^2$ -- $10^3$ s till $10^3$ -- $10^4$ s after the trigger time. The plateau is thought to provide evidence of ongoing central engine activity \citep{nousek2006,zhang2006}. \cite{evans2009} studied 162 GRBs in the {\it Swift} sample identifying a ``canonical'' lightcurve in 42\% of GRB X-ray lightcurves, including 2 (051221A and 060313) out of 11 SGRBs analysed.

Although studies of flares and plateaus are typically conducted for LGRBs, fainter versions are evident in many SGRB X-ray lightcurves suggesting a long lived central engine \citep[e.g.][]{margutti2011}. This is problematic for SGRB progenitor theories as accretion is expected to end within a few seconds \citep[powering relativistic jets;][]{rezzolla2011} and only a small fraction of the merger mass is available \citep[0.01 -- 0.1 M$_{\odot}$ although this is dependant on the NS equation of state, ][]{lee2007}. Additionally, it is thought that the accretion disk gets destroyed after a few seconds \citep[e.g.][]{metzger2008}. There have been studies of fallback accretion, in which the NS is shredded and parts ($\le$ 10\% of the original disk mass) are flung into highly eccentric orbits which accrete onto the central engine at late times giving flares in the X-ray lightcurve \citep{rosswog2007}. Flares may also be caused by Toomre instabilities within the accretion disk \citep{perna2006}, although this does not explain plateau emission or late time flares as the SGRB accretion disks are expected to accrete within the first few seconds. \cite{cannizzo2011} have attempted to explain plateaus by introducing a band of material at a large distance from the central engine. \cite{cannizzo2011} suggest that the required reservoir of material could be provided via the accretion disk moving outwards (due to having a large amount of angular momentum) or ejecta thrown out during the merger in highly eccentric orbits that circularises forming an accretion disk.

An alternative theory is that during some GRBs a millisecond pulsar (magnetar) may be formed with enough rotational energy to prevent gravitational collapse \citep{usov1992, duncan1992, dai1998a, dai1998b, zhang2001}. The rotational energy is released as gravitational waves and electromagnetic radiation, causing the magnetar to spin down. If the magnetar is sufficiently massive it may reach a critical point at which differential rotation is no longer able to support it, resulting in collapse to a BH. Assuming constant radiative efficiency, the energy injection from the magnetar would produce a plateau in the X-ray light curve \citep{zhang2001} and would be followed by a steep decay if the magnetar collapses to a BH. The progenitor of this system is typically thought to be a collapsar and LGRB candidates have been identified by \cite{troja2007} and \cite{lyons2009}. However, it has also been proposed that such a magnetar could be formed by the merger of two neutron stars \citep{dai1998a, dai2006, yu2007} or via the accretion induced collapse (AIC) of a white dwarf (WD) \citep{nomoto1991, usov1992, levan2006, metzger2008b}. A candidate event for this is GRB 090515 with an unusual X-ray plateau followed by a steep decay \citep{rowlinson2010b}. The likelihood of producing this event is dependent on the equation of state of neutron stars. \cite{morrison2004} studied the effect that the equation of state of a NS and rotation would have on the remnant of a compact merger, i.e. whether a NS or a BH is formed \citep[see also][]{shibata2006}. They showed that, even for the harder nuclear equations of state, the rotation of the NS could increase the maximum mass by $\sim50\%$ and hence mergers could often result in a NS. Considering the parameters of 6 known Galactic NS binaries and a range of equations of state, \cite{morrison2004} predict that the majority of mergers of the known binaries will form a NS.

The recent discovery of an 1.97 M$_{\odot}$ NS \citep{demorest2010} provides further supporting evidence of the possibility that high mass magnetars can be formed from NS mergers (the maximum mass of NSs is dependent on the very uncertain NS equation of state, so this is a conservative lower limit on the maximum mass of a NS). \cite{ozel2010b} show that, for a maximum non-rotating NS mass of $M_{\rm max} =$2.1 M$_{\odot}$, the merger of two NSs with a total mass $\le 1.4M_{\rm max}$ will have a delayed collapse to a BH (i.e. a magnetar phase). They also predict a regime in which the merged remnant does not collapse to form a BH, in this case the total mass is $\le 1.2M_{\rm max}$. If the maximum NS mass is 2.1 M$_{\odot}$, then the merger of two NSs of masses up to 1.3 M$_{\odot}$ would result in a stable magnetar and the merger of two NSs with larger masses (up to 1.5-1.7 M$_{\odot}$) would form an unstable magnetar. As the majority of observed NSs have masses $\sim$1.4 M$_{\odot}$, it seems reasonable to predict that many NS mergers could result in a magnetar. The stability of the final magnetar is dependent on the maximum possible mass of a NS which is still uncertain. Its lifetime depends both on the rate that additional mass (if any) is accreted after formation, as well as the rate at which angular momentum is extracted by e.g. gravitational waves or magnetic torques \citep[e.g.][]{shibata2006, oechslin2007}.

In this paper, we consider all {\it Swift} detected SGRBs, T$_{90} \le$ 2 s, observed until May 2012 with an X-ray afterglow or which were promptly slewed to and observed by the X-ray Telescope \citep[XRT;][]{burrows2005}. This allows the inclusion of SGRBs without an X-ray afterglow but which do have a constraining upper limit. For all the SGRBs, we analysed the BAT \citep[Burst Alert Telescope;][]{barthelmy2005b} data by creating lightcurves using a variety of binning in signal-to-noise ratios and time looking for evidence of extended emission at the 3$\sigma$ level where we consistently saw extended emission over more than 30 s (the SGRBs with identified extended emission are 050724, 050911, 051227, 060614, 061006, 061210, 070714B, 071227, 080123, 080503, 090531B, 090715A, 090916 and 111121A). This procedure recovered all of the extended emission bursts identified by \cite{norris2010}. Hence, our selection criteria excludes SGRBs with extended emission, which may share a common progenitor to SGRBs but this remains uncertain. This sample is used to identify those with a plateau phase in their lightcurves suggesting ongoing central engine activity. These results are discussed in section 2. A sub-sample with sufficient data are then studied for the signature of a magnetar (with or without collapse to a BH) which may signify the coalescence of two NSs. If found, this would provide additional support to this popular progenitor theory although forming a magnetar via the AIC of a WD is not ruled out. The magnetar model is considered in section 3; with a description of the model and sample used and analysis of the available data. A discussion of the implications, e.g. for gravitational waves, is given in section 4 and our conclusions are given in section 5. Throughout this work, we adopt a cosmology with $H_0 =  71$ km\,s$^{-1}$\,Mpc$^{-1}$, $\Omega_m = 0.27$, $\Omega_\Lambda = 0.73$. Errors are quoted at 90\% confidence for X-ray data and at 1$\sigma$ for fits to the magnetar model.

\section{Plateau phases in SGRB lightcurves}

\begin{figure}
\centering
\includegraphics[width=7.5cm]{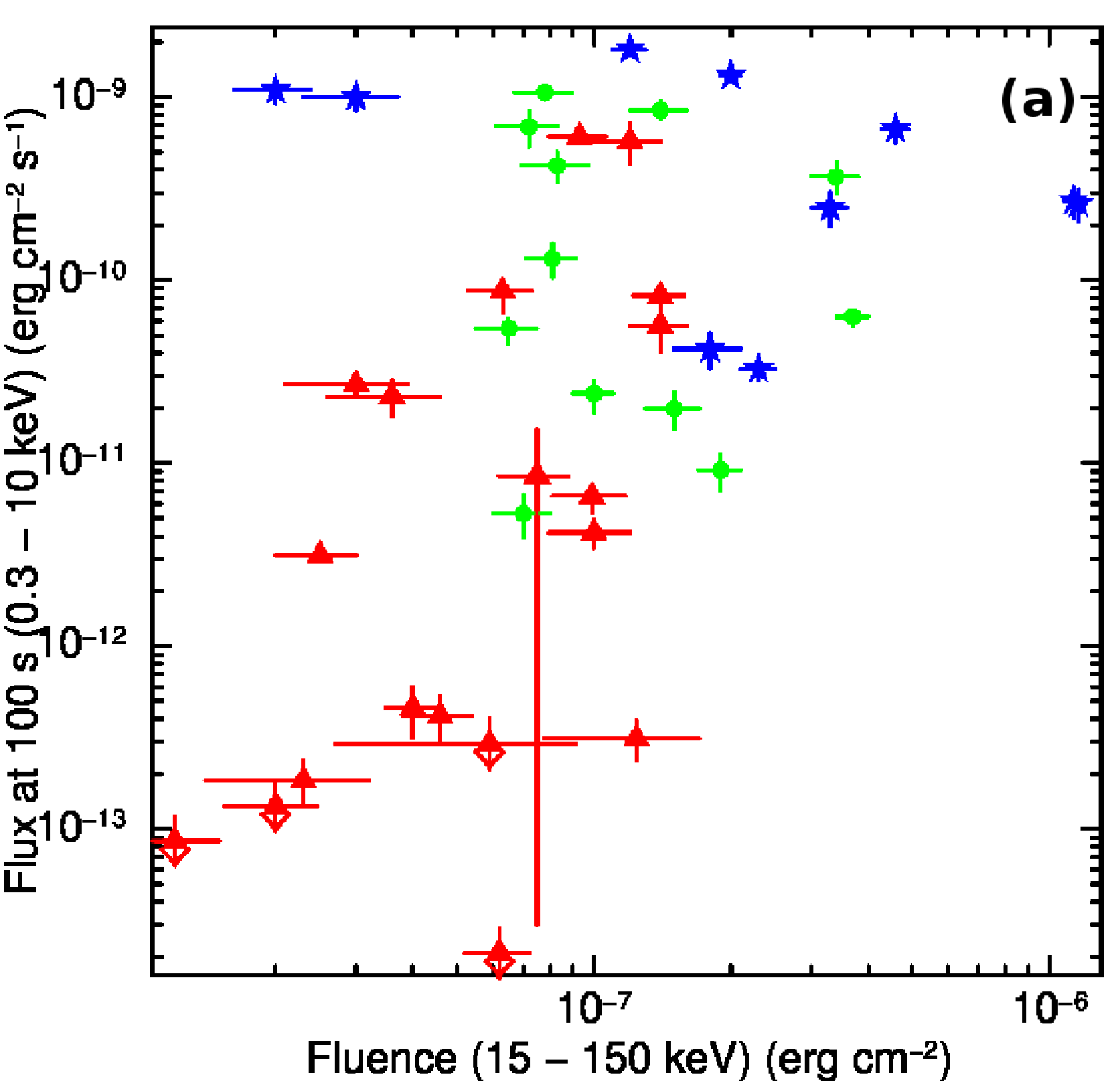}
\includegraphics[width=7.5cm]{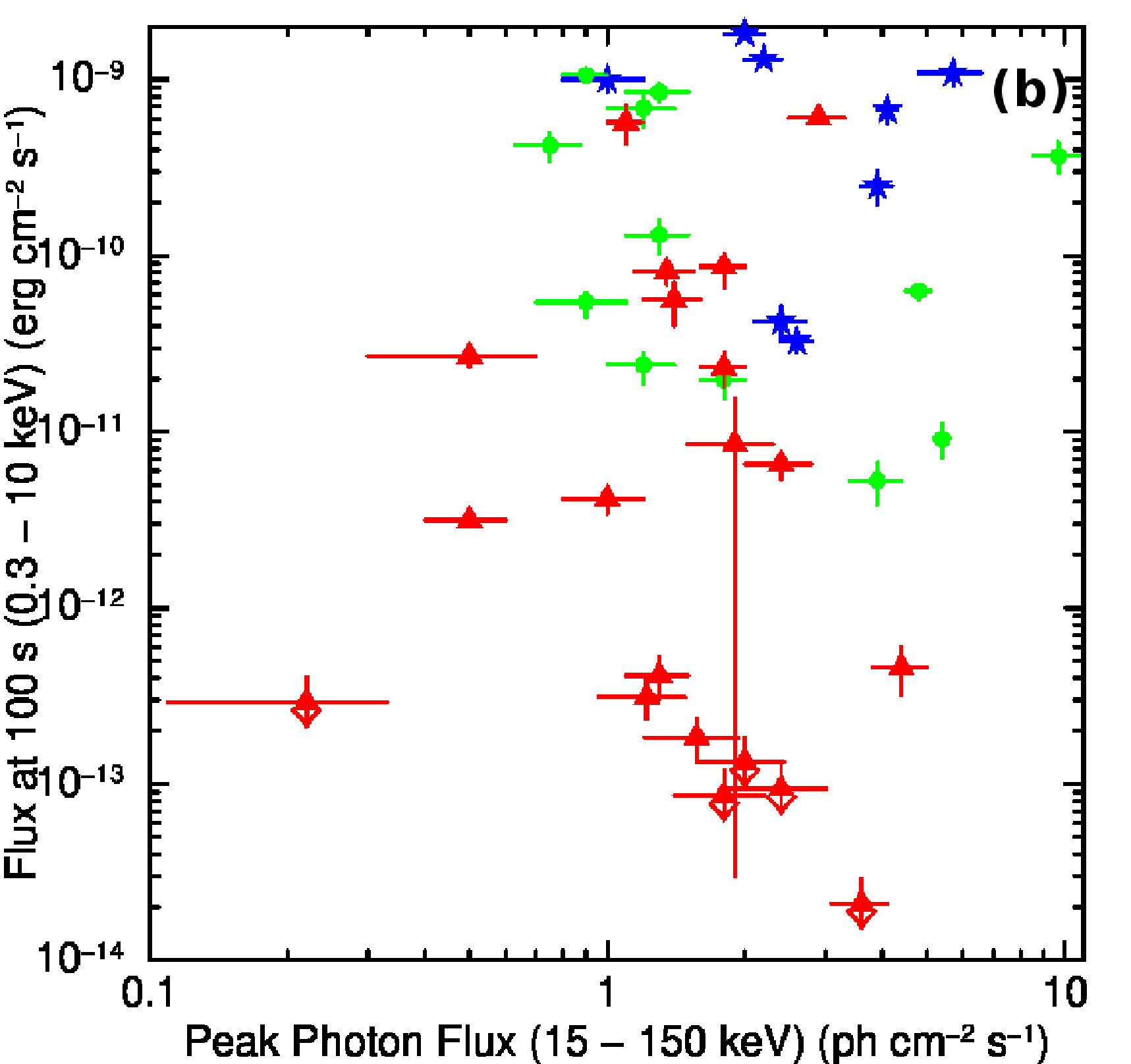}
\caption{(a) The BAT fluence (15 -- 150 keV) plotted against the XRT unabsorbed flux at 100 s (0.3 -- 10 keV). Blue stars have 2 or more significant breaks in their lightcurves, green circles have 1 break and red triangles have no significant breaks in their lightcurves. (b) The BAT peak photon flux (15-150 keV) against the XRT unabsorbed flux at 100 s in the observer frame (0.3 -- 10 keV). Symbols are as in (a).}
\label{fig0.2}
\end{figure}

\begin{figure}
\centering
\includegraphics[width=7.5cm]{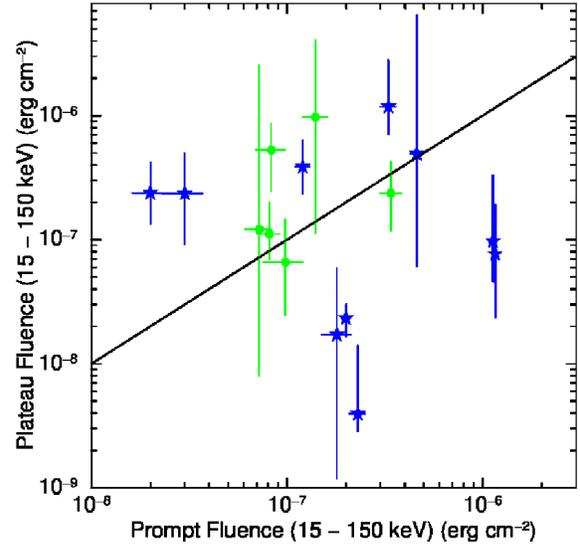}
\caption{The prompt BAT 15 -- 150 keV fluence in comparison to the shallow decay phase unabsorbed X-ray fluence extrapolated to the 15 -- 150 keV energy band. Symbols are as defined in Figure \ref{fig0.2} and the black line shows where the shallow decay phase fluence is equal to the prompt fluence.}
\label{fig0.8}
\end{figure}

\begin{figure}
\centering
\includegraphics[width=7cm]{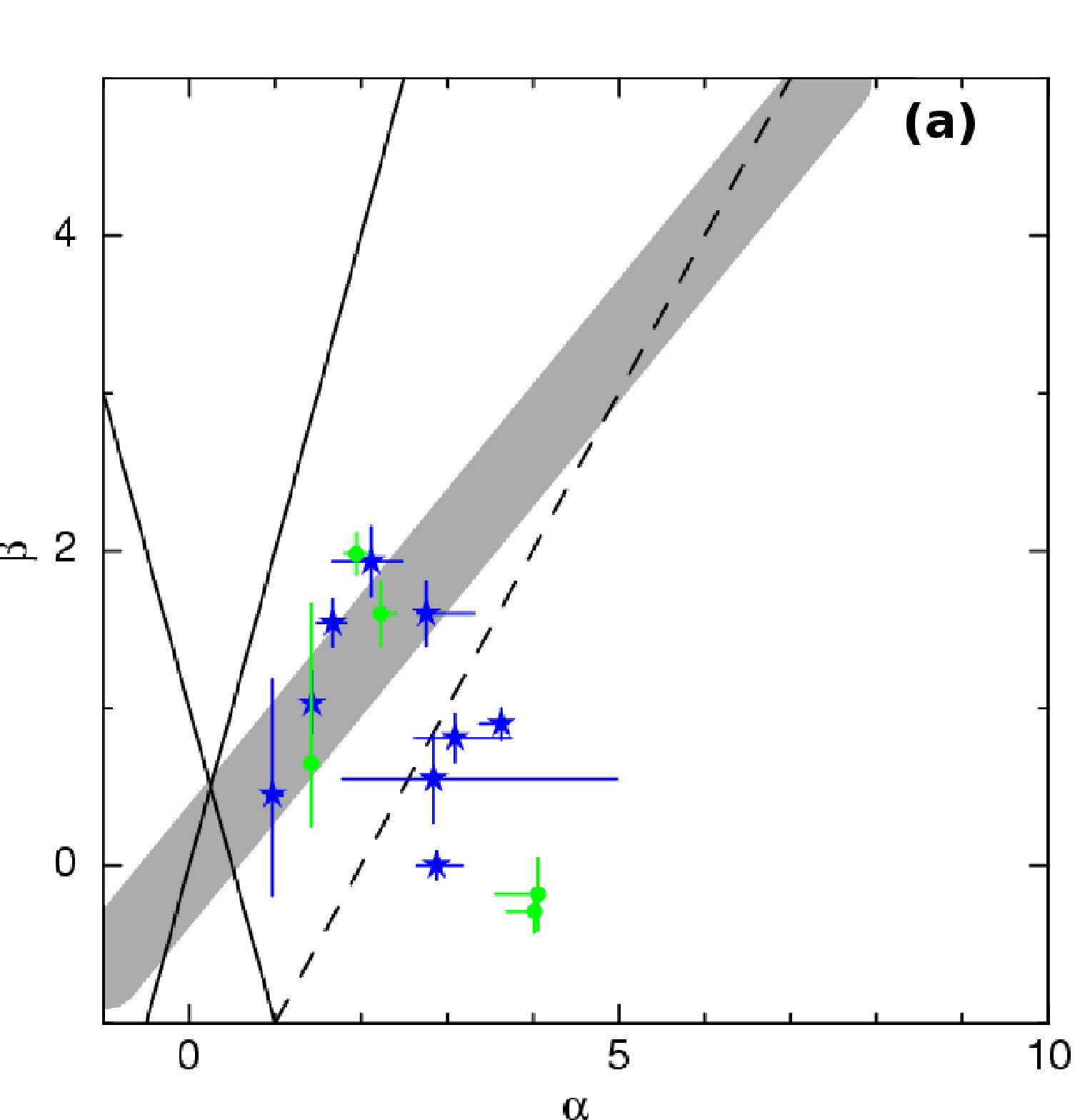}
\includegraphics[width=7cm]{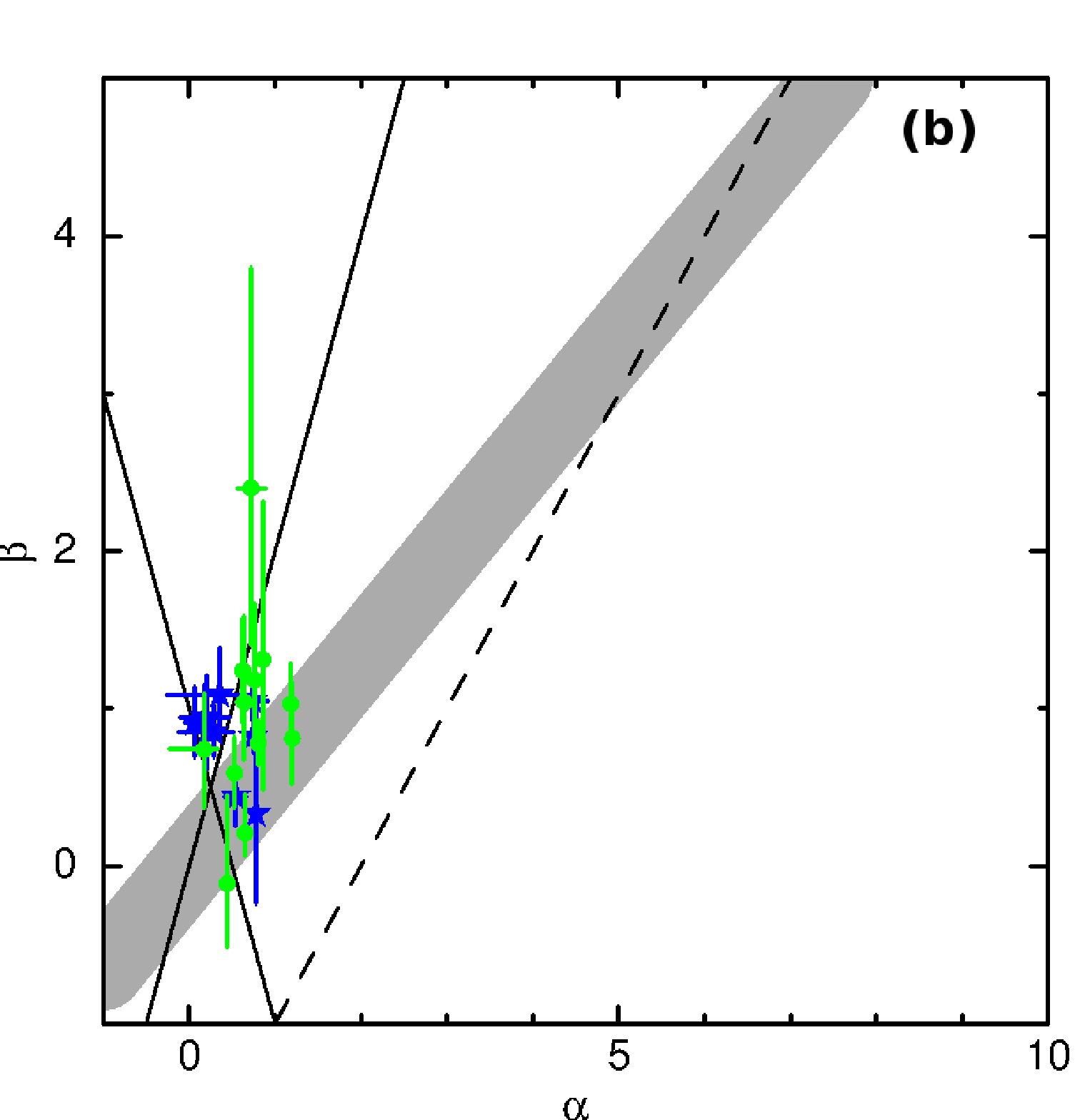}
\includegraphics[width=7cm]{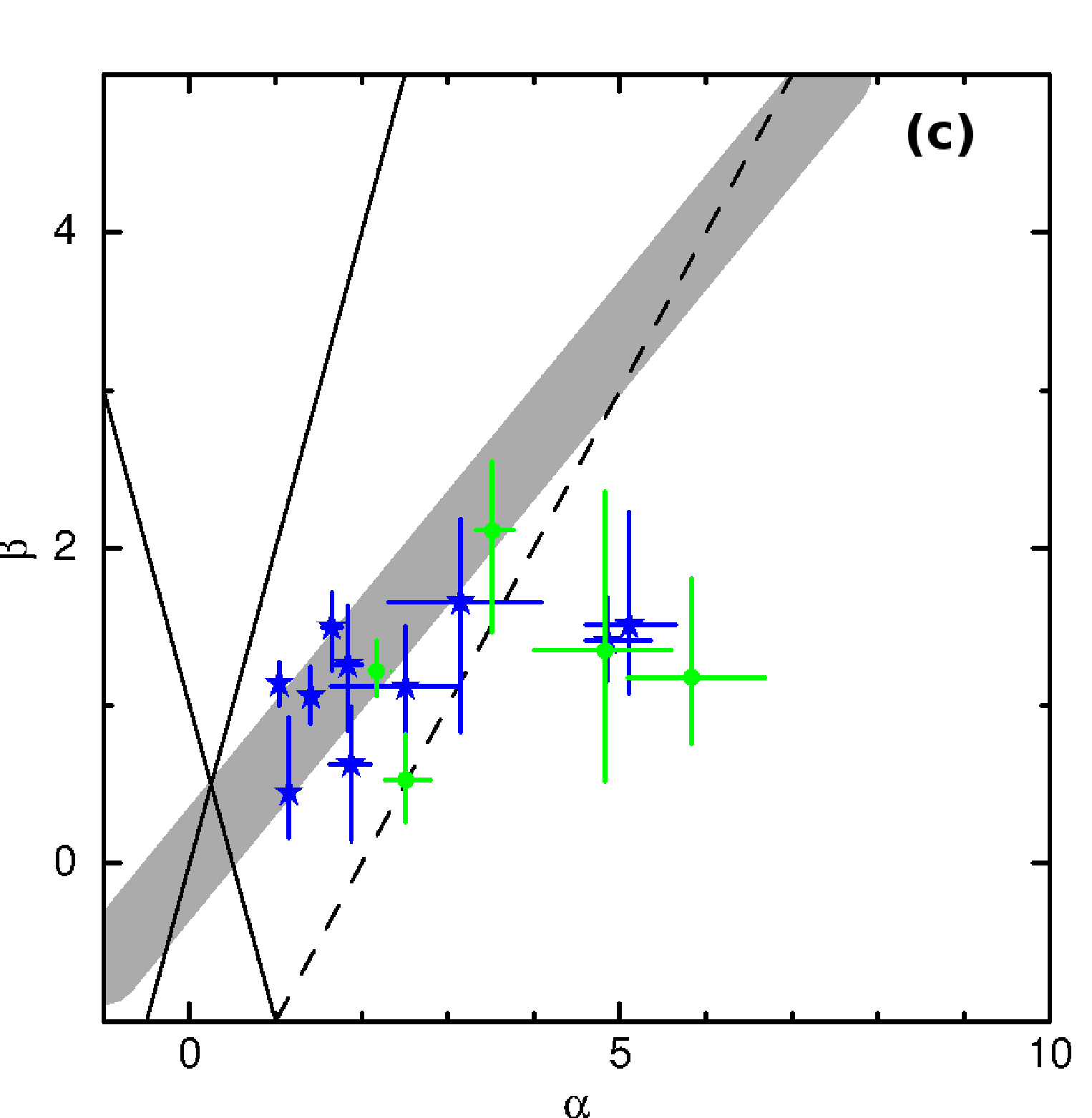}
\caption{The spectral index $\beta$ versus the temporal index $\alpha$ for the three regimes of the lightcurves with a plateau phase: (a) steep decay phase, (b) plateau phase and (c) standard afterglow phase. Where there is no XRT spectrum available for the steep decay phase, the BAT spectrum is used. All symbols are as defined in Figure \ref{fig0.2}, the solid lines and grey regions show the closure relations as defined by \citet{zhang2004}, and the black dashed line shows where $\alpha = \beta + 2$.}
\label{fig0.9}
\end{figure}

\begin{figure}
\centering
\includegraphics[width=7.5cm]{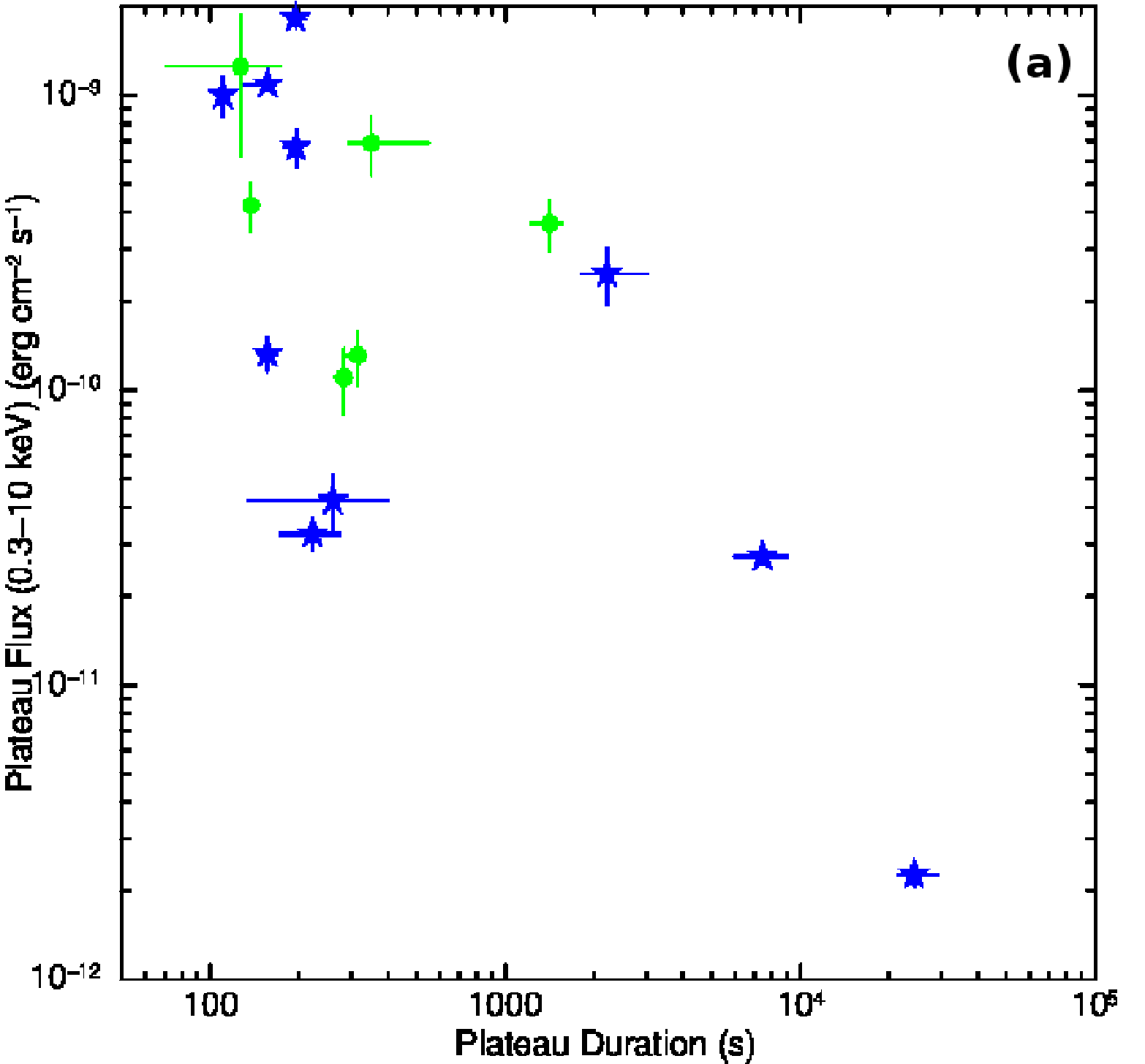}
\includegraphics[width=7.5cm]{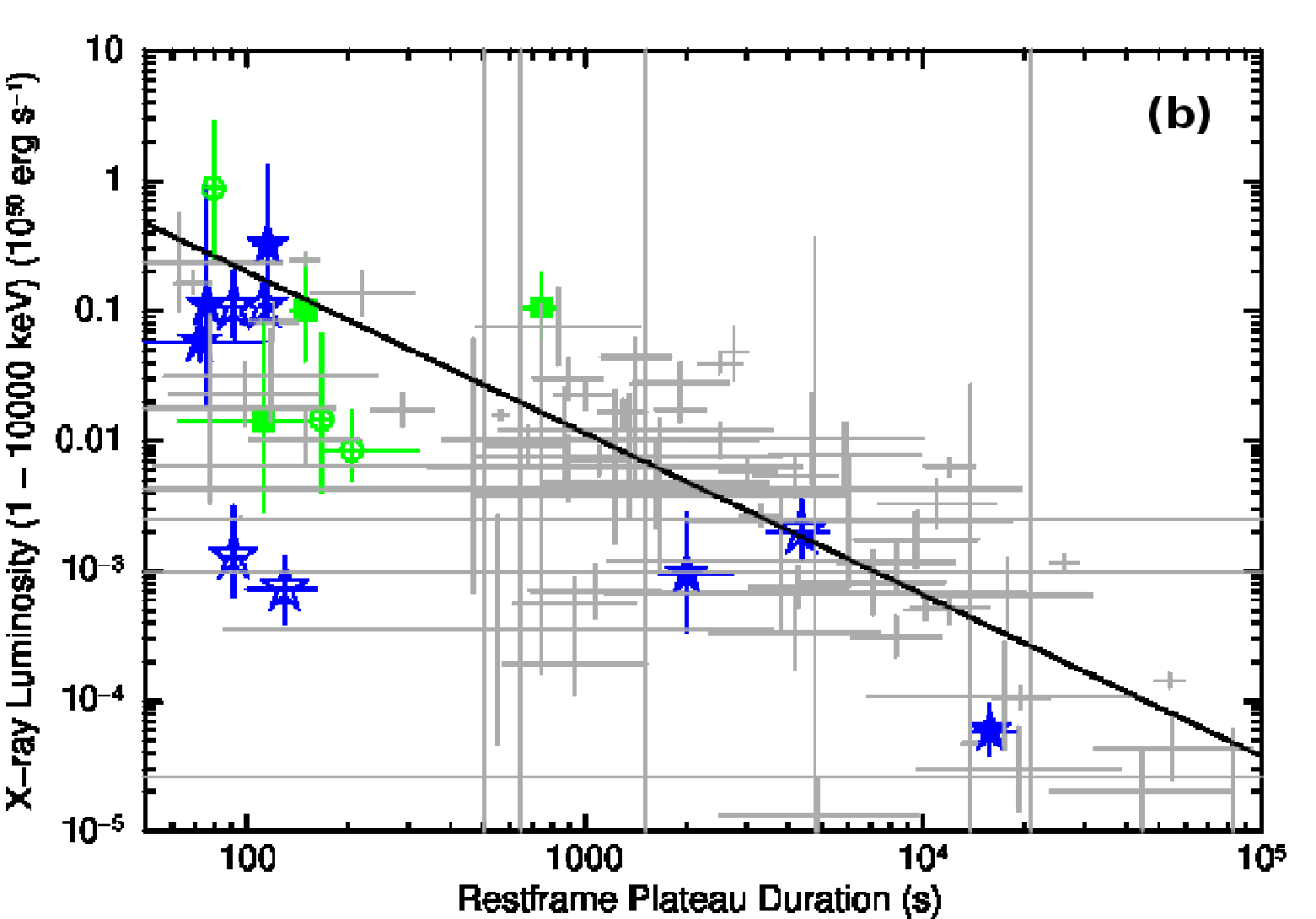}
\caption{(a) The plateau phase unabsorbed flux versus the duration of this phase. Symbols are as defined in Figure \ref{fig0.2}. (b) The plateau phase luminosity, using published redshifts (filled symbols) or the average redshift (open symbols), versus the restframe duration of this phase. The light grey data points are the \citet{dainotti2010} sample of LGRBs. The black line shows the correlation between the luminosity and duration for the SGRB and LGRB samples which is consistent with the relationship found by \citet{dainotti2010}.}
\label{fig0.7}
\end{figure}

Out of our sample of 43 SGRBs, shown in Table \ref{lcfits}, only 6 did not have a detected X-ray afterglow (GRBs 050906, 051105, 070209, 070810B 071112B and 081101). Hence, $\sim$86\% of {\it Swift} SGRBs with a prompt slew have detected X-ray afterglows. The observed properties of the SGRB sample are given in Table \ref{candidates}.

The 0.3 -- 10 keV observed flux X-ray lightcurves were obtained from the automated analysis page for each individual SGRB from the UK {\it Swift} Science Data Centre website \citep{evans2007, evans2009}. The Burst Alert Telescope (BAT) lightcurves were created using standard pipelines in the {\sc Heasoft} package with 3$\sigma$ significance bins. The 15 -- 150 keV BAT spectra were fitted in {\sc XSpec} for each SGRB and then extrapolated to obtain the flux at 0.3 -- 10 keV. Using this extrapolated flux and the net count-rate in the BAT spectrum, each count-rate data point in the BAT lightcurve was scaled to an 0.3 -- 10 keV flux using a simple power law spectral model. These were combined with the XRT lightcurves to make the BAT-XRT lightcurves used in this analysis. For later comparison with the magnetar model, these lightcurves were converted into unabsorbed flux lightcurves and then into restframe 1-10000 keV luminosity lightcurves using a k-correction \citep{bloom2001} giving an approximation to a bolometric lightcurve. The range of k-corrections obtained are typically consistent with the range 0.4-7 as obtained by \cite{bloom2001}. However, there are a small number of large k-corrections (particularly for GRBs 070809, 080905A and 101219A), showing the spectrum may be poorly understood at the frequency range we extrapolate to. If no redshift is known the mean SGRB redshift is used, $z\sim0.72$ \citep[excluding the redshift for GRB 061201 as the host galaxy association remains uncertain;][Tunnicliffe et al. in prep]{stratta2007}, and the implications of choosing this average redshift are discussed in Section 3.3. All the SGRB observed BAT-XRT lightcurves were fitted with multiple power laws from the final decay phase in the BAT prompt emission throughout the total X-ray afterglow using {\sc QDP}\footnote{https://heasarc.gsfc.nasa.gov/docs/software/ftools/others/qdp/qdp.html}. These fits were then used to identify those with a ``canonical'' like lightcurve. An XRT spectrum was created for each region of the lightcurve using the automatic data products on the UK {\it Swift} Science Data Centre website \citep{evans2007, evans2009}. The SGRB lightcurves are shown in Figure \ref{fig0.1}. 

We assume that $F_{\nu} \propto \nu^{-\beta} t^{-\alpha}$ where $\beta=\Gamma-1$ is the spectral index, $\Gamma$ is the photon index ($\Gamma_{\gamma}$ is the photon index measured using BAT and $\Gamma_{x}$ is the photon index measured using XRT) and $\alpha$ is the temporal index. We define the steep decay phase following the prompt emission to have a power law decay of $\alpha_{1}$, after which the decay can break to a decay of $\alpha_{2}$ and a further break to $\alpha_{3}$. We always define $\alpha_{2}$ to be the shallowest decay phase, as this allows the direct comparison of all plateau phases in the subsequent analysis. All SGRBs with 1 or more breaks in their lightcurves have a plateau phase. In three cases there are more than two breaks in the lightcurve. GRB 070724 has a third break at $T_3 = 152^{+18}_{-5}$ ($\alpha_{4} = 1.15^{+0.07}_{-0.06}$ and $\Gamma_{4} = 1.45^{+0.48}_{-0.29}$), GRB 090515 has a third break at $T_4=241^{+8}_{-10}$ ($\alpha_4=10^{+0}_{-0.97}$) and GRB 101219A has a break at $T_3 = 241^{+15}_{-13}$ ($\alpha_3 = 1.88^{+0.23}_{-0.25}$ and $\Gamma_4 = 1.63^{+0.37}_{-0.49}$). The 6 GRBs which were undetected by XRT are fitted with lower limits for $\alpha_{1}$ using the shallowest decay allowed by the BAT data and the XRT upper limit. In Table \ref{lcfits}, we provide the lightcurve fits for all the SGRBs in the sample. An F-test was conducted using the $\chi^2$ and degrees of freedom for each fit to determine the probability that the fit is a chance improvement on a simpler model (i.e. an F-test between the model provided in Table \ref{lcfits} and a model with 1 less break in the lightcurve). We utilise the method described in \cite{evans2009} to determine the best fit, i.e. the model that has the most breaks and the probability of being a chance improvement on a simpler model as $\le$0.3\%.

There are several caveats which need to be considered with the results in this Section and for the magnetar fits in Section 5.3.2. As SGRB afterglows are often faint and fade rapidly, these lightcurves and spectra can be poorly sampled giving large errors on the values in Table \ref{lcfits}. This could also cause breaks in the lightcurve to be missed due to large bin sizes \citep[bins typically contain 20 photons in PC mode data so bins could have long durations;][]{evans2007}. Additionally, the {\it Swift} satellite slews to observe GRBs after detection, leading to a characteristic gap between the BAT data and the XRT data, and XRT can only observe for short windows, due to Earth occultation during orbits, giving further gaps in the lightcurves which could also hide features in the lightcurves.

Using the broken powerlaw fit method, we find that 22 SGRBs ($\sim$50\%) are consistent with having a plateau phase in their lightcurves, although the plateau phase is not always directly observed due to the gap in the lightcurve prior to the XRT observations. It is hard to rule out plateau phases in other cases (since the plateau phase could be missed by the sampling or lost due to the faintness of the afterglow). Those which were undetected by XRT do not require extreme decay slopes relative to the rest of the sample of SGRBs. The break times of the SGRBs with plateaus are typically occuring orders of magnitude earlier than for the canonical LGRBs (as shown in Figure \ref{fig0.1c}). This may be caused by our use of BAT and XRT data whereas \cite{evans2009} use only the XRT data and is only able to find plateaus at times after XRT has started observing. However, it is very rare for XRT to not observe the steep decay phase for LGRBs so the inclusion of BAT data does not affect the plateau fits \citep[e.g.][]{obrien2006, willingale2007}. Additionally, \cite{evans2009} discussed whether their type b and type c LGRBs can be cannonical (i.e. those which are steep then shallow or shallow then steep). They conclude that they are not based on the plateau decay rates and break times for type b and the relative BAT versus XRT fluxes for type c. Whereas for SGRBs, the BAT observations need including in order to identify the steep decay phase. Histograms showing the various SGRB decay slopes for those with plateau phases are shown in Figure \ref{fig0.1b} with the values for canonical LGRBs, determined by \cite{evans2009}. The values for $\alpha_1$ and $\alpha_2$ are consistent with the LGRB sample, but the final decay phase ($\alpha_3$) is typically steeper than for the LGRB counterparts \citep[consistent with the results obtained by ][ where the overall decay of SGRBs is typically found to be steeper than LGRBs, however their sample of SGRBs used for this is dominated by SGRBs with extended emission]{margutti2012}. Using a Kolmogorov-Smirnov test between the values for LGRBs and SGRBs, $\alpha_1$ is consistent with being drawn from the same distribution (p-value = 0.07), although the values for $\alpha_2$ are unlikely to be from the same distribution (p-value = 0.003) and $\alpha_3$ are highly likely to be drawn from completely different distributions (p-value = 0.00007). In the following analysis we consider SGRBs with 2 or more breaks in their lightcurves (blue stars,  the GRBs in Figure \ref{fig0.1}a), 1 break (green circles, the GRBs in Figure \ref{fig0.1}b) and those with no breaks in their lightcurves (red triangles, the GRBs in Figure \ref{fig0.1}c and d).

The BAT fluence (15 -- 150 keV)  of these GRBs is plotted against their 0.3 -- 10 keV flux at 100 s in Figure \ref{fig0.2}a. Those GRBs with a plateau tend to be clustered at somewhat higher fluences and their X-ray fluxes are significantly higher at 100 s ($\sim 10^{-11}$ -- $10^{-9}$ erg cm$^{-2}$ s$^{-1}$). The GRBs which do not have a plateau phase in their lightcurves tend to have faint X-ray afterglows at 100 s ($\le 2 \times 10^{-11}$ erg cm$^{-2}$ s$^{-1}$) and relatively low fluences ($\le 2 \times 10^{-7}$ erg cm$^{-2}$). Figure \ref{fig0.2}b shows there is a wide variation in XRT flux at 100 s for SGRBs with similar prompt fluxes.

\cite{obrien2006} and \cite{willingale2007} found that the prompt fluence is comparable to the plateau fluence for LGRBs. In order to compare this result to our sample, we took the average flux for the plateau phase and multiplied it by the time at which the decay broke to a more typical afterglow (assuming this component started at the initial trigger time) giving the 0.3 -- 10 keV fluence. This fluence was then converted to a 15 -- 150 keV fluence using the spectral index and fitted absorption. Figure \ref{fig0.8} shows the prompt and plateau fluence are generally comparable, which is consistent with the result obtained for LGRBs. There are four significant outliers (GRBs 061201, 070724A, 080905A and 090515), lying significantly above the one-to-one line, whose plateaus are significantly more energetic than their prompt emission.

Figure \ref{fig0.9} shows the spectral indicies plotted against the temporal indicies for the lightcurves with a plateau phase (values all tabulated in Table \ref{lcfits}). Also plotted are the closure relations for the slow cooling regime (grey band) and the fast cooling regime (solid black lines) \citep{zhang2004}. These show the same behaviour identified by \cite{evans2009} for the canonical sample of LGRBs. In particular Figure \ref{fig0.9}b shows evidence of energy injection during the plateau phase as described by \cite{evans2009}. These figures can be compared to updated values for the whole GRB sample using the UK Swift Science Data Centre \citep[www.swift.ac.uk/xrt{\_}live{\_}cat;][]{evans2009}.

\cite{dainotti2010} identified a correlation between the plateau phase luminosity and duration for LGRBs with a canonical lightcurve. Using redshifts where available or the average SGRB redshift ($z\sim0.72$) and a k-correction \citep{bloom2001} we calculated the luminosity and restframe durations for the SGRB sample (XRT fluxes used are the observed values which have not been corrected for absorption). These results are plotted in Figure \ref{fig0.7} and the luminosity -- duration correlation is identified. The fitted correlation for the SGRB and LGRB sample, $b=-1.29\pm0.12$, $log(a)= 48.74\pm0.44$, intrinsic scatter $\sigma_{V}=9\times10^{-11}\pm 0.01$ \citep[where $L_X=aT_{\rm plateau}^b$ and the uncertainties on each datapoint and an intrinsic scatter are accounted for in the fit using the method described in ][]{dagostini2005}, is consistent with that for obtained the LGRB sample \citep[$-1.06\pm0.28$, $51.06\pm1.02$, although they did not account for the intrinsic scatter;][]{dainotti2010}. The SGRB plateau phases are typically more luminous and the plateau is shorter in duration than the LGRB counterparts. This may be a selection affect due to the inclusion of BAT observations in this analysis, and hence finding earlier plateaus. However, when BAT data are included in LGRB analysis the plateau properties do not significantly change \citep{obrien2006, willingale2007} additionally, there is a shortage of long duration plateaus observed in the SGRB sample. \cite{cannizzo2011} argue that the relationship identified by \cite{dainotti2010} is dominated by selection affects at z$>1.5$ as there is an observational bias against faint plateaus due to the limiting XRT flux. However, SGRBs are typically at lower redshift (the SGRBs with an observed redshift in our sample have an average redshift of z$\sim$0.72) so our sample lies well within the region which is not dominated by selection affects.

The plateau phases of GRB lightcurves are typically explained as ongoing central engine activity, for example on going accretion onto the central BH. However, ongoing accretion is problematic for NS-NS and NS-BH merger theories as there is insufficient surrounding material to maintain this accretion \citep{lee2007}. Fallback accretion from material on highly eccentric orbits has been postulated to resolve this \citep{rosswog2007, kumar2008, cannizzo2011}, however it is unclear how to produce the required reservoir of material at a fixed radius. In the remainder of this paper, we suggest that the plateau phases could be powered by a magnetar formed via the merger of two NSs.

\section{Magnetar model}
\begin{table*}
\begin{tabular}{cccccccc}
\hline
GRB	& E$_{iso}$ & P$_{-3}$ & B$_{15}$ & $\alpha_1 = \Gamma_{\gamma} + 1$ & Collapse time & Plateau Luminosity & Plateau Duration\\       
        & (erg)        & (ms)      & ($10^{15}$ G)         &                                  & (s)           & (erg s$^{-1}$)     & (s) \\
\hline
\multicolumn{8}{l|}{Magnetar candidates}\\
051221A  & 1.83$^{+0.45}_{-0.35}\times$10$^{52}$ & 7.79$^{+0.31}_{-0.28}$   & 1.80$^{+0.14}_{-0.13}$    & (1.39$^{+0.01}_{-0.02}$) & -   & 8.8$^{+3.0}_{-2.3}\times$10$^{45}$ & 38300$^{+9800}_{-7700}$ \\
060313   & 3.12$^{+1.06}_{-0.79}\times$10$^{53}$ & 3.80$^{+0.15}_{-0.13}$   & 3.58$^{+0.24}_{-0.22}$    & 1.71                     & -   & 6.2$^{+1.9}_{-1.5}\times$10$^{47}$ & 2310$^{+520}_{-420}$\\
060801   & 1.17$^{+1.79}_{-0.71}\times$10$^{53}$ & 1.95$^{+0.15}_{-0.13}$   & 11.24$^{+1.93}_{-1.78}$   & 1.47                     & 326 & 8.7$^{+7.1}_{-4.1}\times$10$^{49}$ & 62$^{+39}_{-23}$\\
070724A  & 1.13$^{+1.87}_{-0.40}\times$10$^{50}$ & 1.80$^{+1.04}_{-0.38}$   & 28.72$^{+1.42}_{-1.29}$   & (1.16$^{+0.10}_{-0.06}$) & 90  & 7.9$^{+14.5}_{-6.7}\times$10$^{50}$ & 8$^{+14}_{-4}$  \\
070809   & 8.87$^{+9.06}_{-3.48}\times$10$^{49}$ & 5.54$^{+0.48}_{-0.43}$   & 2.06$^{+0.48}_{-0.42}$   & (1.68$^{+0.11}_{-0.08}$) & -   & 4.5$^{+5.0}_{-2.5}\times$10$^{46}$ & 14800$^{+12800}_{-6500}$\\
080426   & 3.48$^{+0.67}_{-0.24}\times$10$^{51}$ & 6.17$^{+0.28}_{-0.24}$   & 8.94$^{+1.53}_{-1.17}$   & 2.98                     & -   & 5.5$^{+3.3}_{-2.0}\times$10$^{47}$ & 976$^{+436}_{-319}$ \\
080905A  & 6.16$^{+12.3}_{-4.03}\times$10$^{50}$ & 9.80$^{+0.78}_{-0.77}$  & 39.26$^{+10.24}_{-12.16}$ & (0.69$^{+0.05}_{-0.10}$) & 274 & 1.8$^{+2.0}_{-1.1}\times$10$^{48}$ & 128$^{+185}_{-60}$ \\
080919   & 5.18$^{+9.34}_{-3.26}\times$10$^{51}$ & 7.68$^{+0.91}_{-0.44}$  & 37.36$^{+13.92}_{-14.67}$ & 2.10                     & 421 & 4.0$^{+5.6}_{-3.1}\times$10$^{48}$ & 87$^{+207}_{-46}$  \\
081024   & 5.65$^{+7.53}_{-3.16}\times$10$^{51}$ & 2.30$^{+0.12}_{-0.11}$   & 31.04$^{+2.82}_{-2.35}$   & 2.33                     & 125 & 3.4$^{+1.5}_{-1.0}\times$10$^{50}$ & 11$^{+3}_{-3}$ \\
090426   & 3.98$^{+1.30}_{-0.03}\times$10$^{52}$ & 1.89$^{+0.08}_{-0.07}$   & 4.88$^{+0.88}_{-0.90}$    & 2.93                     & -   & 1.9$^{+1.2}_{-0.8}\times$10$^{49}$ & 310$^{+190}_{-110}$\\
090510   & 5.76$^{+6.86}_{-3.10}\times$10$^{52}$ & 1.86$^{+0.04}_{-0.03}$   & 5.06$^{+0.27}_{-0.23}$    & 1.98                     & -   & 2.1$^{+0.4}_{-0.4}\times$10$^{49}$ & 277$^{+40}_{-35}$\\
090515   & 3.44$^{+3.55}_{-1.55}\times$10$^{50}$ & 2.05$^{+0.06}_{-0.05}$   & 12.27$^{+1.14}_{-1.11}$   & 2.60                     & 175 & 8.5$^{+2.7}_{-2.2}\times$10$^{49}$ & 57$^{+16}_{-12}$  \\
100117A  & 1.42$^{+2.08}_{-0.84}\times$10$^{52}$ & 1.13$^{+0.07}_{-0.06}$   & 11.89$^{+0.50}_{-0.52}$   & 1.88                     & -   & 8.7$^{+3.0}_{-2.4}\times$10$^{50}$ & 19$^{+4}_{-3}$  \\
100702A  & 2.28$^{+1.46}_{-0.80}\times$10$^{51}$ & 1.29$^{+0.22}_{-0.12}$   & 19.50$^{+0.24}_{-0.76}$   & 2.54                     & 178 & 1.4$^{+0.7}_{-0.7}\times$10$^{51}$ & 9$^{+4}_{-2}$   \\
101219A  & 1.69$^{+0.79}_{-0.54}\times$10$^{53}$ & 0.95$^{+0.05}_{-0.05}$   & 2.81$^{+0.47}_{-0.39}$    & (1.22$^{+0.03}_{-0.03}$) & 138 & 9.7$^{+6.7}_{-3.8}\times$10$^{49}$ & 234$^{+116}_{-80}$\\
111020A  & 1.98$^{+2.55}_{-0.99}\times$10$^{51}$ & 7.76$^{+1.06}_{-0.69}$  & 2.24$^{+1.13}_{-0.73}$    & (1.44$^{+0.05}_{-0.05}$) & -   & 1.4$^{+3.9}_{-1.0}\times$10$^{46}$ & 24600$^{+45300}_{-16300}$  \\
120305A  & 2.02$^{+0.10}_{-0.10}\times$10$^{52}$ & 2.22$^{+0.09}_{-0.04}$   & 10.22$^{+0.35}_{-0.27}$    & (6.26$^{+0.17}_{-0.16}$) & 182 & 4.3$^{+0.6}_{-0.8}\times$10$^{49}$ & 97$^{+14}_{-10}$\\
120521A  & 8.42$^{+12.19}_{-4.95}\times$10$^{51}$ & 4.88$^{+0.63}_{-1.10}$   & 15.04$^{+8.42}_{-7.93}$    & 1.98 & 207 & 4.0$^{+23.0}_{-3.4}\times$10$^{48}$ & 216$^{+1015}_{-163}$\\
\hline
\multicolumn{8}{l|}{Possible candidates}\\
050509B  & 3.82$^{+16.9}_{-2.87}\times$10$^{49}$ & 80.32$^{+24.98}_{-17.91}$& 21.85$^{+16.44}_{-11.98}$ & 2.5                      & -   & 1.2$^{+8.5}_{-1.1}\times$10$^{44}$ & 27700$^{+206000}_{-22300}$ \\
051210   & 5.98$^{+13.5}_{-4.05}\times$10$^{51}$ & 0.68$^{+0.03}_{-0.03}$   & 7.68$^{+0.44}_{-0.39}$    & 2.1                      & 225 & 2.8$^{+0.9}_{-0.7}\times$10$^{51}$ & 16$^{+3}_{-3}$   \\
061201   & 1.42$^{+1.67}_{-0.69}\times$10$^{51}$ & 14.52$^{+0.59}_{-0.52}$  & 19.00$^{+1.75}_{-1.44}$   & 1.57                     & -   & 8.1$^{+3.1}_{-2.2}\times$10$^{46}$ & 1200$^{+320}_{-260}$ \\
070714A  & 3.28$^{+3.08}_{-1.48}\times$10$^{51}$ & 10.77$^{+1.04}_{-1.06}$  & 16.21$^{+4.29}_{-4.04}$   & 3.60                     & -   & 2.0$^{+2.7}_{-1.2}\times$10$^{47}$ & 905$^{+1000}_{-460}$ \\
080702A  & 1.20$^{+4.90}_{-0.90}\times$10$^{51}$ & 13.55$^{+1.39}_{-1.10}$  & 36.18$^{+12.25}_{-8.32}$ & 2.34                     & -   & 3.9$^{+5.9}_{-2.3}\times$10$^{47}$ & 290$^{+300}_{-150}$ \\
090621B  & 1.31$^{+2.07}_{-0.80}\times$10$^{52}$ & 26.65$^{+5.44}_{-3.42}$ & 23.05$^{+10.79}_{-6.6}$ & (4.72$^{+0.04}_{-0.05}$) & -   & 1.0$^{+2.9}_{-8.0}\times$10$^{46}$ & 2700$^{+5100}_{-1800}$\\
091109B  & 5.25$^{+3.95}_{-2.27}\times$10$^{52}$ & 13.60$^{+1.61}_{-1.24}$  & 9.16$^{+2.75}_{-2.33}$   & (3.16$^{+0.45}_{-0.53}$) & -   & 2.5$^{+3.6}_{-1.6}\times$10$^{46}$ & 4500$^{+5600}_{-2300}$ \\
100625A  & 3.27$^{+1.76}_{-1.15}\times$10$^{52}$ & 23.08$^{+3.59}_{-3.92}$  & 168.40$^{+32.78}_{-25.72}$& (4.09$^{+1.52}_{-0.73}$) & -   & 1.0$^{+2.0}_{-0.6}\times$10$^{48}$ & 38$^{+33}_{-20}$ \\
110112A  & 2.91$^{+5.85}_{-0.17}\times$10$^{50}$ & 13.14$^{+0.93}_{-0.75}$  & 18.85$^{+3.48}_{-2.52}$   & 3.14                     & -   & 1.2$^{+0.9}_{-0.5}\times$10$^{47}$ & 996$^{+530}_{-370}$\\
111117A  & 4.78$^{+5.71}_{-2.58}\times$10$^{52}$ & 17.73$^{+2.08}_{-2.47}$  & 68.69$^{+20.17}_{-17.39}$  & 1.65                     & -  & 5.5$^{+11.6}_{-3.5}\times$10$^{47}$ & 127$^{+160}_{-72}$\\
\end{tabular}
\caption{The SGRB magnetar sample used with their magnetar fits. E$_{iso}$, 1--10000 keV, is calculated using the fluences and redshifts in Table \ref{candidates}, a simple power law model and a k-correction \citet{bloom2001}. The values for $\alpha$ are input into the model unless they are bracketed - in this case the values are fit within the model. If there is a steep decay phase, we assume the magnetar collapses to form a BH and the model determines the collapse time. The values for P$_{-3}$ and B$_{15}$ are fitted from the model assuming isotropic emission. Using the values of P$_{-3}$ and B$_{15}$ obtained from the model, we derive the plateau luminosity and duration using equations \ref{luminosity} and \ref{period}. The derived plateau duration is from the initial formation of the magnetar (i.e. the time of the GRB) and the point at which the X-ray emission from the magnetar starts to turn over from the plateau phase to a powerlaw decay phase.}
\label{table:log}
\end{table*}

The magnetar model predicts a plateau phase in the X-ray lightcurve which is powered by the spin down of a newly formed magnetar. This section fits the model directly to the restframe SGRB lightcurves. The magnetar component is expected to be an extra component to the typical lightcurve. Therefore, we assume there is a single power law decay, $\alpha_{1}$, underlying the magnetar component. This value has been set to $\alpha_{1} = \Gamma_{\gamma} + 1$, where $\Gamma_{\gamma}$ is the photon index of the prompt emission, assuming that the decay slope is governed by the curvature effect \citep{kumar2000}, i.e. that the surrounding medium is very low density as might be expected for neutron star mergers. We note that the simple curvature effect assumed here does not account for any spectral evolution (for example as the peak energy moves through the observation band) however this is to keep the number of free parameters fitted in the model low. The normalisation of the power law decay fit is constrained using the last decay from the prompt emission. In a small number of cases, the decay slope is significantly different from prediction and we allow $\alpha_{1}$ to vary. It is important to note that the underlying lightcurve could be similar to other GRBs with a more complex afterglow light curve, but this work assumes that these are naked bursts (i.e. no surrounding ISM for neutron star mergers) and only the curvature effect is important.

Also, we expect there to be flares also overlying the powerlaw decay and magnetar component \citep[e.g.][]{margutti2011}. Due to the limited statistics in SGRB lightcurves, we do not attempt to exclude possible flares from the lightcurve fits (except GRB 060313, which has multiple flares early in the lightcurve) and the underlying flares will slightly affect the fit parameters.

\subsection{Theory}

The model used here is as described in \cite{zhang2001} and was suggested to explain GRB 051221A with a long lived magnetar \citep{fan2006}, for several LGRBs \citep{troja2007, lyons2009, bernardini2012} and for the short GRB 090515 \citep{rowlinson2010b}. This model is consistent with the late time residual spin down phase driving a relativistic magnetar wind as described in \cite{metzger2010}. We use the equations below with an underlying powerlaw component. Previously, the plateau duration and luminosity were calculated and then input into the equations. In this work, the equations are fit directly to the rest-frame light curves, taking into account the shape of the lightcurve \citep[this is a comparable method to that used by][who fitted a stable magnetar to the lightcurves of 4 LGRBs]{dallosso2011, bernardini2012}. We can then use the values of the magnetic field and spin period obtained to derive the luminosity and plateau duration.

\begin{eqnarray}
T_{em,3}=2.05~(I_{45}B^{-2}_{p,15}P^2_{0,-3}R^{-6}_6)\label{period}\\
L_{0,49}\sim(B^2_{p,15}P^{-4}_{0,-3}R^6_6)\label{luminosity}\\
B^{2}_{p,15}=4.2025 I_{45}^{2}R^{-6}_{6}L_{0,49}^{-1}T_{em,3}^{-2}\label{b^2}\\
P^{2}_{0,-3}=2.05 I_{45}L_{0,49}^{-1}T_{em,3}^{-1}\label{p^2}
\end{eqnarray}

\begin{figure*}
\centering
\includegraphics[width=5.5cm]{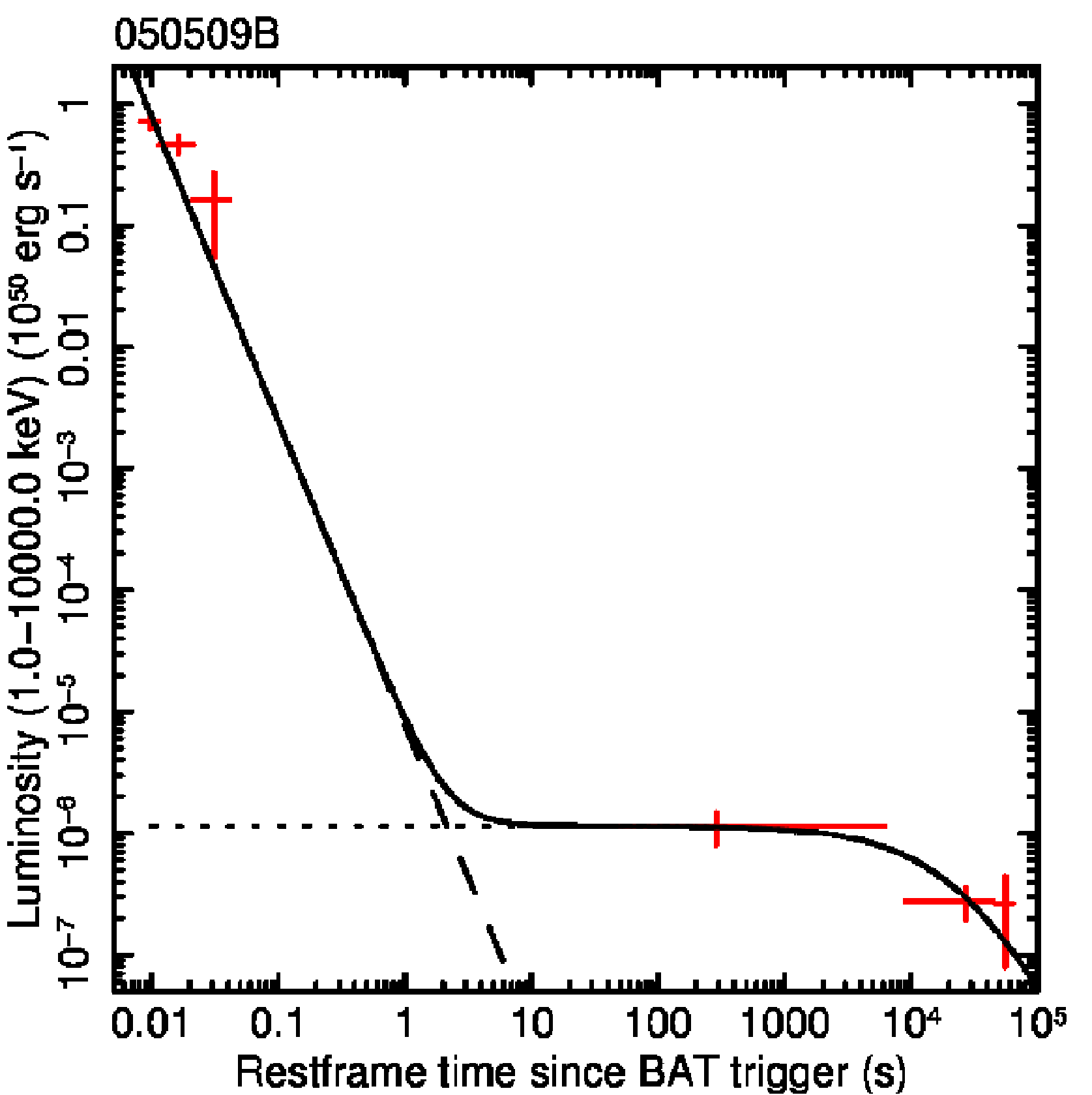}
\includegraphics[width=5.5cm]{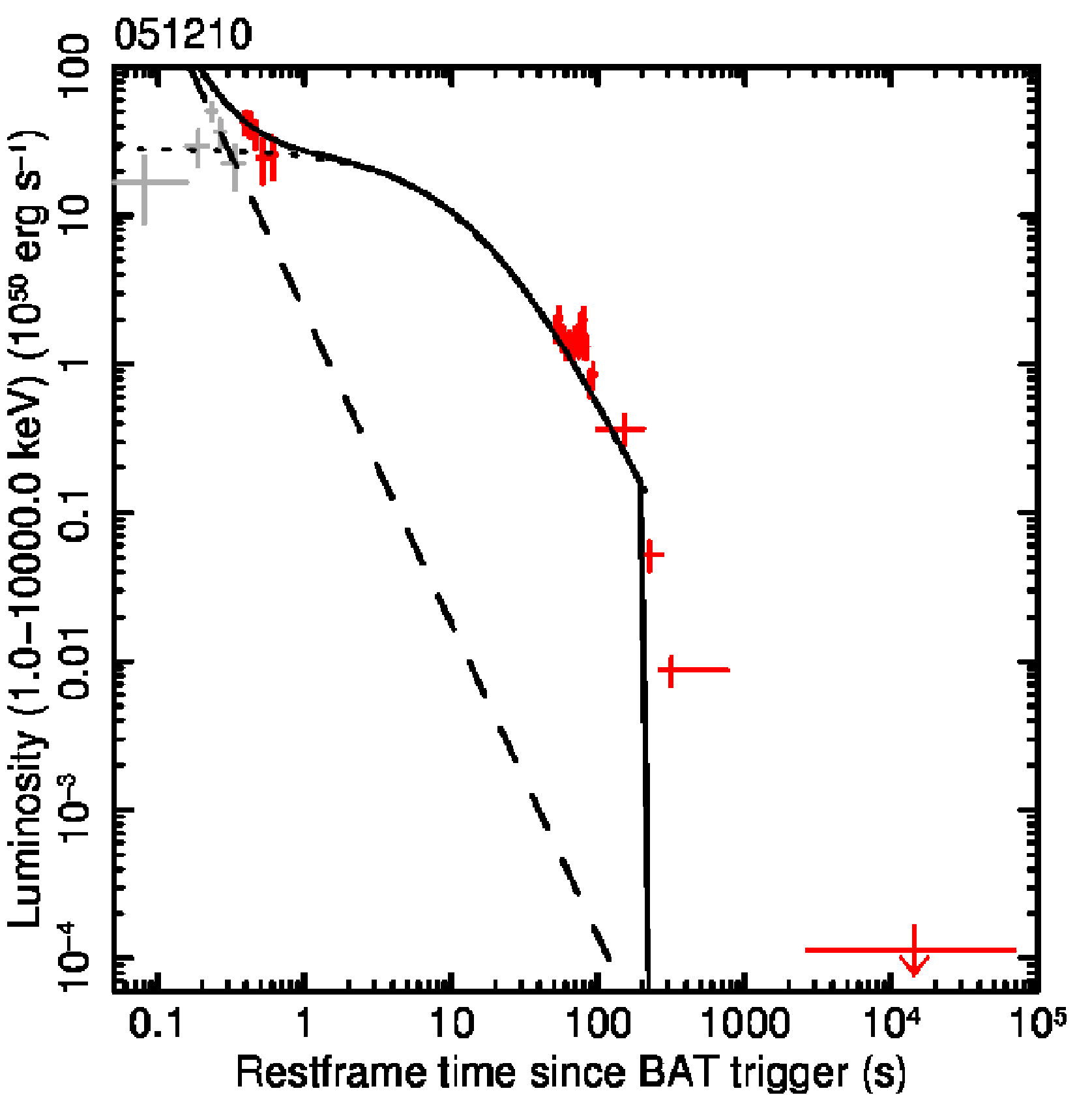}
\includegraphics[width=5.5cm]{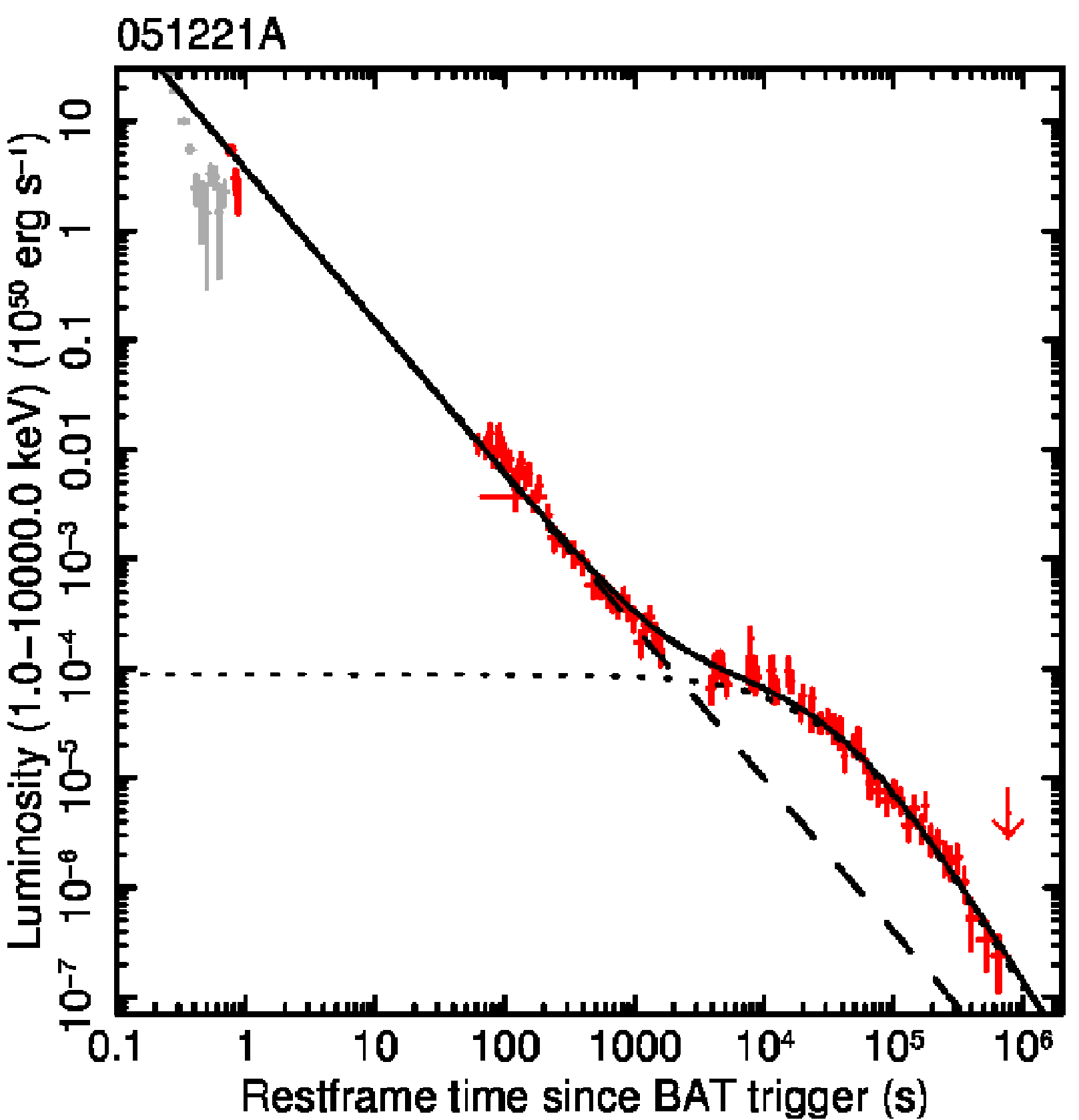}
\includegraphics[width=5.5cm]{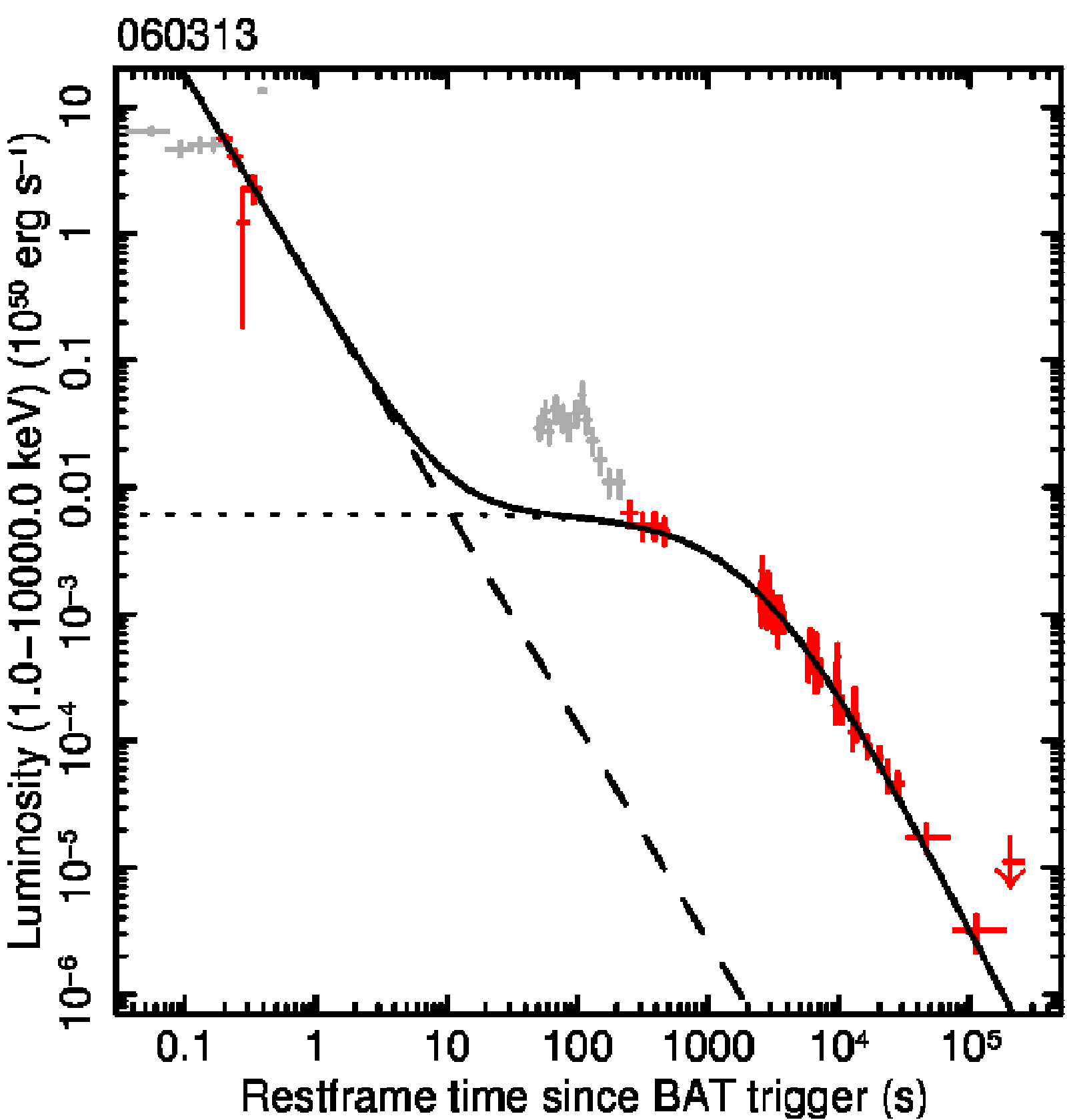}
\includegraphics[width=5.5cm]{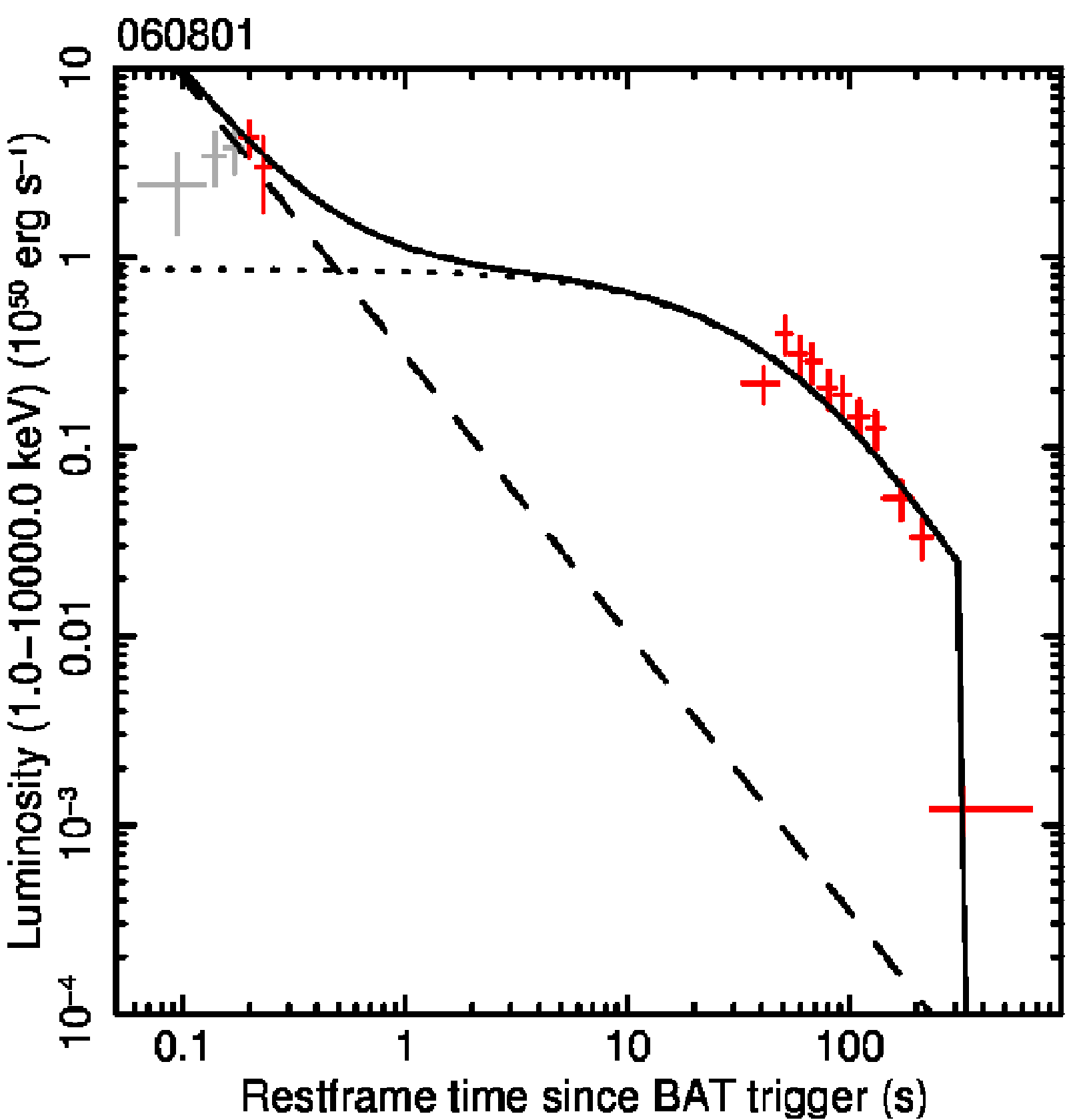}
\includegraphics[width=5.5cm]{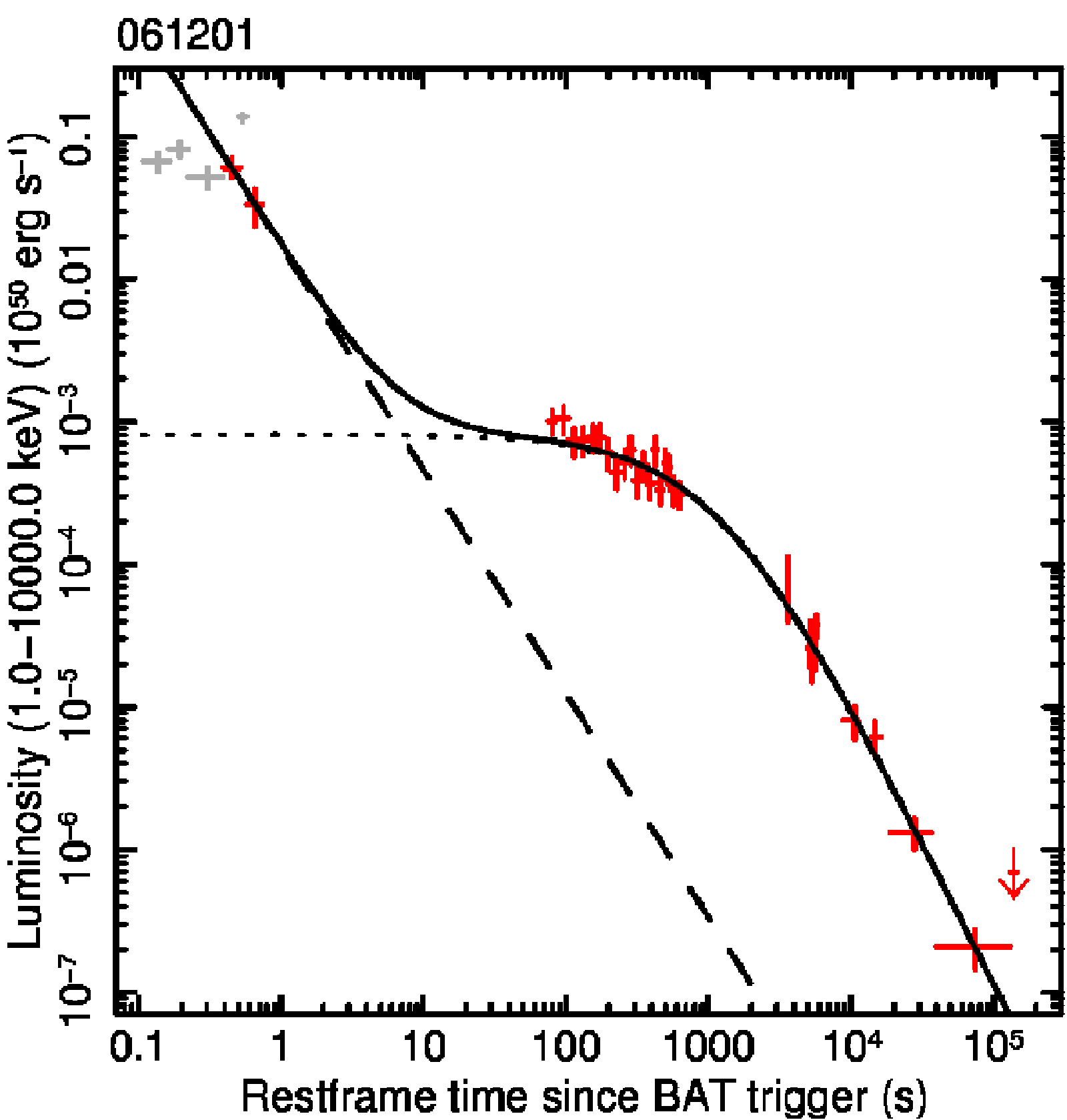}
\includegraphics[width=5.5cm]{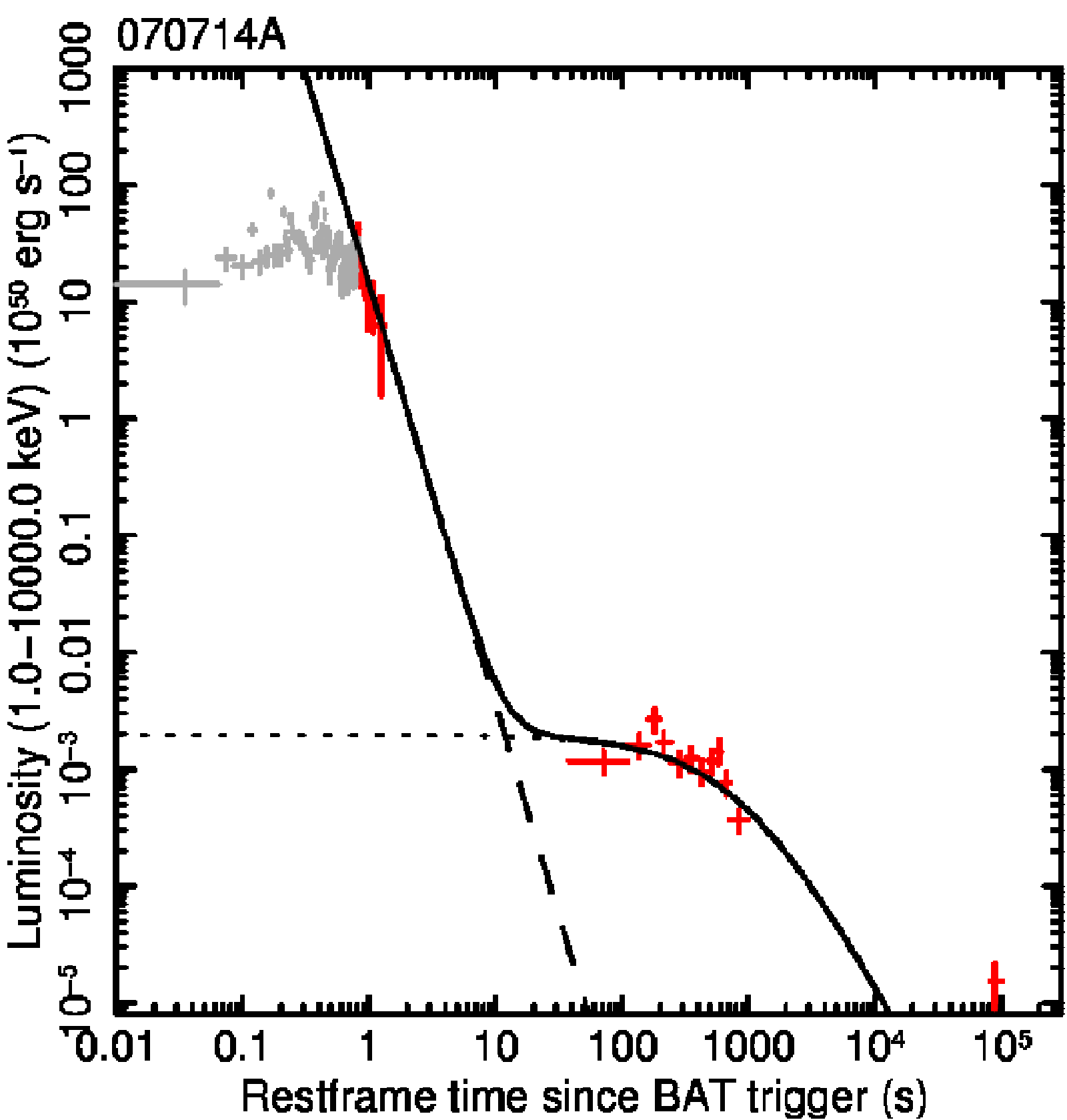}
\includegraphics[width=5.5cm]{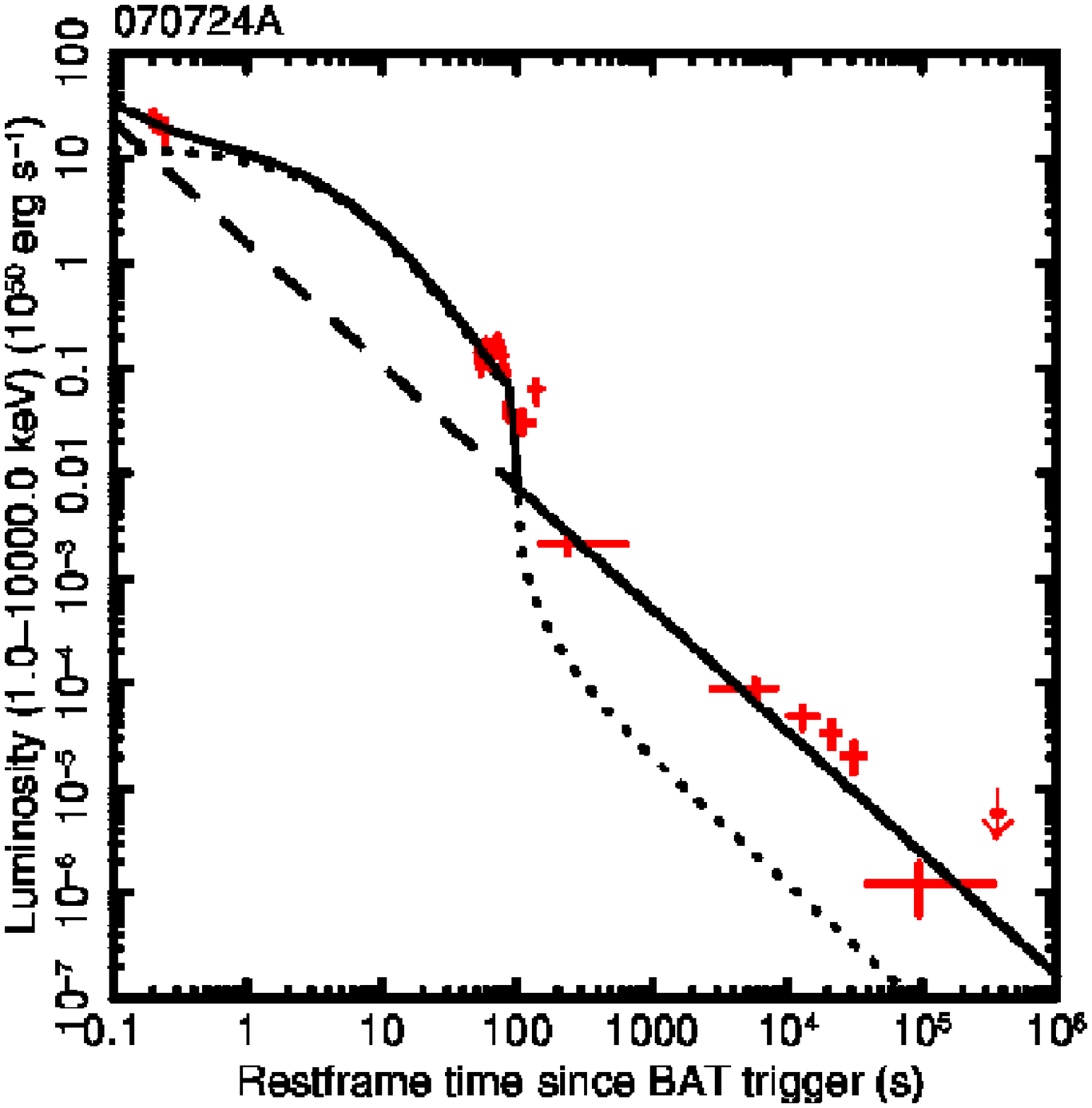}
\includegraphics[width=5.5cm]{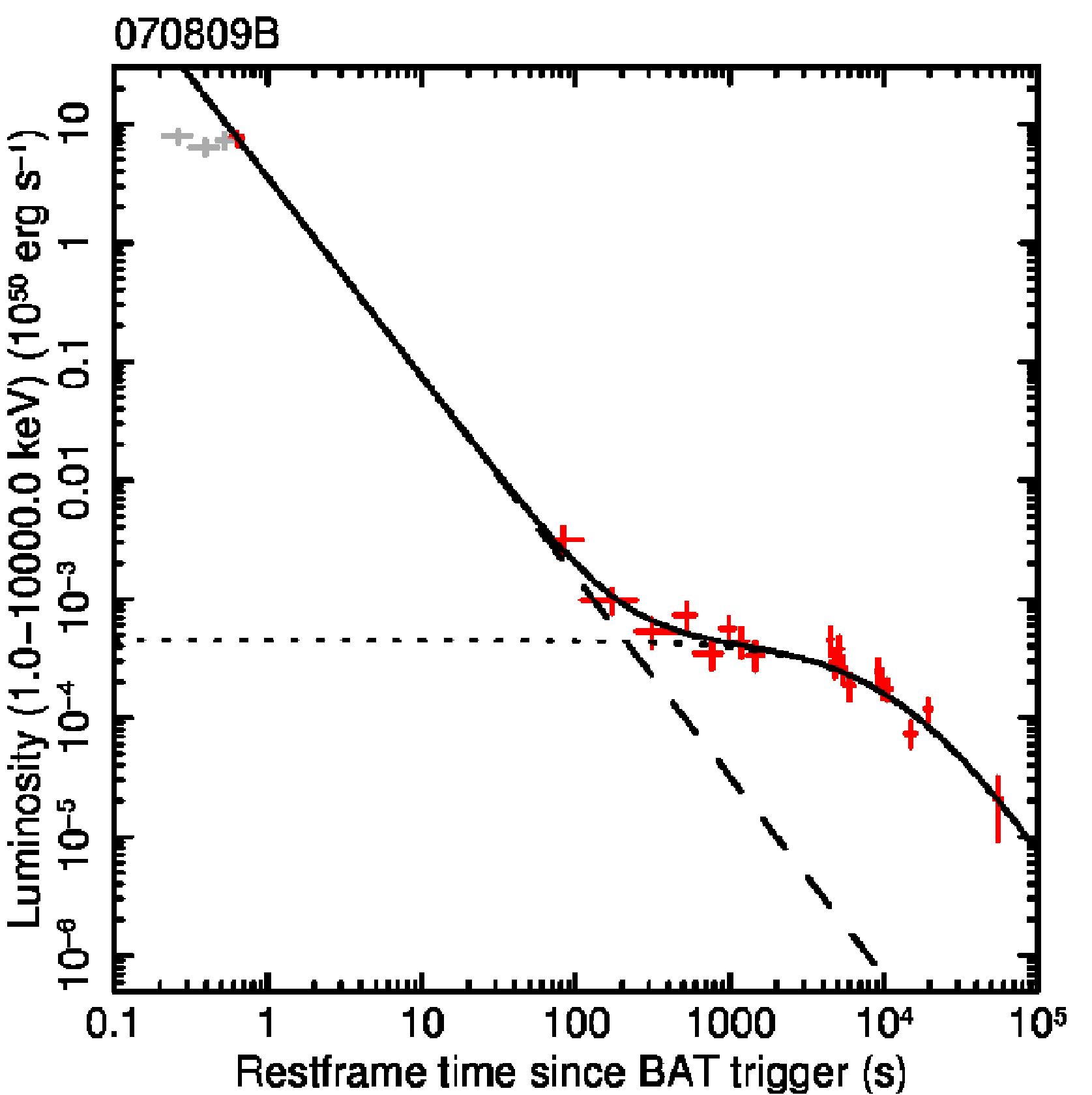}
\includegraphics[width=5.5cm]{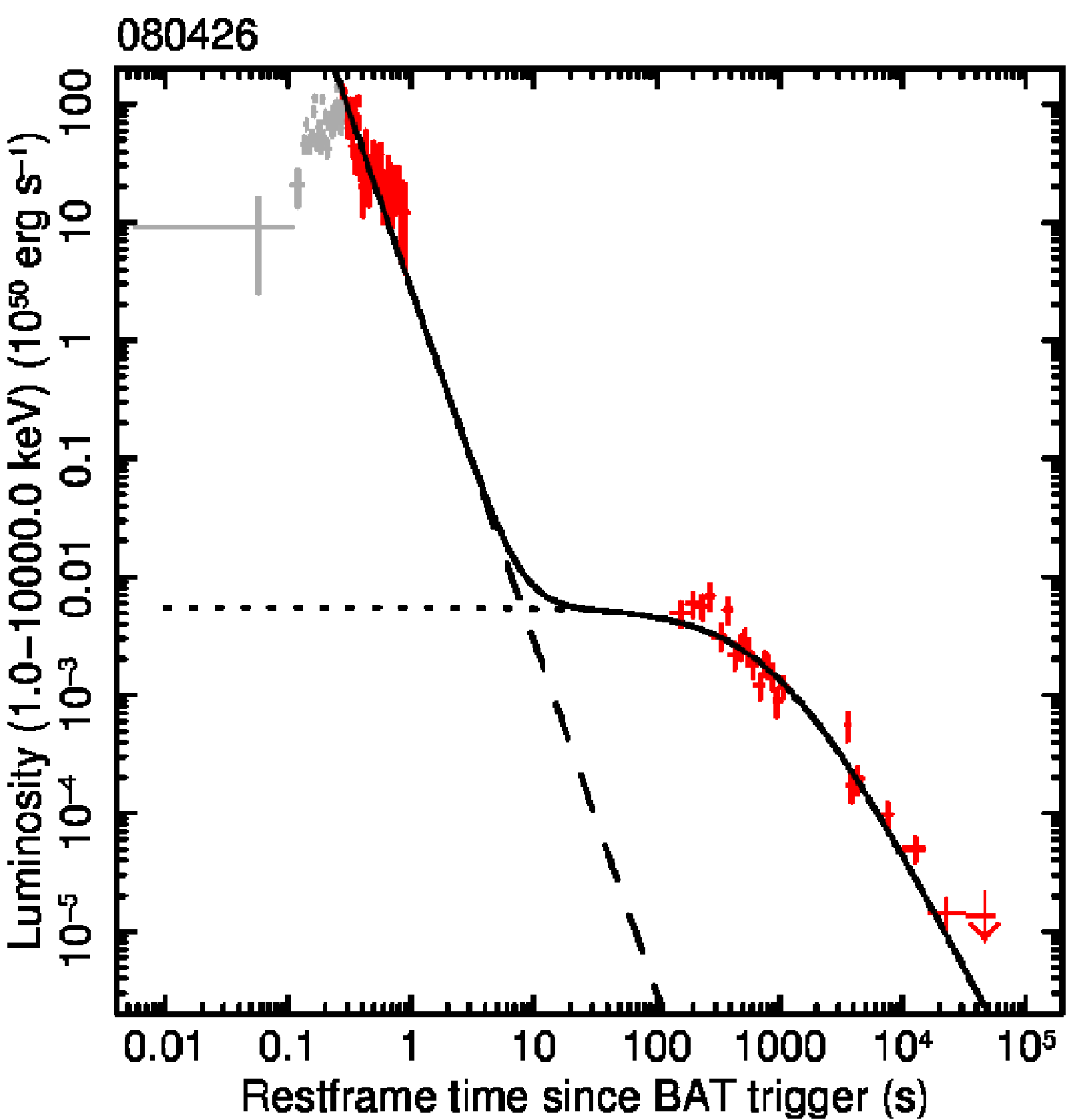}
\includegraphics[width=5.5cm]{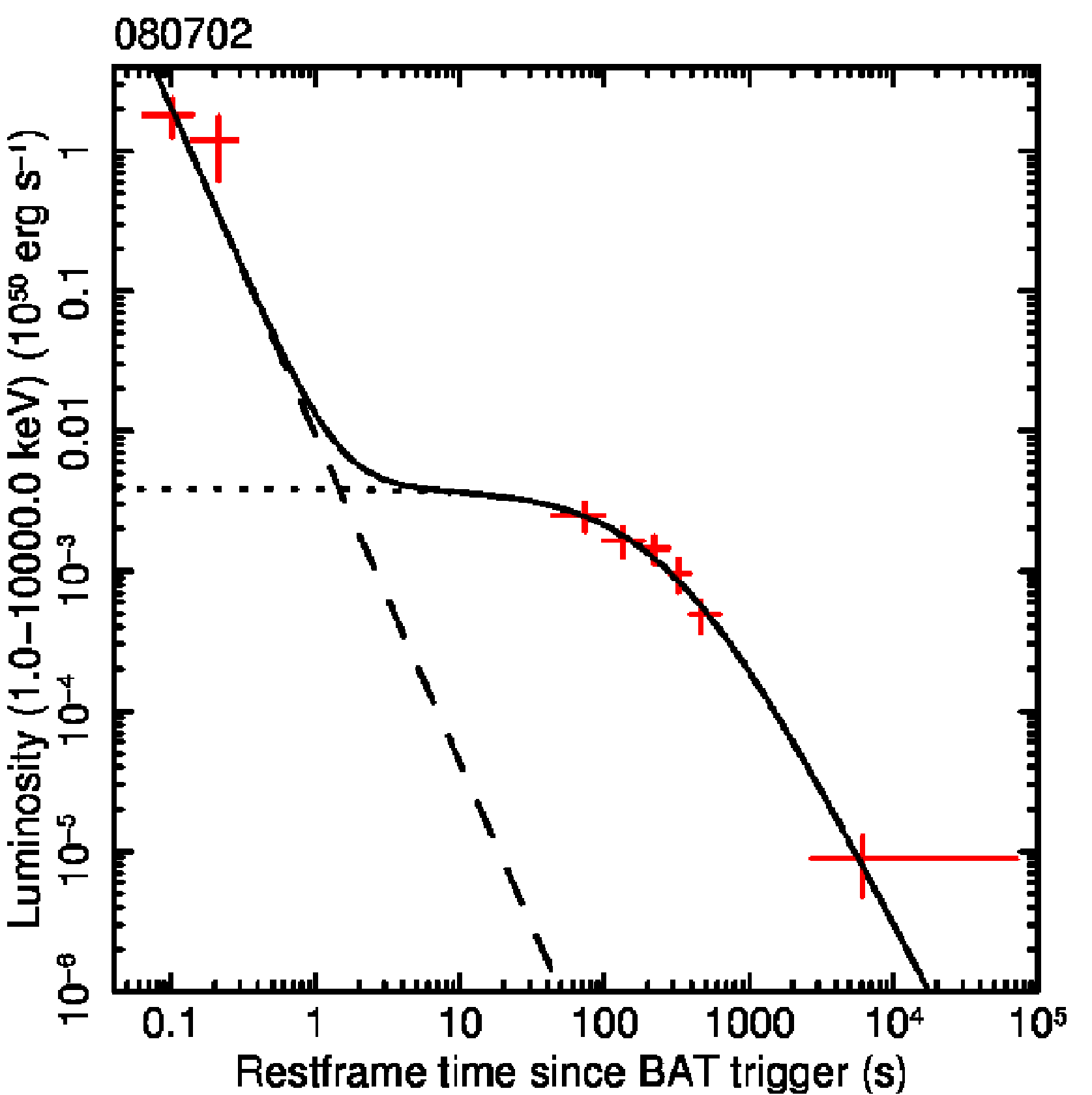}
\includegraphics[width=5.5cm]{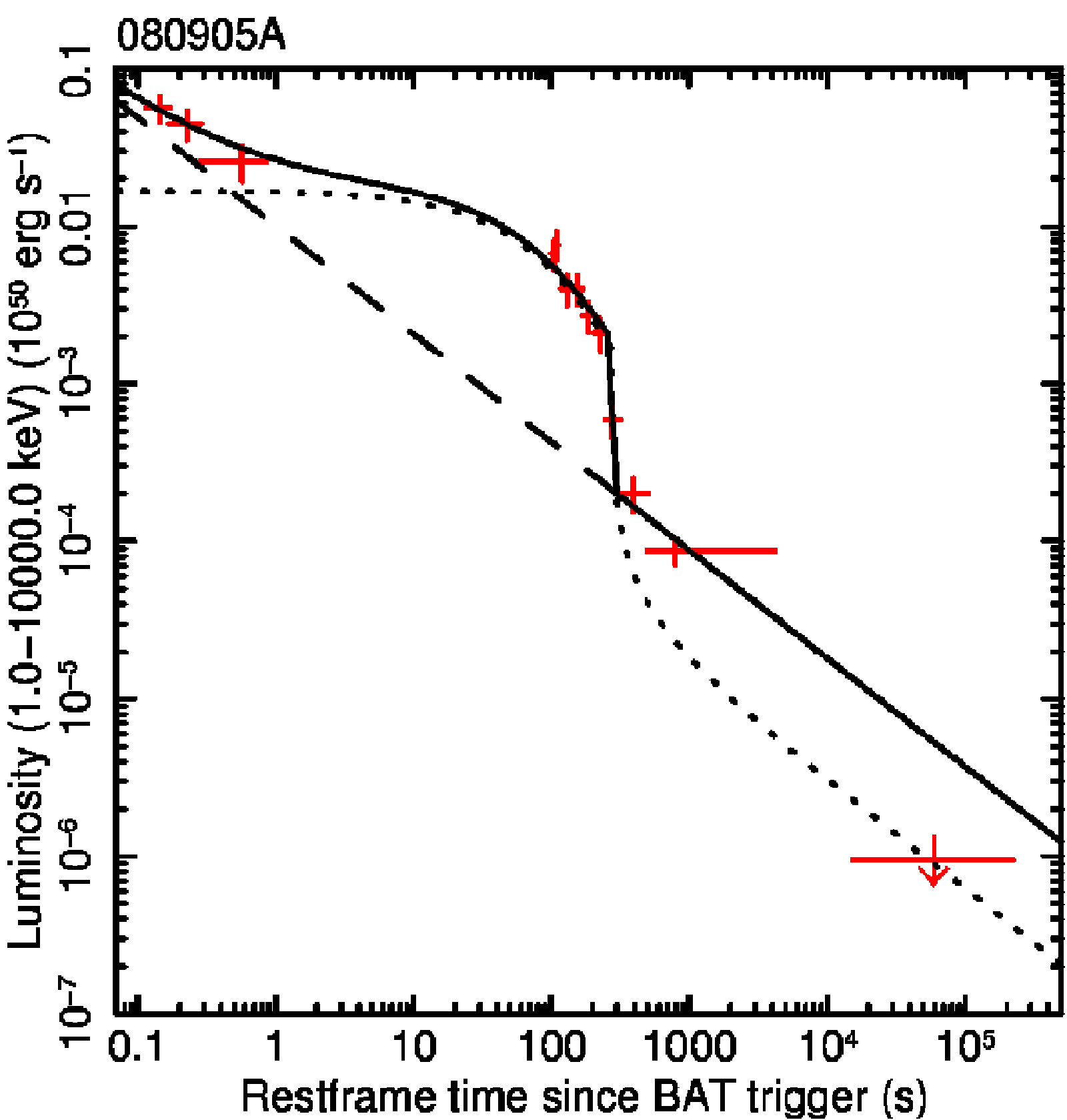}
\caption[SGRB lightcurves fit with the magnetar model]{SGRB BAT-XRT restframe lightcurves fit with the magnetar model. The light grey data points have been excluded from the fit. The dashed line shows the power-law component and the dotted line shows the magnetar componenet.}
\label{fig1}
\end{figure*}

\begin{figure*}
\centering
\includegraphics[width=5.5cm]{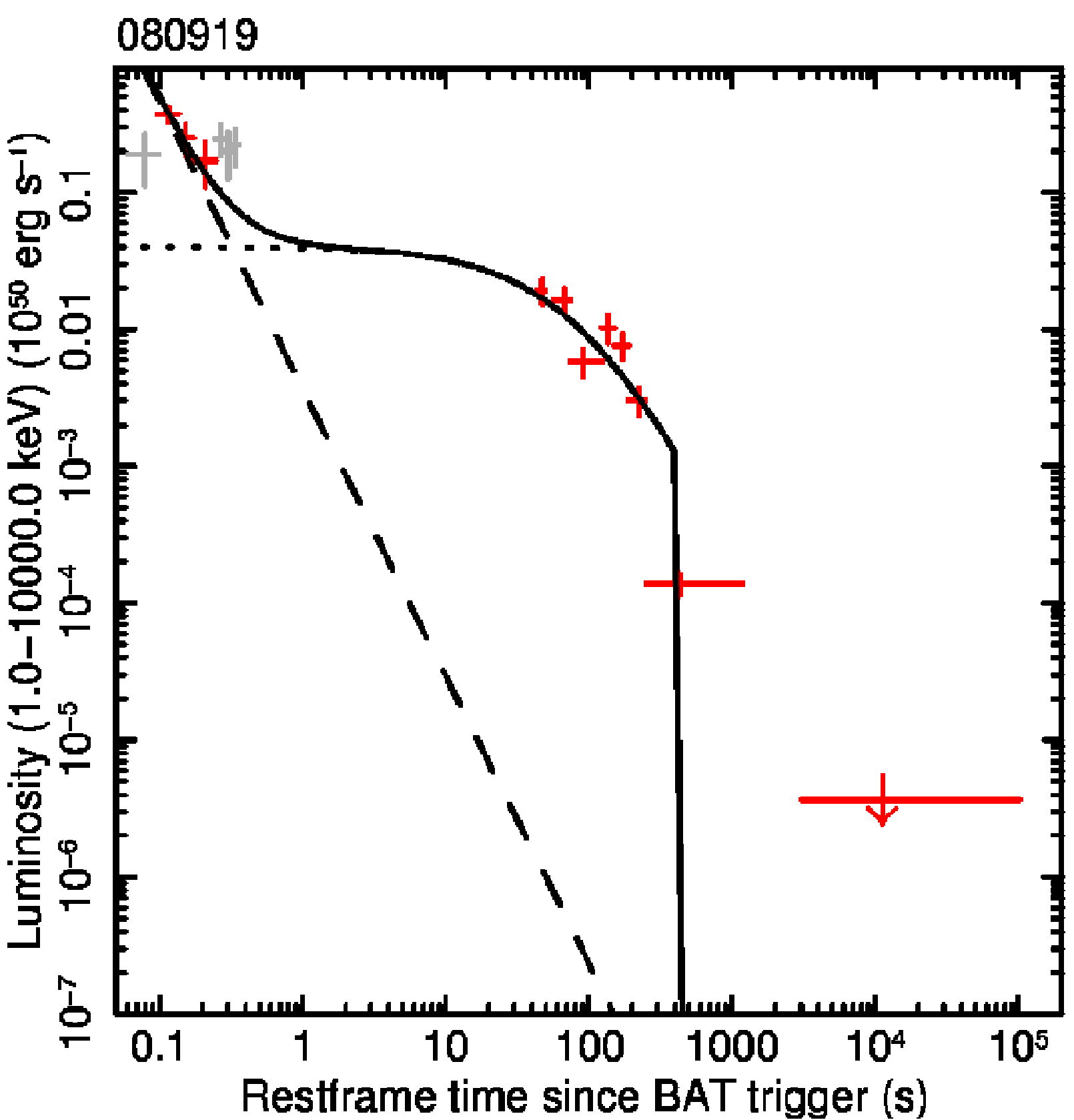}
\includegraphics[width=5.5cm]{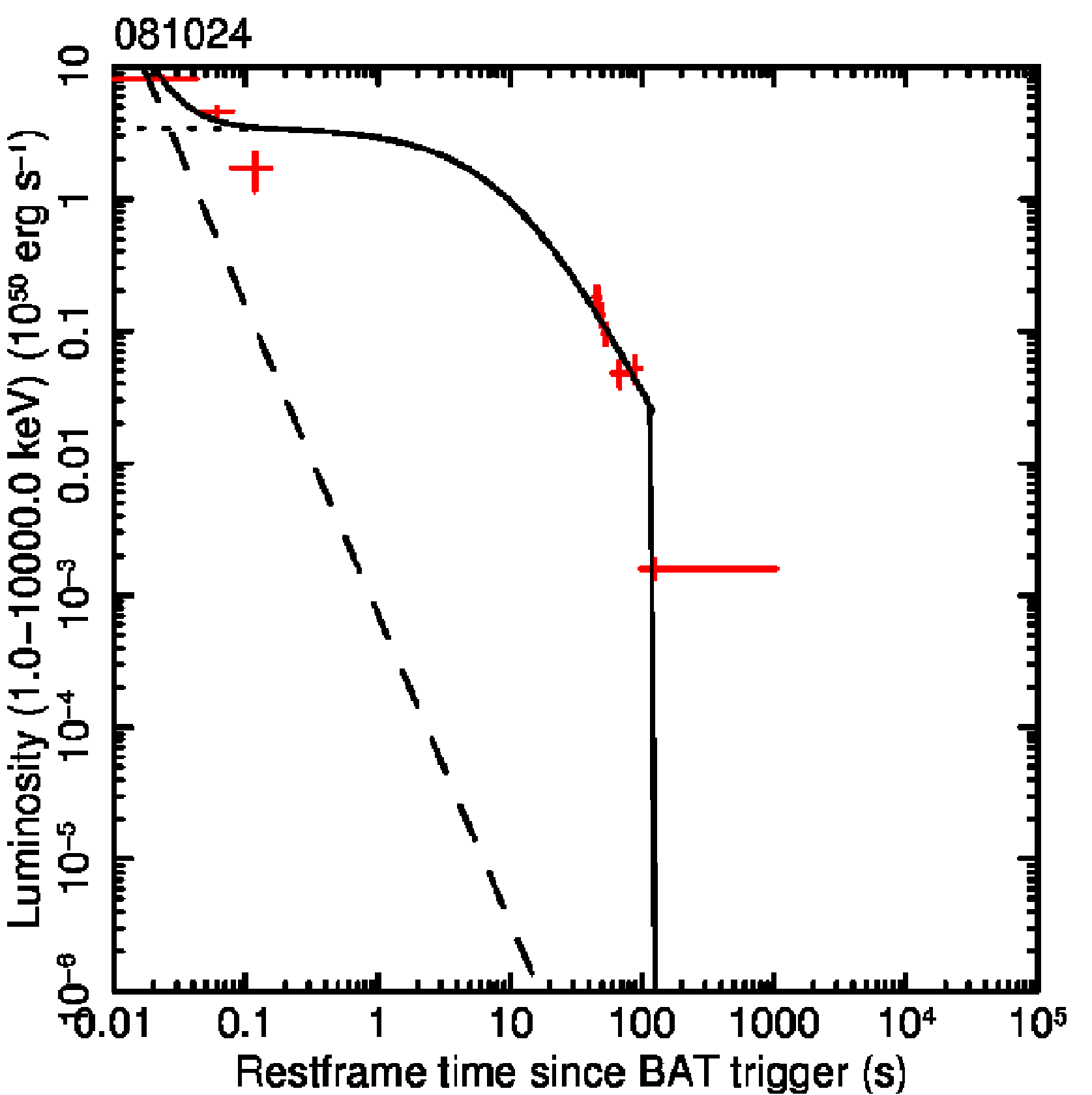}
\includegraphics[width=5.5cm]{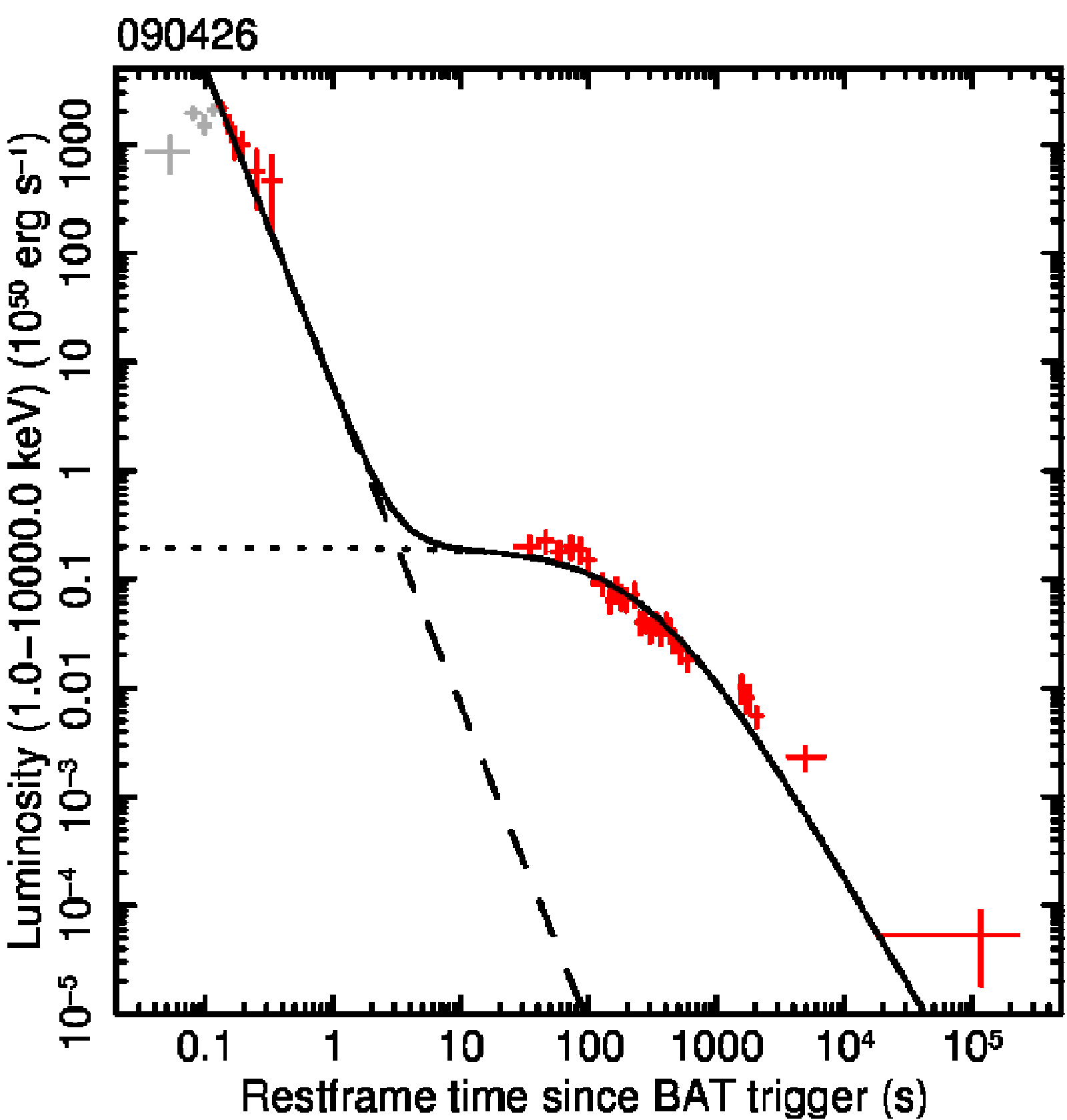}
\includegraphics[width=5.5cm]{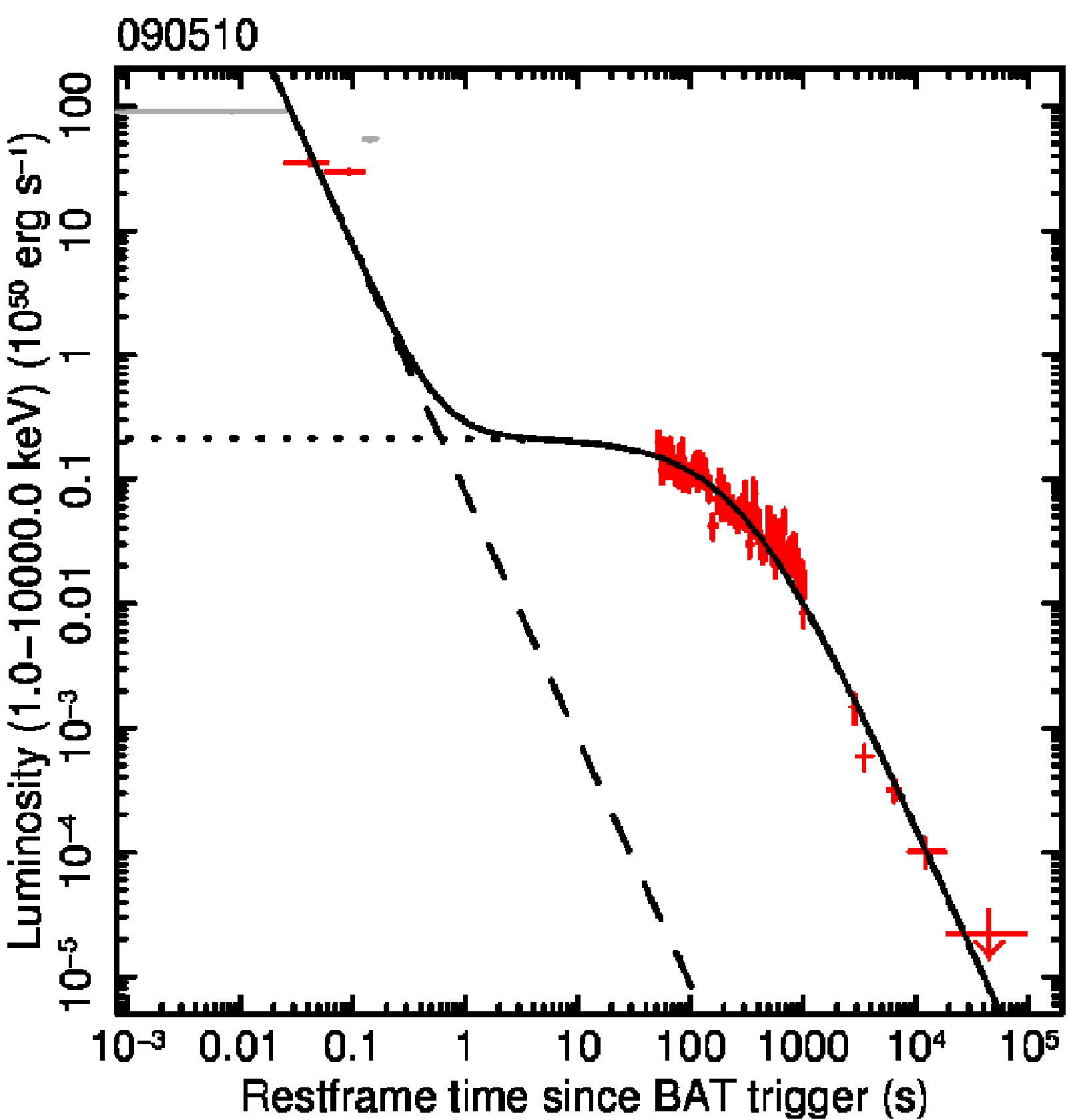}
\includegraphics[width=5.5cm]{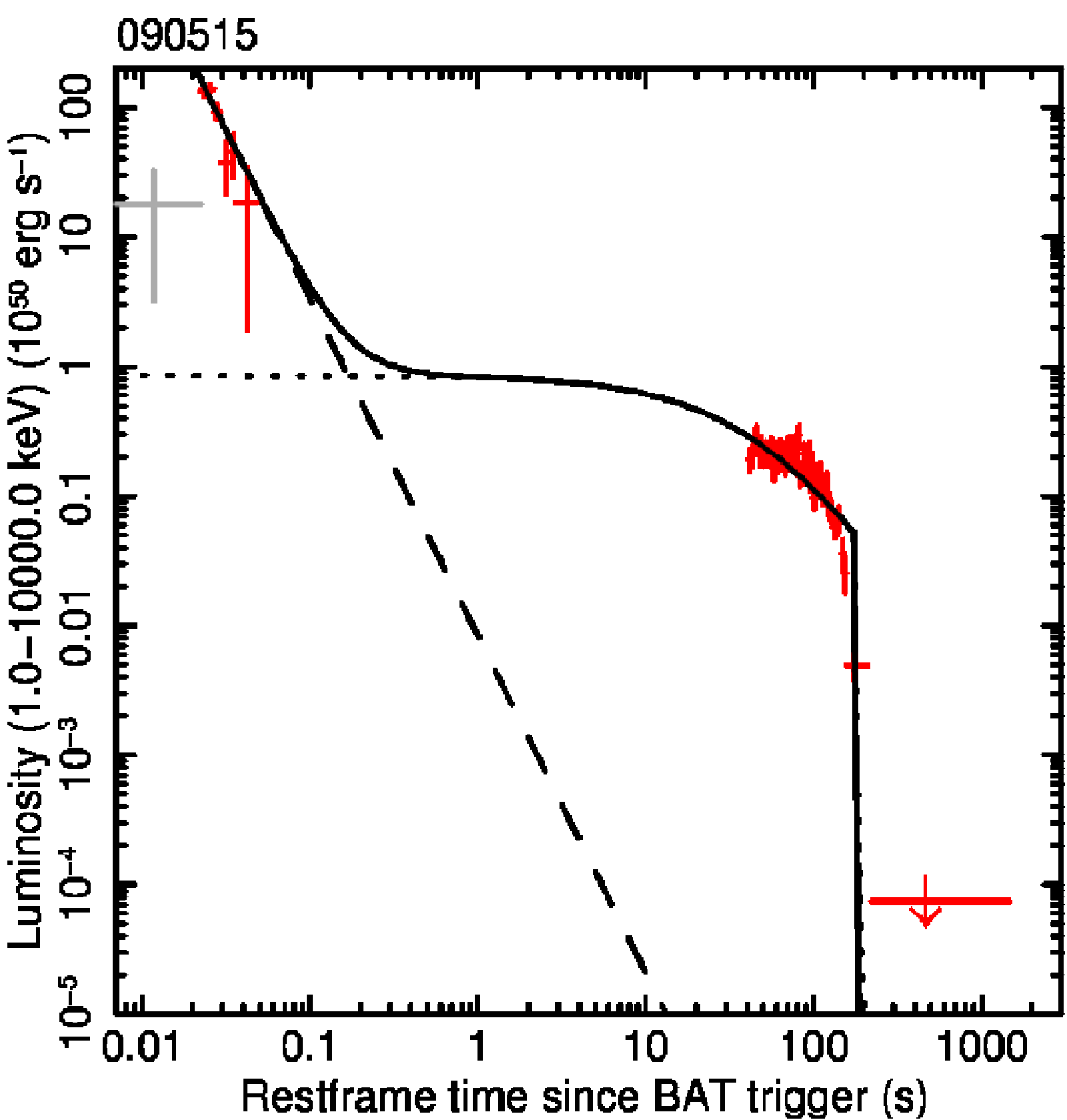}
\includegraphics[width=5.5cm]{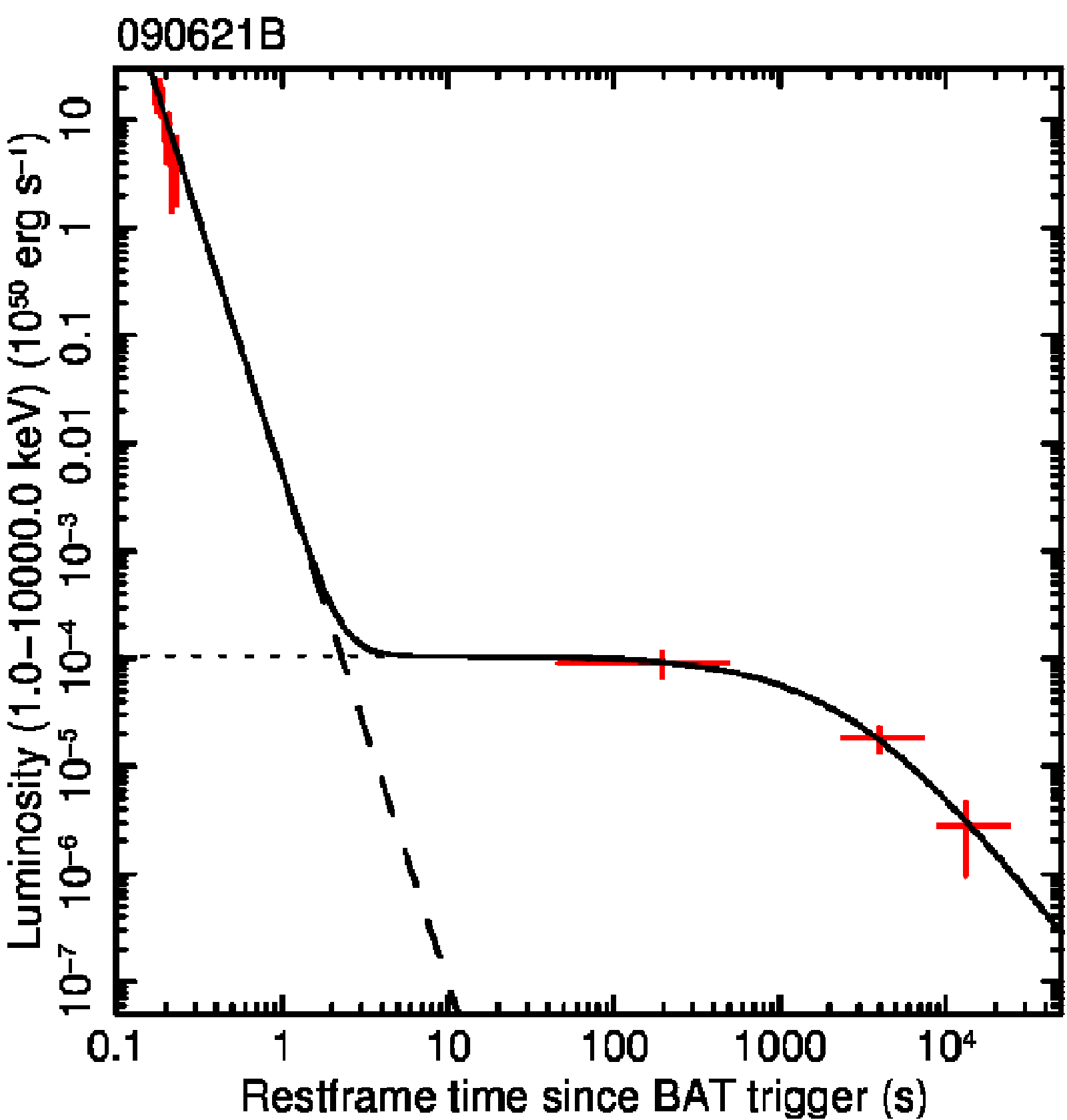}
\includegraphics[width=5.5cm]{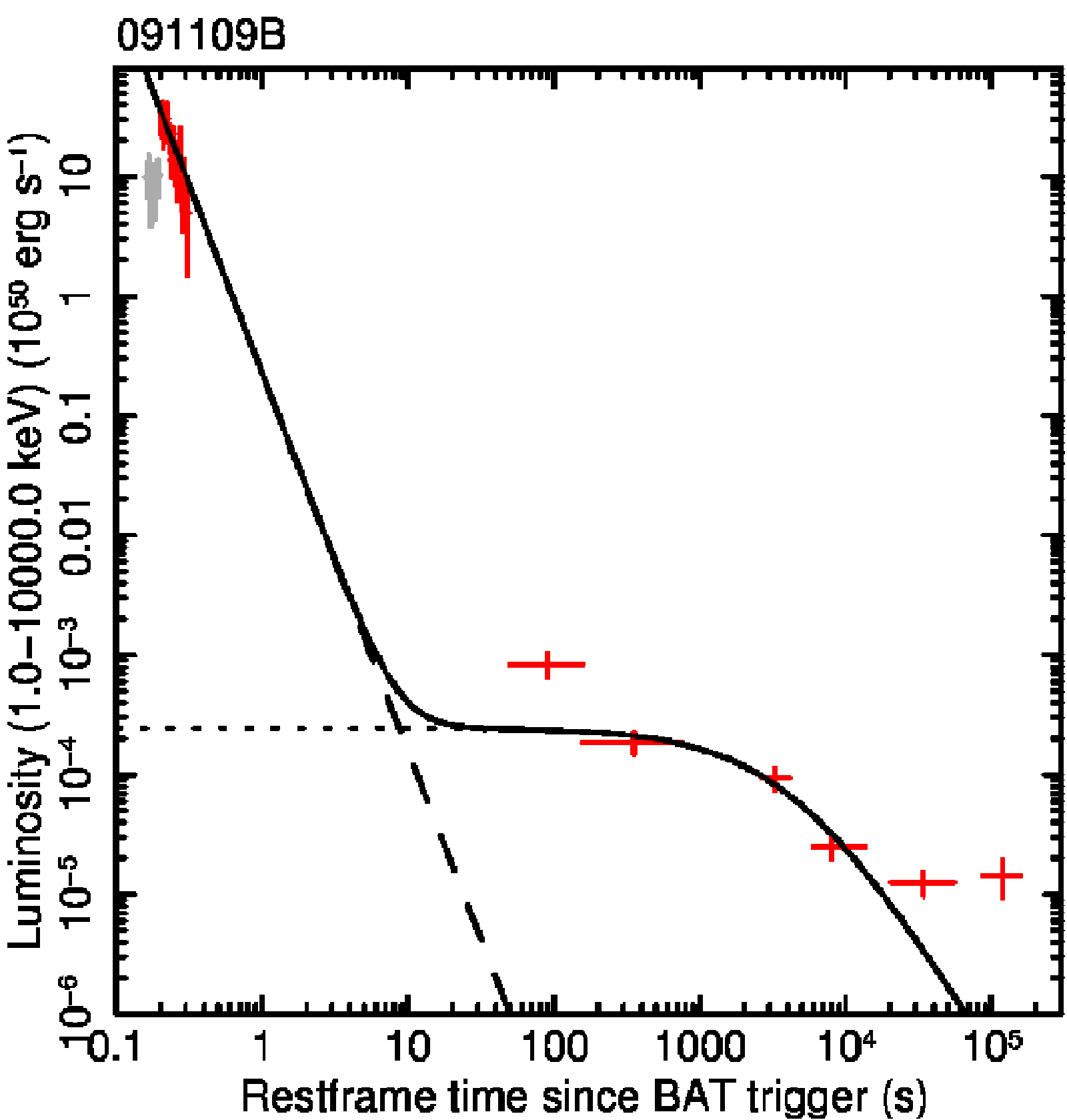}
\includegraphics[width=5.5cm]{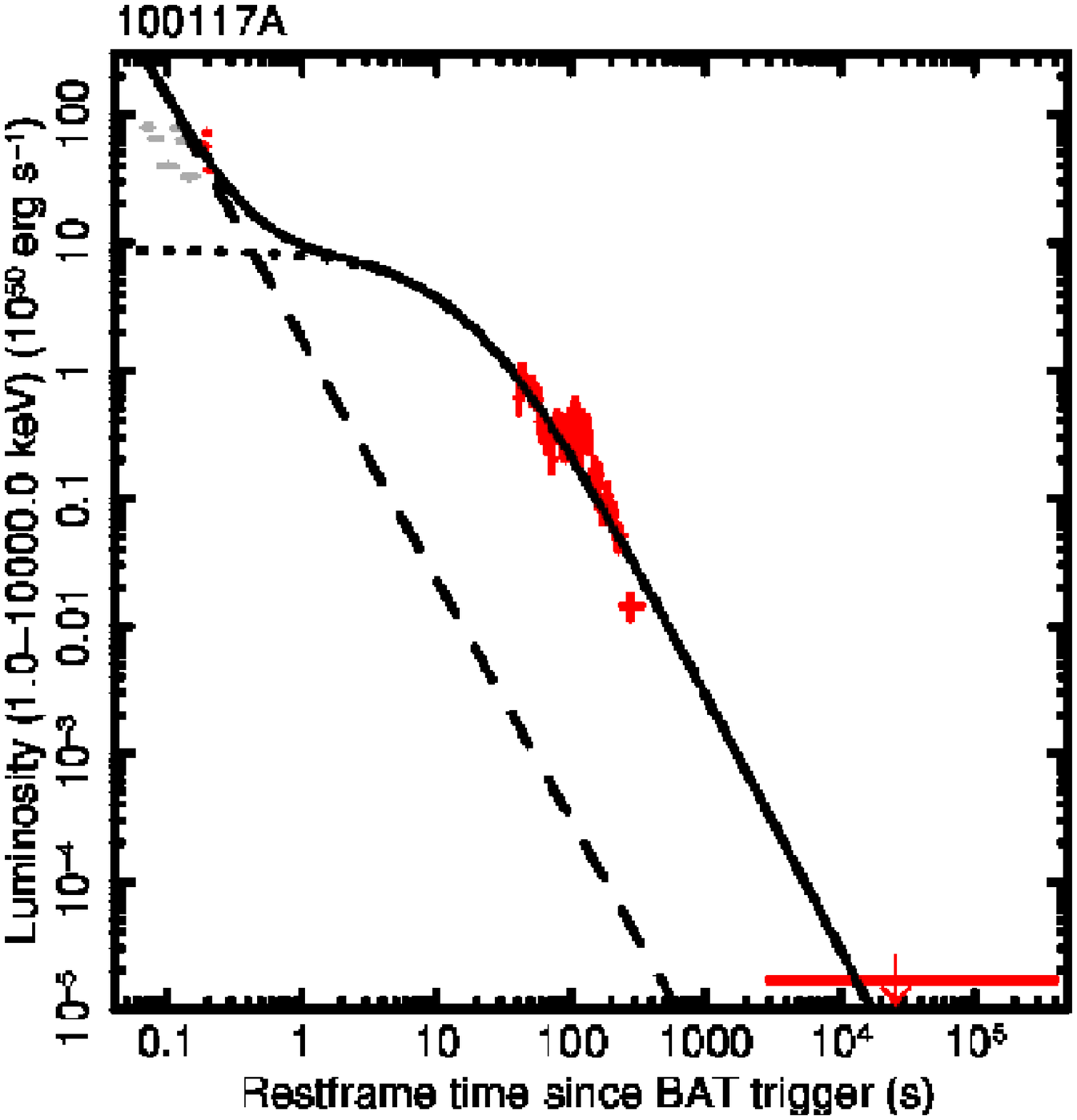}
\includegraphics[width=5.5cm]{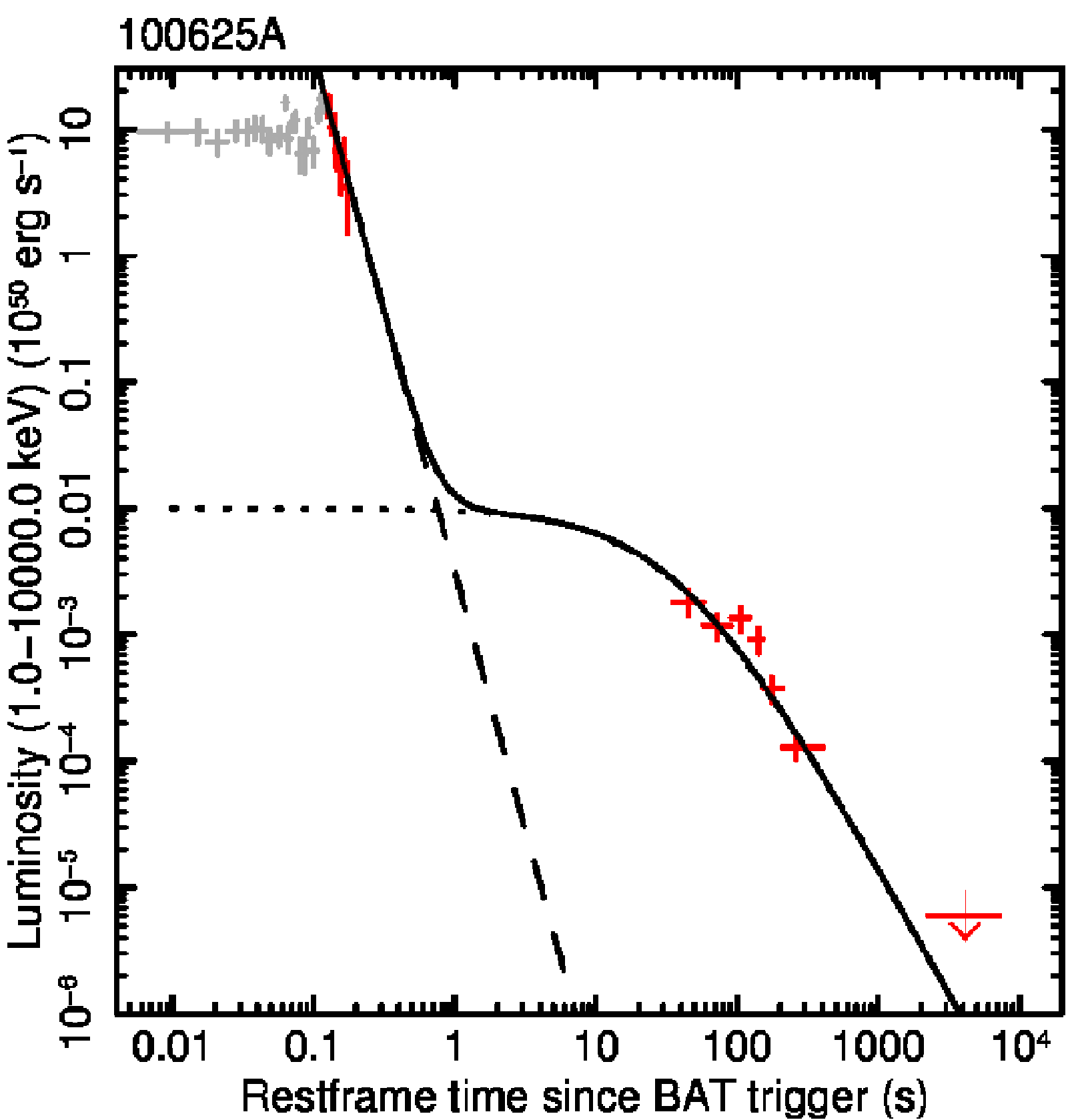}
\includegraphics[width=5.5cm]{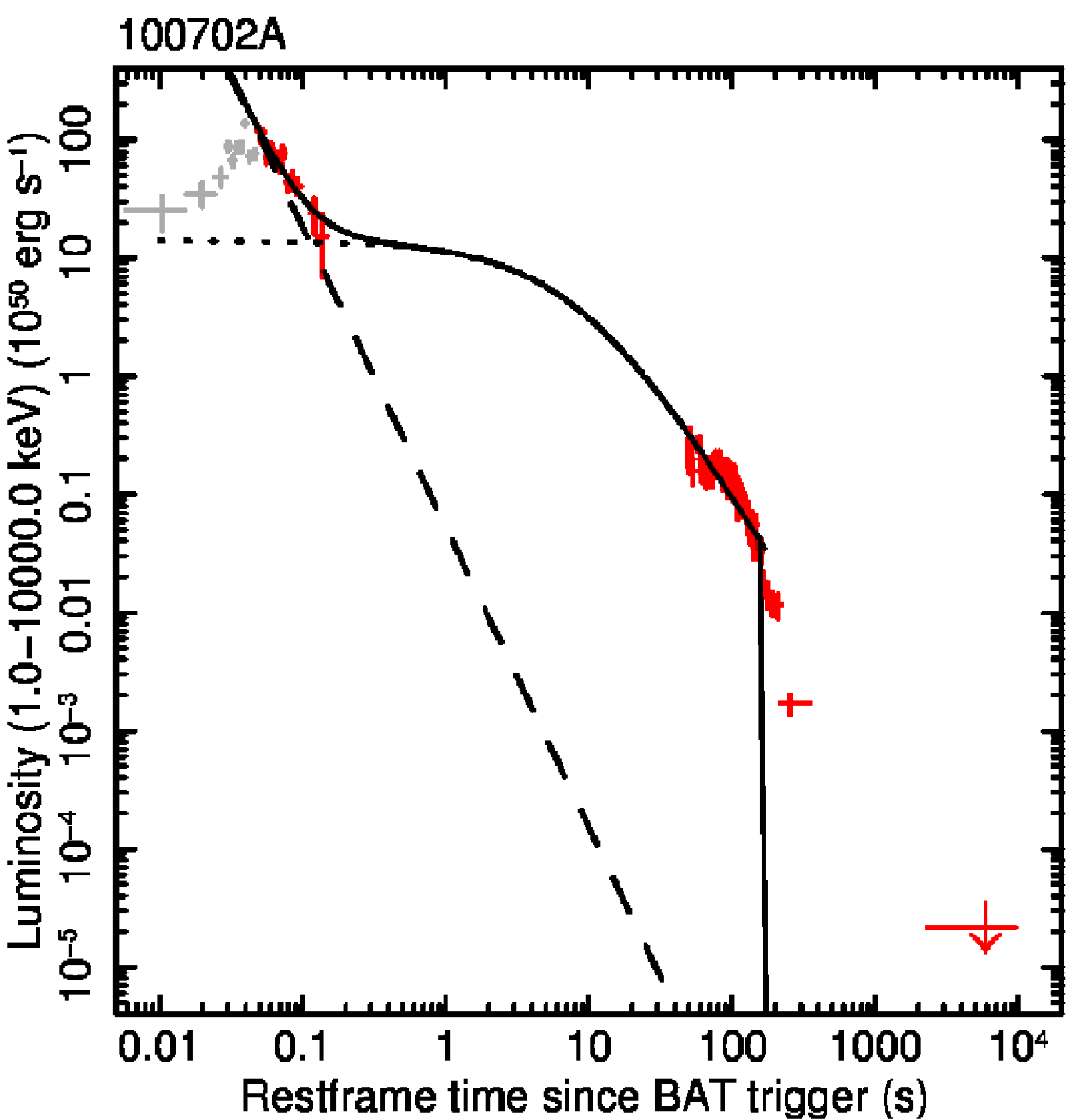}
\includegraphics[width=5.5cm]{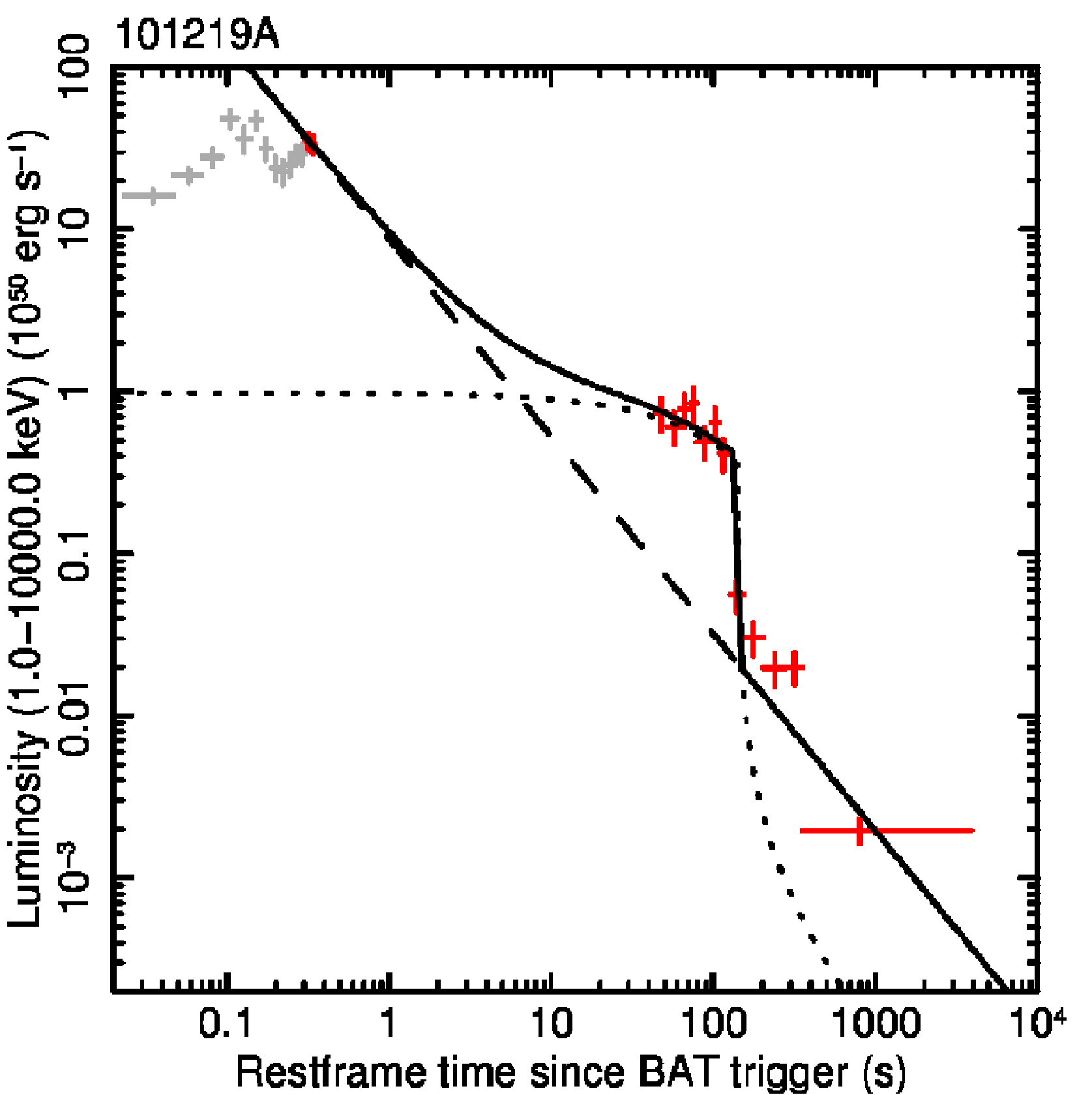}
\includegraphics[width=5.5cm]{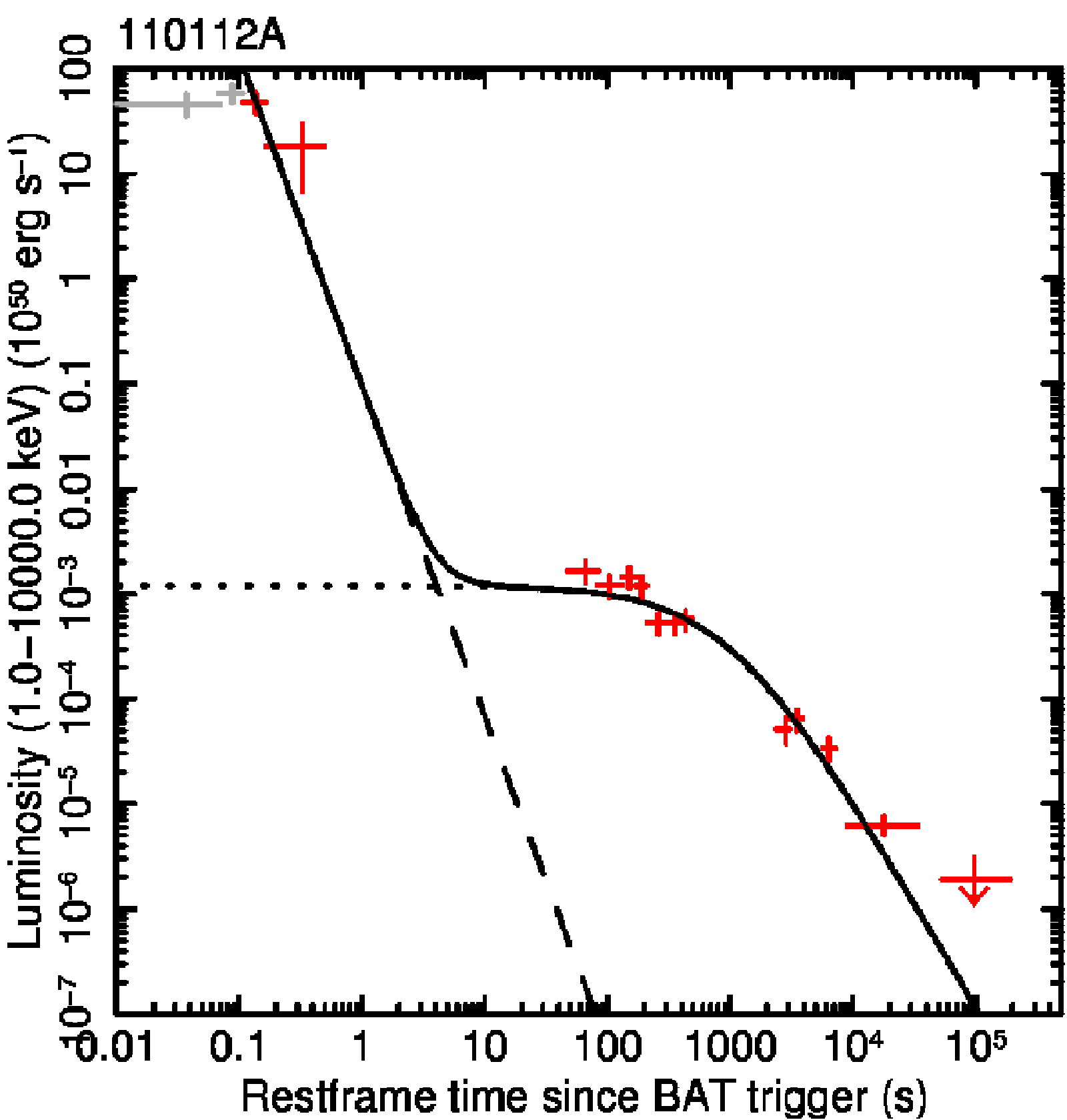}
\contcaption{}
\end{figure*}

\begin{figure*}
\centering
\includegraphics[width=5.5cm]{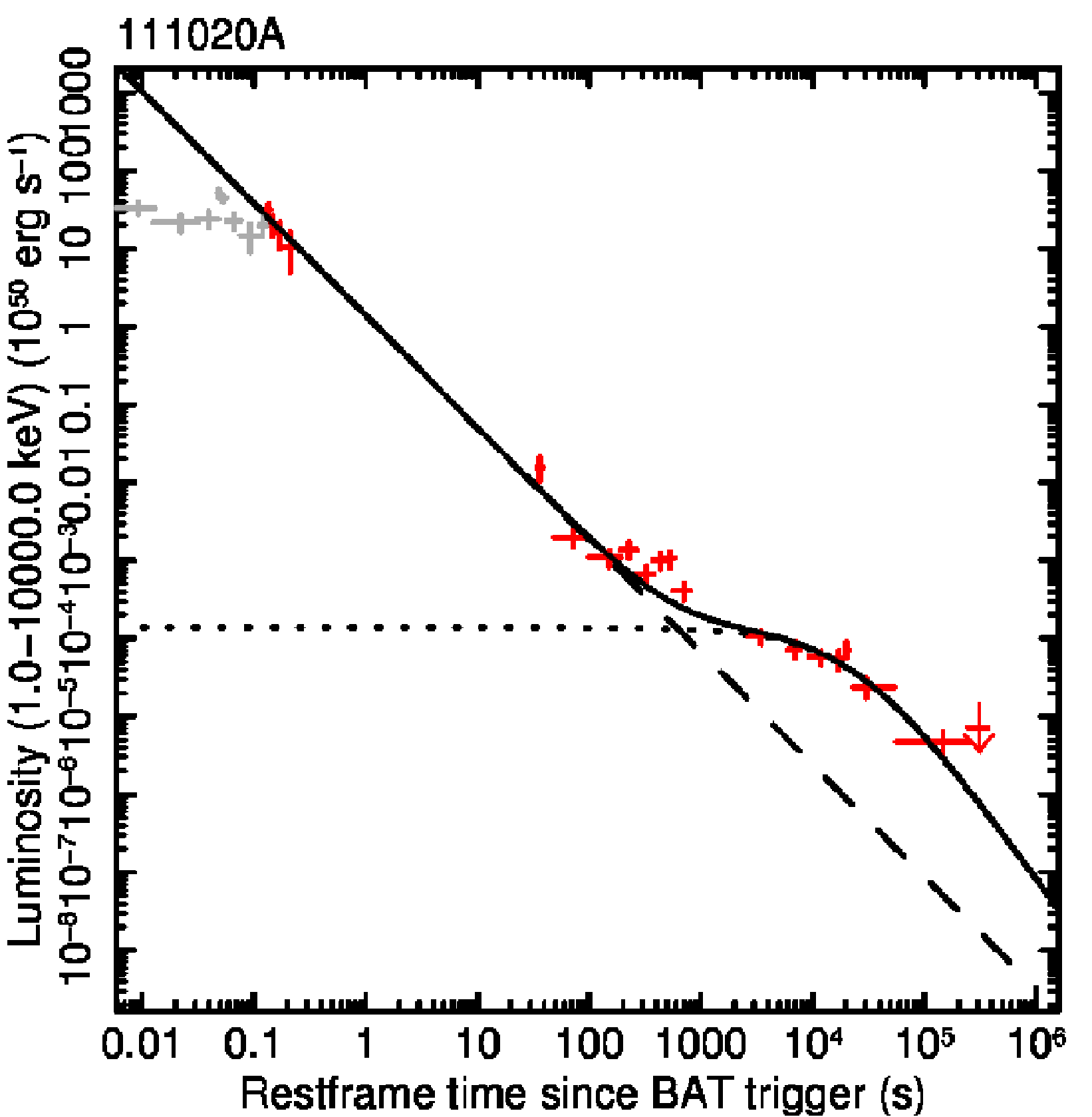}
\includegraphics[width=5.5cm]{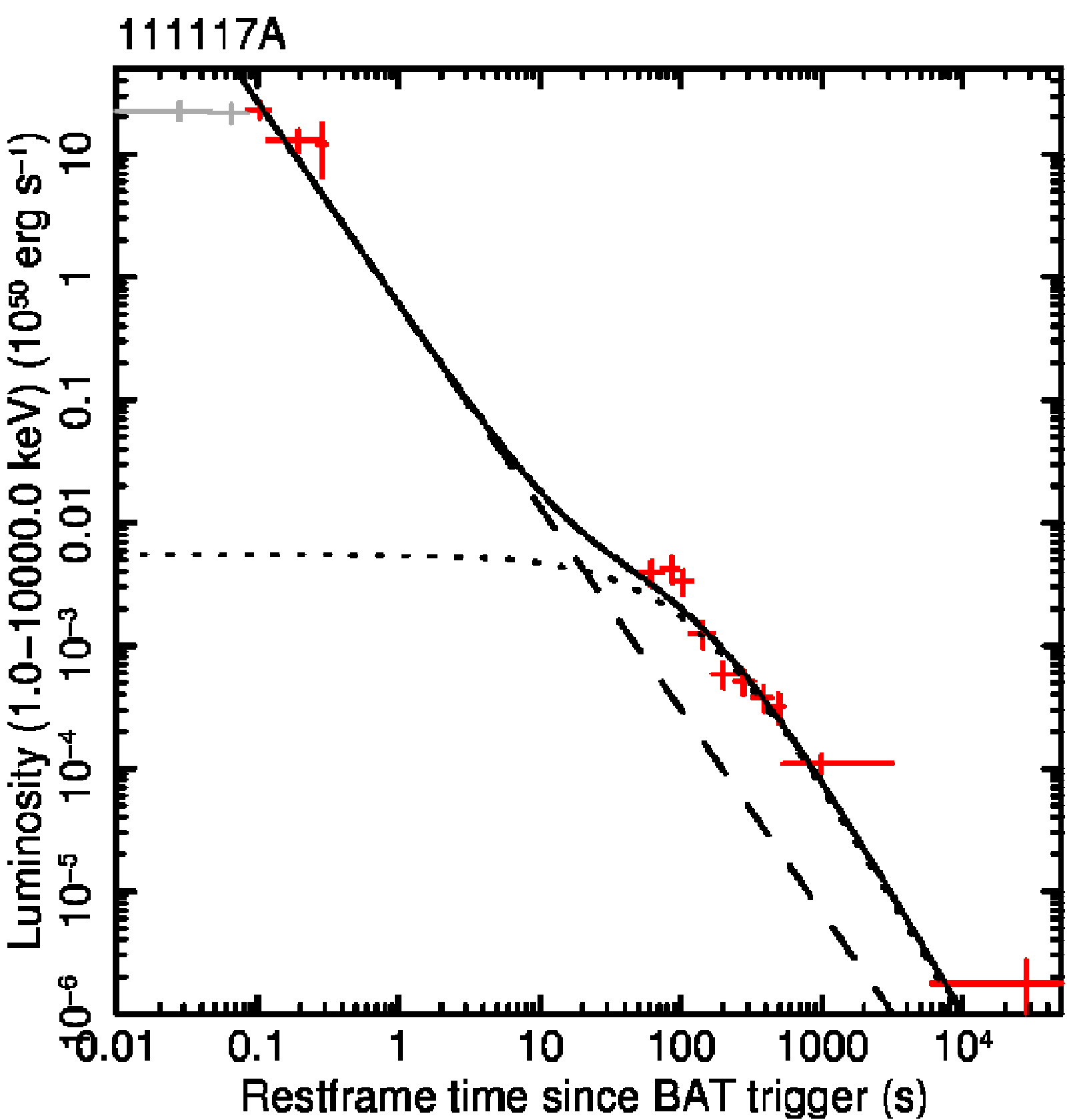}
\includegraphics[width=5.5cm]{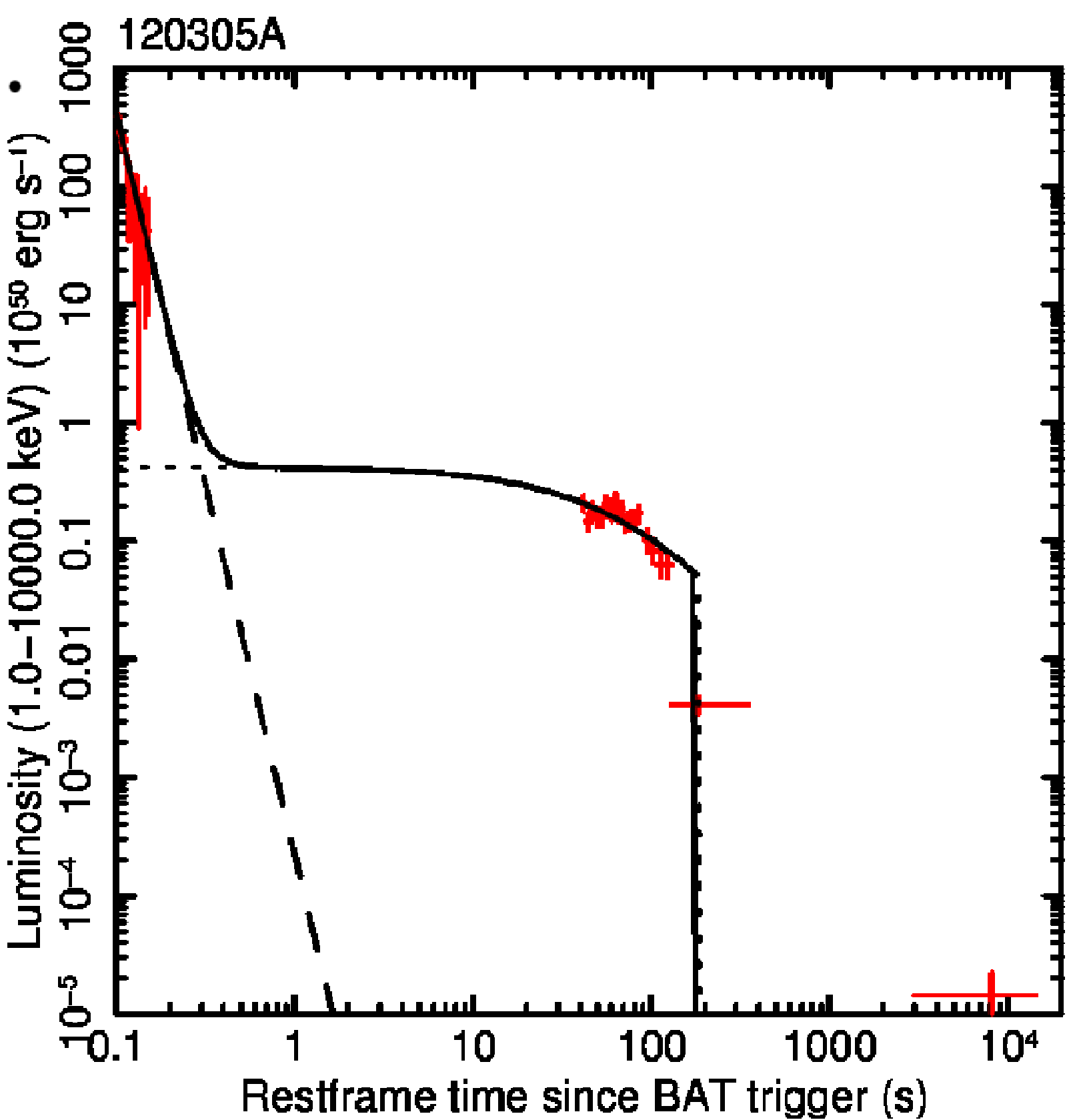}
\includegraphics[width=5.5cm]{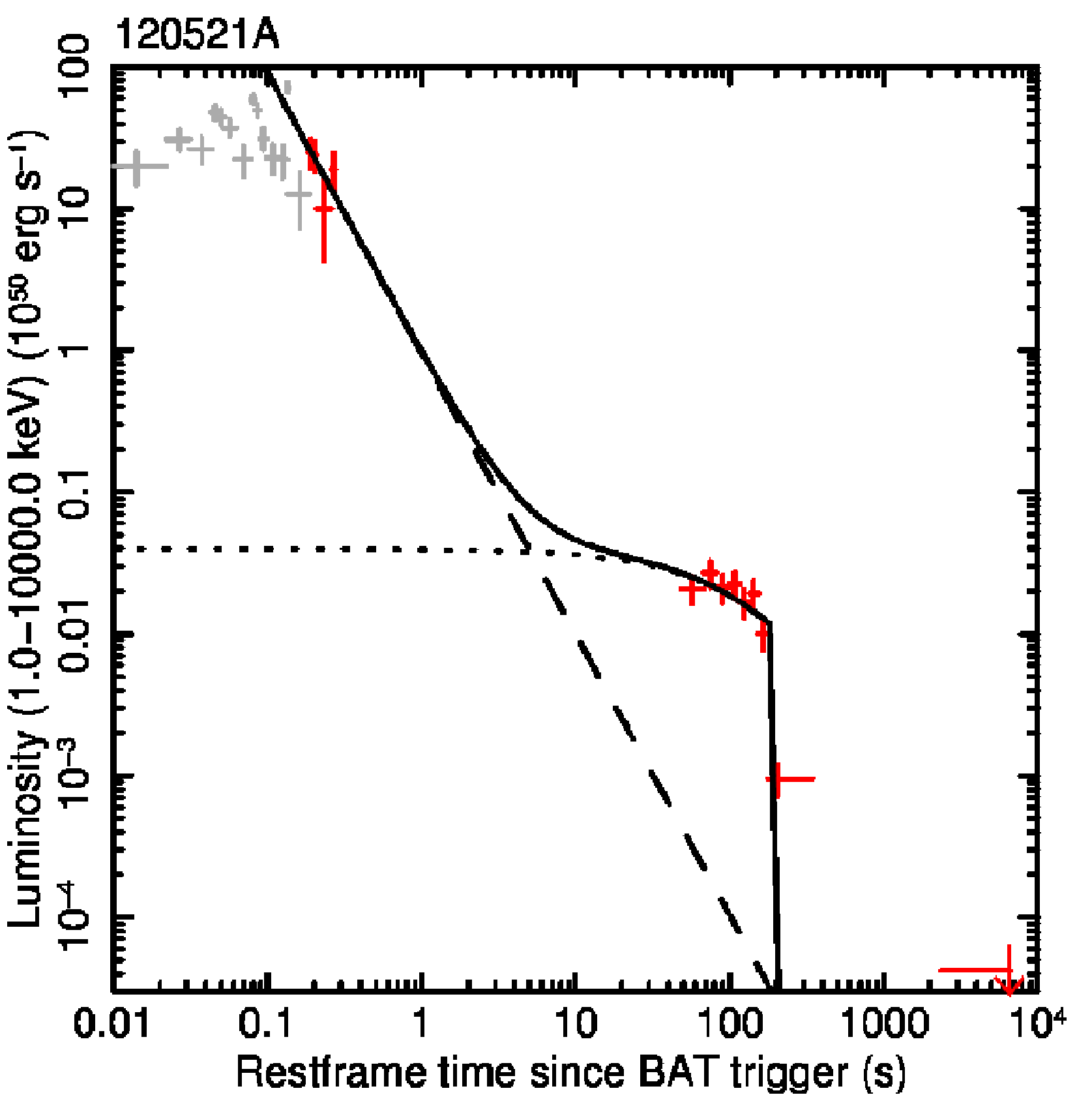}
\contcaption{}
\end{figure*}

\begin{figure*}
\centering
\includegraphics[width=8cm]{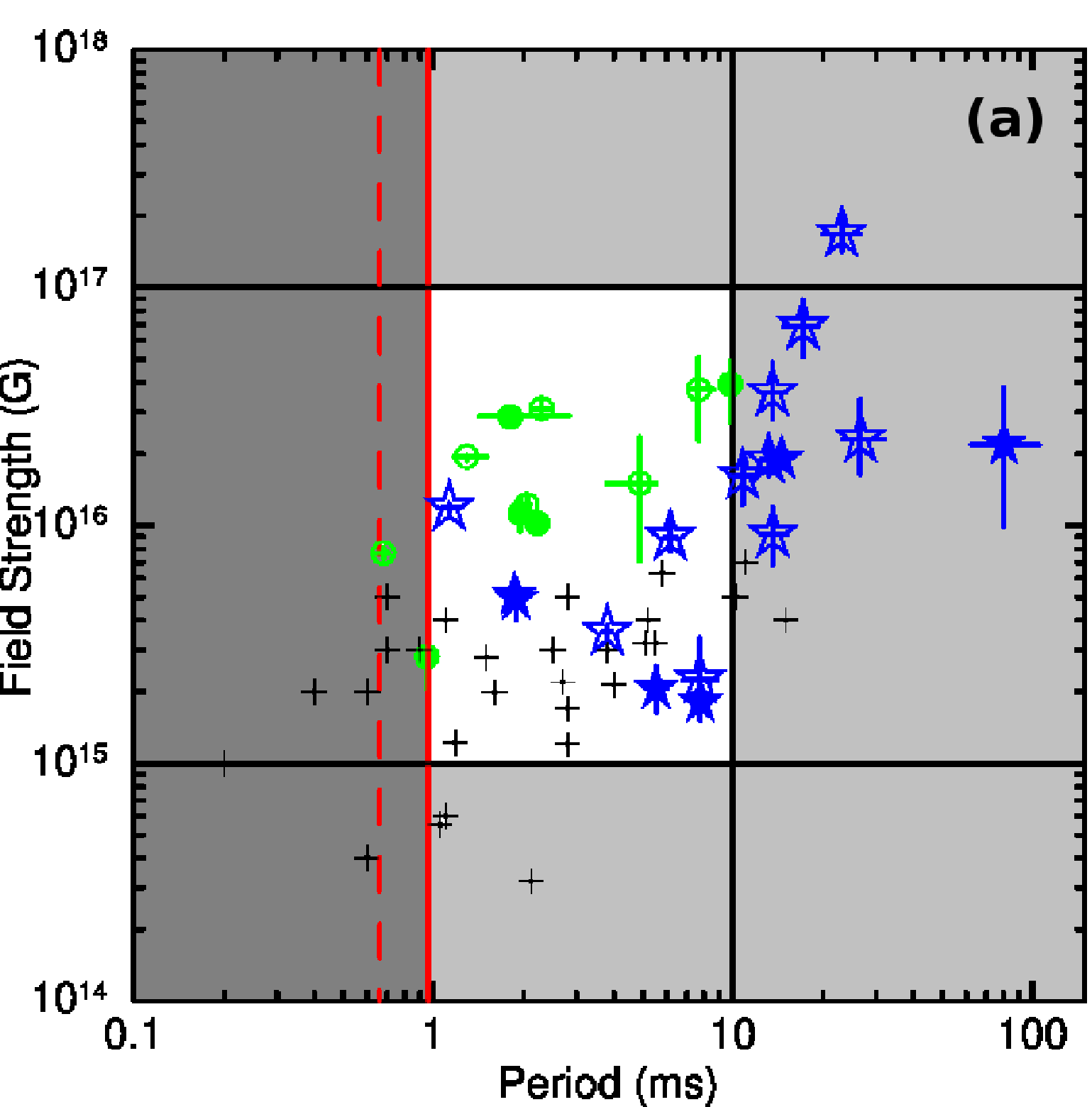}
\includegraphics[width=8cm]{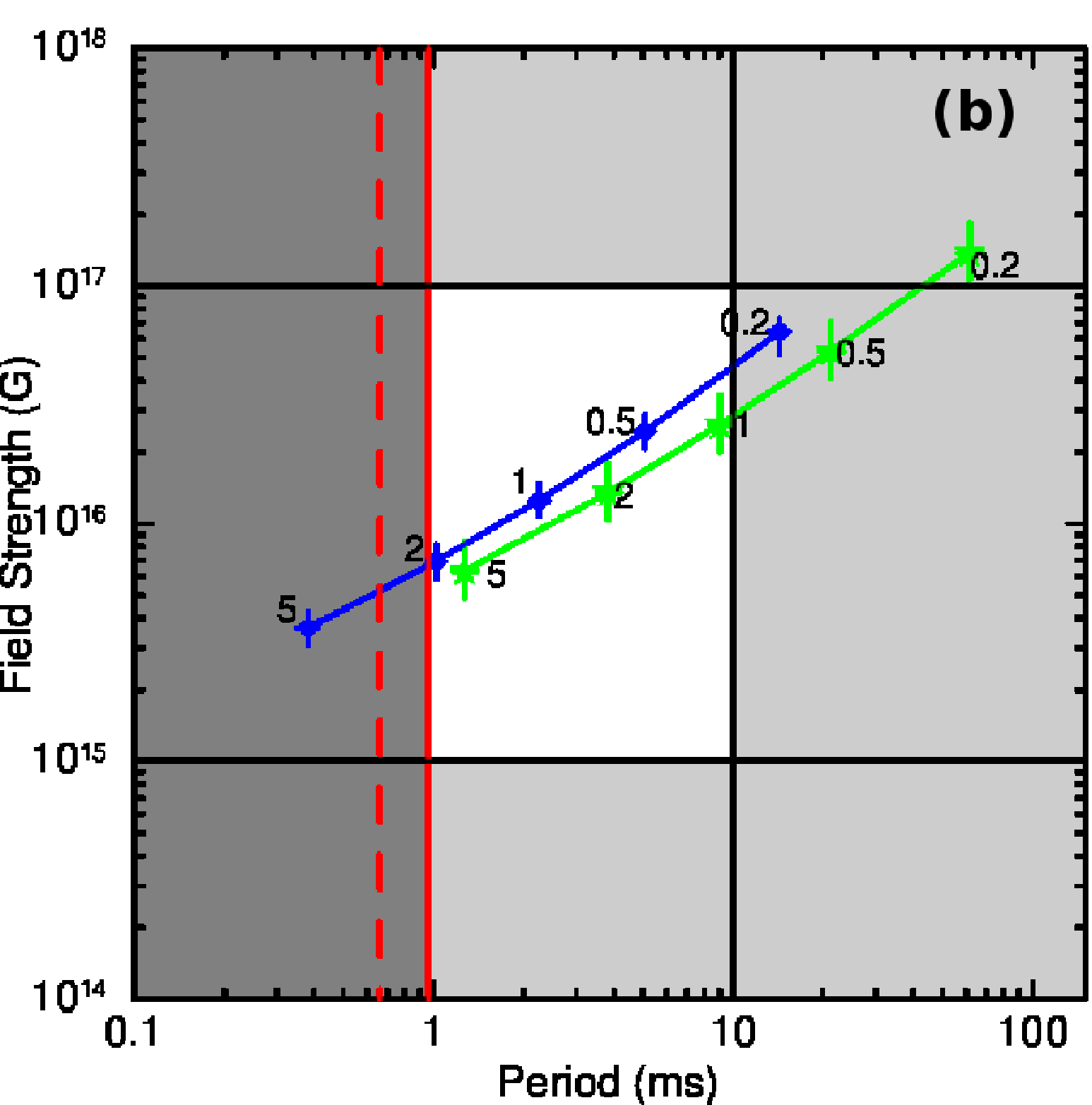}
\caption{(a) A graph showing the magnetic field and spin period of the magnetar fits produced. The solid (dashed) red line and dark shaded area represent the spin break up period for a collapsar (binary merger) progenitor \citep{lattimer2004} and the unshaded region shows the expected region for an unstable pulsar, as defined in \citet{lyons2009} and \citet{rowlinson2010b}. The initial rotation period needs to be $\le$10 ms \citep{usov1992} and the lower limit for the magnetic field is $\ge$10$^{15}$ G \citep{thompson2007}.  Blue stars = stable magnetar and Green circles = unstable magnetar which collapses to form a BH. The black '+' symbols are the LGRB candidates identified by \citet{lyons2009, dallosso2011, bernardini2012}. Filled symbols have observed redshifts whereas open symbols use the average SGRB redshift. (b) This graph is as (a) but focusing on the fits for two GRBs at different redshifts. The number below each data point is the corresponding redshift. GRB 060801 in blue is an unstable magnetar which collapses to form a BH whereas GRB 080702A forms a stable magnetar. As expected, the paths of these lines are consistent with the predictions for GRB 090515 \citep{rowlinson2010b}.}
\label{fig5}
\end{figure*}

\begin{figure*}
\centering
\includegraphics[width=7.8cm]{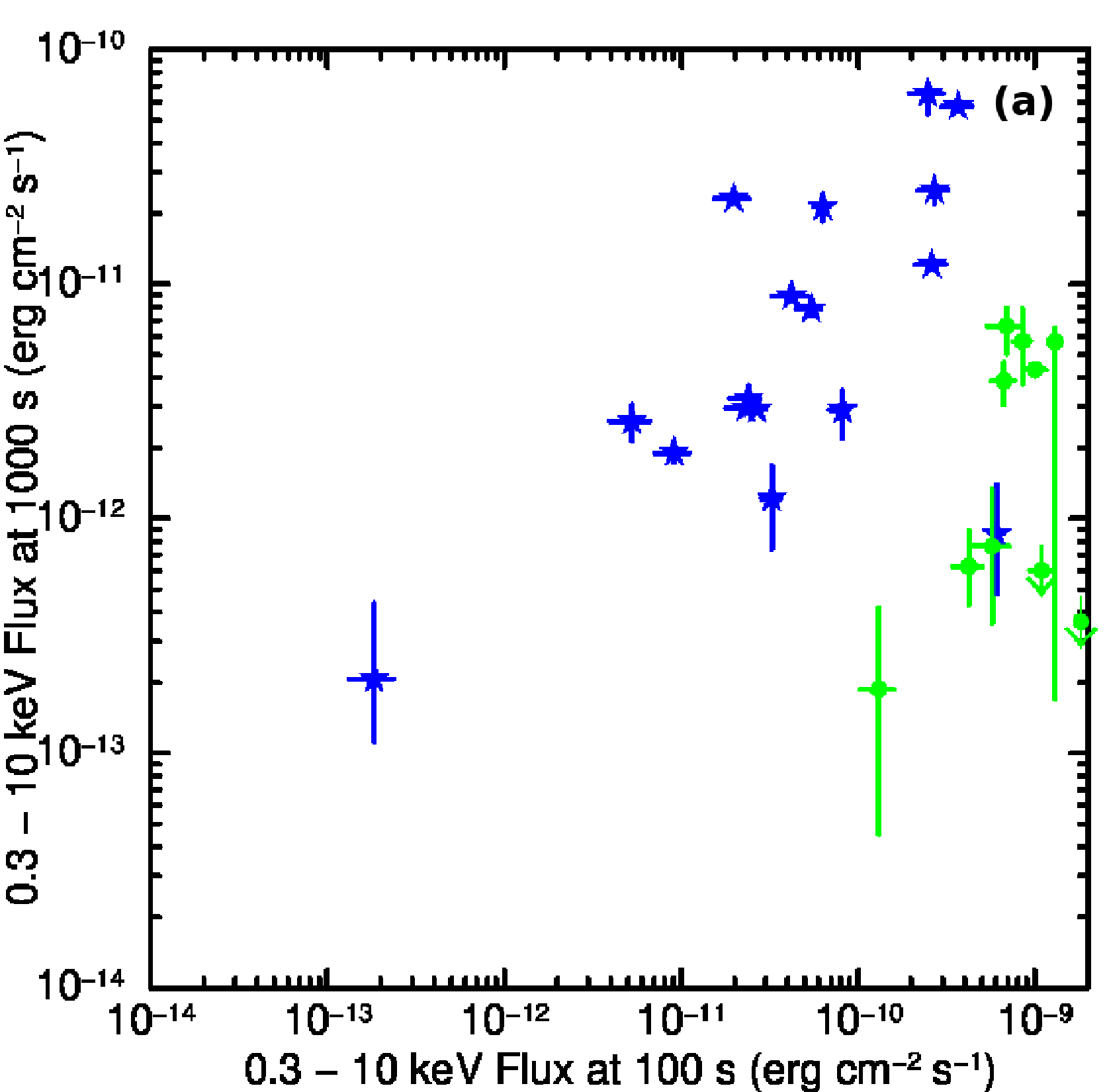}
\includegraphics[width=7.8cm]{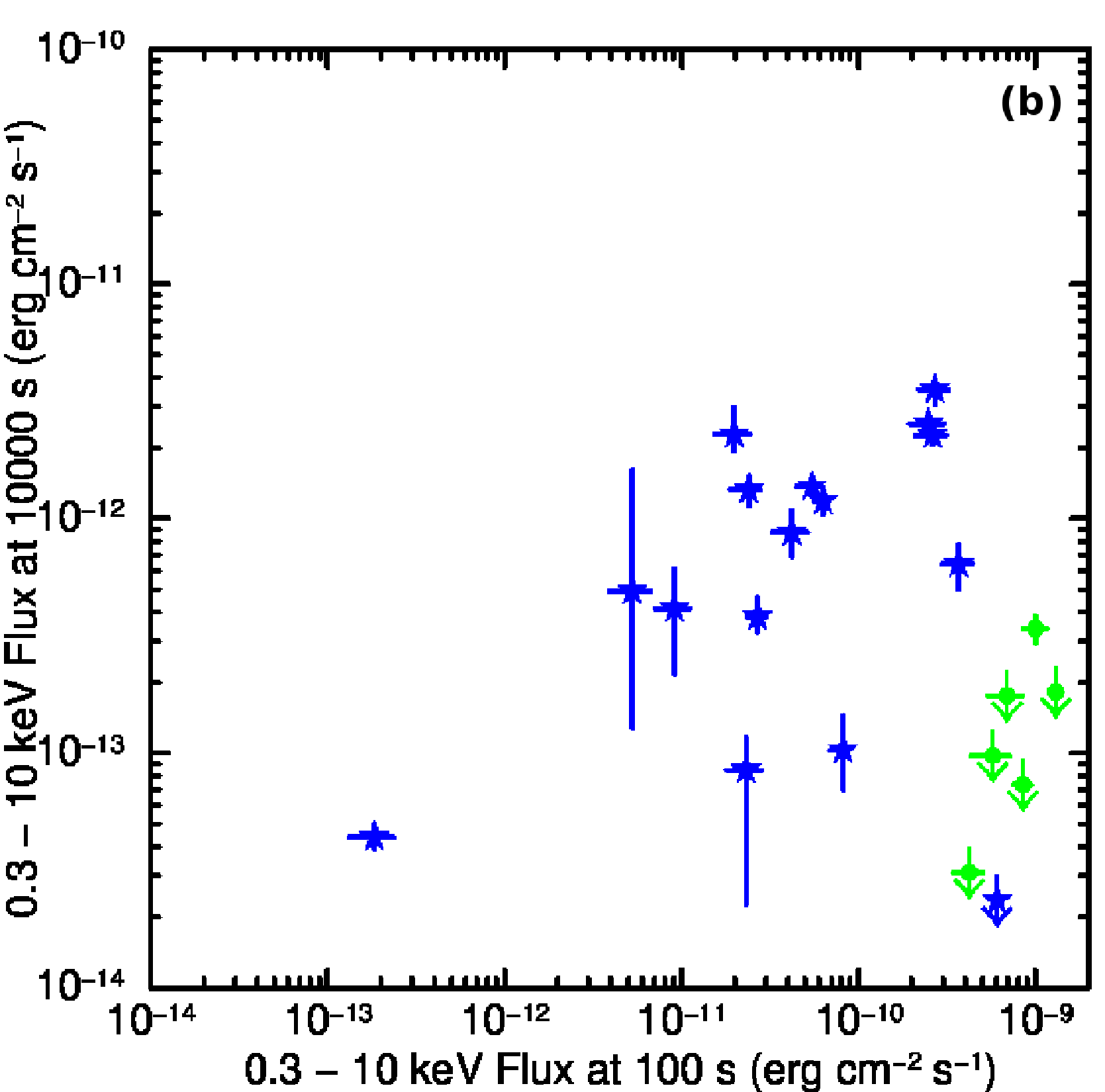}
\includegraphics[width=7.8cm]{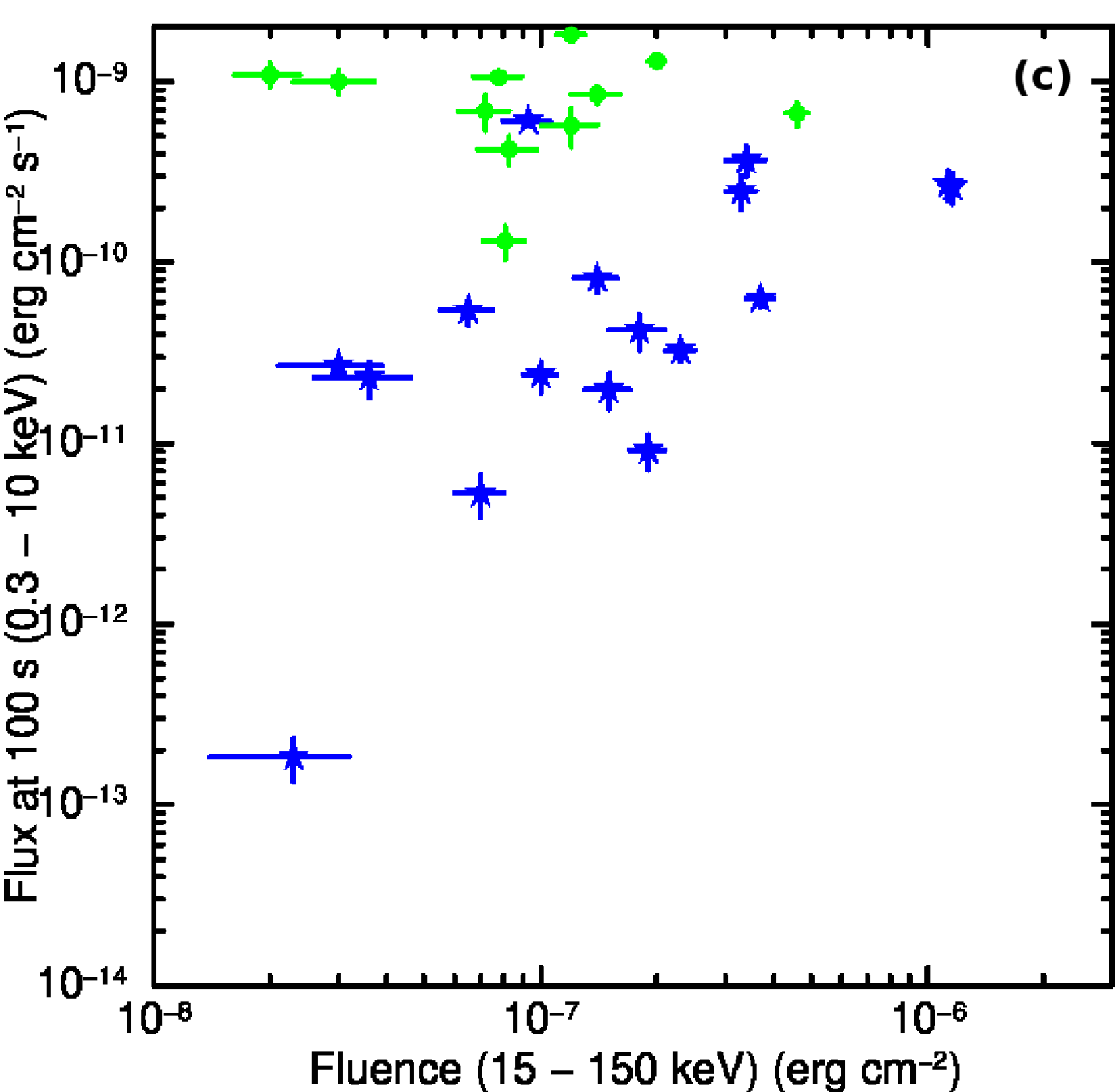}
\includegraphics[width=7.8cm]{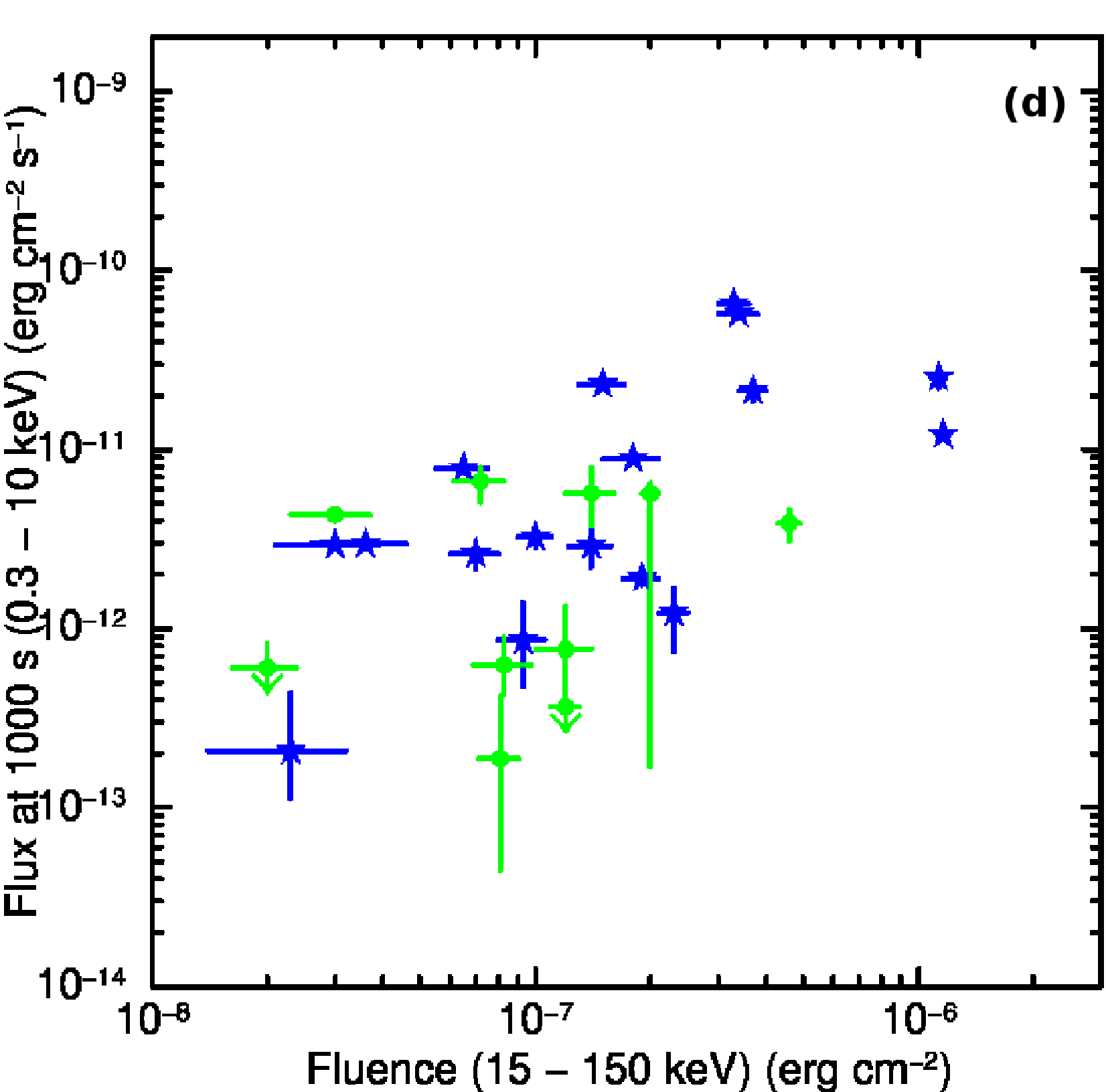}
\caption{(a) The 0.3 -- 10 keV unabsorbed flux at 100 s versus 1000 s. (b) The 0.3 -- 10 keV unabsorbed flux at 100 s versus 10000 s. (c) The 15 -- 150 keV flunce versus the 0.3 -- 10 keV unabsorbed flux at 100 s. (d) The 15 -- 150 keV flunce versus the 0.3 -- 10 keV unabsorbed flux at 1000 s. Symbols are as in Figure \ref{fig5}.}
\label{fig7a}
\end{figure*}

where $T_{em,3}$ is the plateau duration in $10^{3}$ s, $L_{0,49}$ is the plateau luminosity in $10^{49}$ erg s$^{-1}$, $I_{45}$ is the moment of inertia in units of $10^{45}$g cm$^{2}$, $B_{p, 15}$ is the magnetic field strength at the poles in units of $10^{15} G$, $R_{6}$ is the radius of the neutron star in $10^{6}$cm and $P_{0,-3}$ is the initial period of the compact object in milliseconds. These equations apply to the electromagnetic dominated spin down regime, as the gravitational wave dominated regime would be extremely rapid and produce a negligble electromagnetic signal. We have assumed that the emission is 100\% efficient and isotropic as the beaming angle and emission mechanism remains very uncertain (see however section 3.4.4). The equations of vacuum dipole spin-down given above neglect the enhanced angular momentum losses due to neutrino-driven mass loss, which are important at early times after the magnetar forms \citep{metzger2010}. Nevertheless, these expressions reasonably approximate the spin-down of very highly magnetized neutron stars of most relevance in this paper. Isotropic emission is also a reasonable assumption for relatively powerful magnetar winds, since (unlike following the collapse of a massive star) the magnetar outflow cannot be confined efficiently by the relatively small quantity of surrounding material expected following a NS merger or AIC \citep{bucciantini2011}.

We use equation \ref{inertia} to obtain the mass dependence of the model, where $M_{1.4} = 1.4 M_{\odot}$, and equation \ref{time_dep} \citep[from ][]{zhang2001} to determine the time dependence of the magnetar emission.

\begin{eqnarray}
I_{45} \sim M_{1.4}R_{6}^2 \label{inertia}\\
L_{em,49}(T) = L_{0,49}\left( 1 + \frac{T}{10^{-3}T_{em,3}} \right)^{-2} \label{time_dep}
\end{eqnarray}

If there is a steep decay phase after the plateau, it is assumed the magnetar has collapsed to a BH at the start of the steep decay (giving the Collapse Time parameter). The decay after collapse to a BH assumes the same powerlaw decay from the curvature effect, but starting at $t_0 = t_{collapse}$. 

This model was then written into a {\sc QDP} cod file (COmponent Definition file, used to generate new models within {\sc QDP} which can then be fitted to data sets). In this analysis, the mass ($M_{1.4}$) and radius ($R_6$) of the neutron star are constrained to be equal to 1 to reduce the number of free parameters in our model. These canonical values are consistent with the values determined by observations of three typical neutron stars, namely $M\le2 M_\odot$ and $7\le R \le 11$ km \citep{ozel2010}. As the model considers an extreme neutron star, we note that the mass and radius may differ from these results. However, this only has a relatively small affect on the magnetic fields and spin periods calculated \citep[as shown in][]{rowlinson2010b} and so it is a reasonable approximation as we are just demontrating the plausibility of the magnetar model fitting the SGRB lightcurves. When this model is fit to the restframe lightcurves it produces B$_{p,15}$, P$_{0,-3}$, $\alpha_{1}$ and the collapse time where appropriate. 

\begin{table}
\begin{center}
\begin{tabular}{|c|c|c|c|}
\hline
GRB     & $\Gamma_{X}$   & Galactic N$_{H}$           & Restframe Intrinsic N$_{H}$ \\
        &                        & (10$^{20}$ cm$^{-2}$)      & (10$^{20}$ cm$^{-2}$)       \\
\hline
\multicolumn{4}{|l|}{Magnetar candidates}\\
\hline
051221A & 2.04$^{+0.14}_{-0.13}$ & 5.70$\pm$0.37              & 18.0$^{+7.10}_{-6.60}$      \\
060313  & 1.61$^{+0.16}_{-0.13}$ & 5.00$\pm$1.17              & 0.00$^{+5.84}_{-0.00}$      \\
060801  & 1.53$^{+0.47}_{-0.43}$ & 1.40$\pm$0.31              & 29.9$^{+68.8}_{-29.9}$      \\
070809  & 1.73$^{+0.83}_{-0.43}$ & 6.40$\pm$0.17              & 2.95$^{+14.9}_{-2.95}$      \\
080426  & 1.93$^{+0.29}_{-0.27}$ & 37.0$\pm$4.19              & 32.0$^{+31.6}_{-25.5}$      \\
080919  & 2.23$^{+1.02}_{-0.84}$ & 26.0$\pm$3.78              &  105$^{+126}_{-75.8}$      \\
090426  & 2.03$^{+0.19}_{-0.11}$ & 1.50$\pm$0.11              & 0.00$^{+36.0}_{-0.00}$      \\
090510  & 1.56$^{+0.20}_{-0.19}$ & 1.70$\pm$0.11              & 10.0$^{+16.0}_{-10.0}$      \\
090515  & 1.89$^{+0.25}_{-0.24}$ & 1.90$\pm$0.25              & 13.1$^{+11.6}_{-10.5}$      \\
101219A & 1.65$^{+0.32}_{-0.31}$ & 4.90$\pm$0.87              & 56.8$^{+26.7}_{-20.4}$      \\
111020A & 2.56$^{+1.69}_{-1.69}$ & 6.89$\pm$0.48              & 7.94$^{+7.90}_{-7.90}$      \\
120305A & 1.94$^{+0.21}_{-0.20}$ & 11.3$\pm$0.70              &  109$^{+32}_{-26}$          \\
120521A & 1.61$^{+0.36}_{-0.22}$ & 20.80$\pm$1.69             &  1.2$^{+14.2}_{-1.2}$          \\
\hline
\multicolumn{4}{|l|}{Possible candidates}\\
\hline
050509B & 1.92$^{+1.09}_{-0.60}$ & 1.60$\pm$0.04              & 8.00$^{+8.10}_{-8.00}$      \\
061201  & 1.44$^{+0.20}_{-0.19}$ & 5.20$\pm$1.58              & 6.77$^{+4.25}_{-3.88}$      \\
070714A & 2.12$^{+0.37}_{-0.35}$ & 9.20$\pm$1.25              &  214$^{+51.8}_{-45.7}$      \\
080702A & 1.57$^{+0.85}_{-0.76}$ & 15.0$\pm$1.50              &  125$^{+251}_{-121}$      \\
090621B & 2.50$^{+1.60}_{-1.00}$ & 19.0$\pm$1.96              & 42.8$^{+108}_{-42.8}$      \\
091109B & 1.96$^{+0.64}_{-0.43}$ & 9.20$\pm$0.96              & 14.5$^{+27.9}_{-14.5}$      \\
110112A & 2.07$^{+0.46}_{-0.24}$ & 5.50$\pm$0.40              & 7.86$^{+12.7}_{-7.86}$      \\
111117A & 2.13$^{+0.39}_{-0.36}$ & 3.70$\pm$0.15              & 39.8$^{+69.7}_{-31.3}$      \\
\hline
\end{tabular}
\caption[Plateau spectral fits for the SGRB magentar sample]{The 0.3 -- 10 keV spectral fits for the derived plateau durations given in Table \ref{table:log}. These are the SGRBs in the magnetar sample which have X-ray data during the plateau phase. Provided are the photon index, $\Gamma_{X, plateau}$, the Galactic N$_{H}$ and the restframe intrinsic N$_{H}$ using the redshifts provided in Table \ref{candidates}.} 
\label{spectra}
\end{center}
\end{table}

\subsection{The Sample GRBs for magnetar fits}

The selected GRBs are those SGRBs in our sample with sufficient data to produce multiple data points in the X-ray lightcurve, giving a sample of 28 SGRBs. GRBs which have insufficient data to fit the magnetar model are not excluded from being magnetar candidates as it's possible to fit a range of realistic magnetar parameters with the minimal data points and unknown redshift. 68\% of SGRBs in our sample have been investigated for evidence of extended emission by \cite{norris2010} but, of these, none show evidence of extended emission. The remaining SGRBs in our sample have no evidence of extended emission in their lightcurves (using a variety of binning in signal-to-noise ratios and time looking for evidence of extended emission at the 3$\sigma$ level).

The magnetar sample are listed in Table \ref{table:log}. The restframe BAT-XRT lightcurves were fitted using the magnetar model, as shown in Figure \ref{fig1}. The lightcurves are fit over plateau region and the power law decay, including the last decay in the prompt emission and the X-ray observations. This removes the effect of the poorly understood flaring prompt emission not modeled by this method. We also provide the derived plateau luminosity and plateau duration calculated using the magnetic field strengths, the spin periods and equations \ref{period} and \ref{luminosity}. The magnetar candidates fit the model well and the possible candidates are GRBs which may fit the magnetar model if various assumptions are made. There are two potential outcomes: a stable long lived magnetar which does not collapse to form a BH and an unstable magnetar which collapses forming a BH after a short timescale (which have a collapse time in Table \ref{table:log}). The following Sections compare the properties of the stable magnetars (blue stars in the figures) and the unstable magnetars which collapse to form a BH (green circles). We note that the fitted plateaus match the observations well but, due to insufficient data points particularly prior to XRT observations, the plateaus are not always required by the observed data which can be fitted by simple broken power-law models. In some cases, the best fitting magnetar model gives a plateau phase ending prior to the start of the XRT observations (e.g. 060801). In this situation, the fit is being constrained by the curving of the magnetar energy injection from a plateau phase to a powerlaw decline giving a characteristic curvature in the lightcurve (described by Equation 6). Therefore, the fitted model does not rely upon data during the plateau phase but instead uses the whole shape of the lightcurve. This leads to the model predition that those GRBs have a magnetar plateau phase which has not been directly observed, this can be used to test the model if we are able to observe SGRBs much sooner after the prompt emission with future X-ray telescopes.

When fitting GRB 060313, which may show evidence of late time central engine activity \citep{roming2006}, it was noted that the model fits part of the lightcurve extremely well. In this case, we ignored the observations between 50 -- 200 s (the initial X-ray data) in the fit as this duration appears to be dominated by flares. If these data are included in the fit, then the model does not fit the data well. The model fits well to GRB 090515 predicting values similar to those given in \cite{rowlinson2010b}.

In some cases, the model used here under predicts the flux at late times (for example GRBs 091109B, 100702A and 120305A). This shows that our simple power law component, given by a simple curvature effect model, is not sufficient and we should include spectral evolution or there may also be an additional afterglow component which has been neglected in this model.

\subsection{Analysis}

In Figure \ref{fig5}(a) we show the spin periods and magnetic fields determined for our sample of GRBs assuming isotropic emission. We also plot the LGRB candidates identified by \cite{lyons2009}, \cite{dallosso2011} and \cite{bernardini2012}, the SGRB candidates tend to have higher magnetic field strengths and spin periods. In Figure \ref{fig5}(b), we confirm the change in magnetic field strength and spin period caused by uncertainties in redshift expected from previous analysis of GRB 090515 \citep{rowlinson2010b}. 18 of the SGRBs fitted by the magnetar model lie within the expected region of the magnetic field strength and spin periods, these are the magnetar candidates listed in Table \ref{candidates}. 10 GRBs are outside the expected region (the possible candidates in Table \ref{candidates}). These GRBs may be in the expected (unshaded) region if they were at a higher redshift as shown in \cite{rowlinson2010b} and Figure \ref{fig5}(b). Additionally, this region is defined using angular momentum conservation during the AIC of a WD \citep{usov1992} and is not a physically forbidden region. Therefore, the candidates with spin periods $>$10 ms may remain good candidate magnetars. GRB 051210 is included in the possible candidates list as it is spinning faster than is allowed in the models, but it is worth noting that if the NS formed had a mass of 2.1M$_{\odot}$ then it would reside within the allowed region, as more massive NSs are able to spin at a faster rate. It is also worth noting that if GRB 051210 occurred at a lower redshift, as shown in Figure \ref{fig5}(b), or if the emission is significantly beamed then the spin period and magnetic field strengths would be higher and GRB 051210 would not be near to the spin break up period. The unstable magnetar candidates tend to have higher magnetic field strengths for their spin periods than the stable magnetar candidates. The only exceptions are GRBs 100117A, which has been fitted with a stable magnetar model but would also be consistent with forming an unstable magnetar, and GRB 090426. 

\subsubsection{Prompt and X-ray Properties}

In Figures \ref{fig7a}(a) and (b), the 0.3 -- 10 keV flux at 1000 s or 10000s are compared to the flux at 100 s. The stable magnetar candidates tend to have a higher flux at 1000 s than the GRBs which are modelled as collapsing to a BH. This graph can be explained if we assume all SGRBs are occuring in a low density environment, resulting in little afterglow, and the only observed emission results from the curvature effect. The magnetar candidates which collapse to form a BH fade rapidly, whereas the stable magnetars are giving prolonged energy injection giving the higher late time X-ray fluxes. The stable magnetar candidate outlier in Figures \ref{fig7a}(a) and (b) is GRB 100117A and it has already been noted that this GRB would also be fitted well by an unstable magnetar model. This analysis suggests that mergers collapsing straight to BHs have significantly fainter X-ray afterglows, which fade rapidly, and hence there may be a selection bias against these objects in our analysis (as we required sufficient data points to fit the model). In Figures \ref{fig7a}(c) and (d) we plot the flux at 100 s and 1000 s versus the prompt 15 -- 150 keV fluence observed. At 100 s the unstable magnetar candidates clearly have a higher flux than comparable stable magnetar candidates (again GRB 100117A is the outlier) although this separation of the two populations has vanished by 1000 s.

For each GRB in the sample, a 0.3 -- 10 keV XRT spectrum \citep[using the automatic data products on the UK Swift Data Centre website;][]{evans2007,evans2009} for the model derived rest frame plateau duration (converted to observed frame durations) was extracted to compare the spectral properties in the proposed magnetar emission phase. This was not possible for some of the sample as XRT observations started after the plateau phase had ended. Each spectrum was fitted in {\sc XSpec} using a power law, $\Gamma_{X}$, the Galactic N$_{H}$ \citep[neutral hydrogen column density, taken from ][]{kalberla2005} and the intrinsic N$_{H}$ at the redshift provided in Table \ref{candidates}. The spectral fits are provided in Table \ref{spectra}.

The majority of the SGRBs are consistent with having negligible intrinsic N$_{H}$ observed in their spectra suggesting they are likely to have occured in low density environments. Recently, \cite{margutti2012} have compared the distribution of intrinsic N$_{H}$ observed in SGRBs to LGRBs finding that they are typically consistent with the lower end of the LGRB distribution consistent with the higher end of our distribution and we find several candidates with negligible intrinsic absorption. Some of the sample have significant N$_{H}$ values, but it is important to note that detailed observations have shown that the optical absorptions found for GRB afterglows can be orders of magnitude less than that expected from the X-ray N$_{H}$ values \citep{schady2010, campana2010}.

\subsubsection{Optical Afterglows}

\begin{figure*}
\centering
\includegraphics[width=8.5cm]{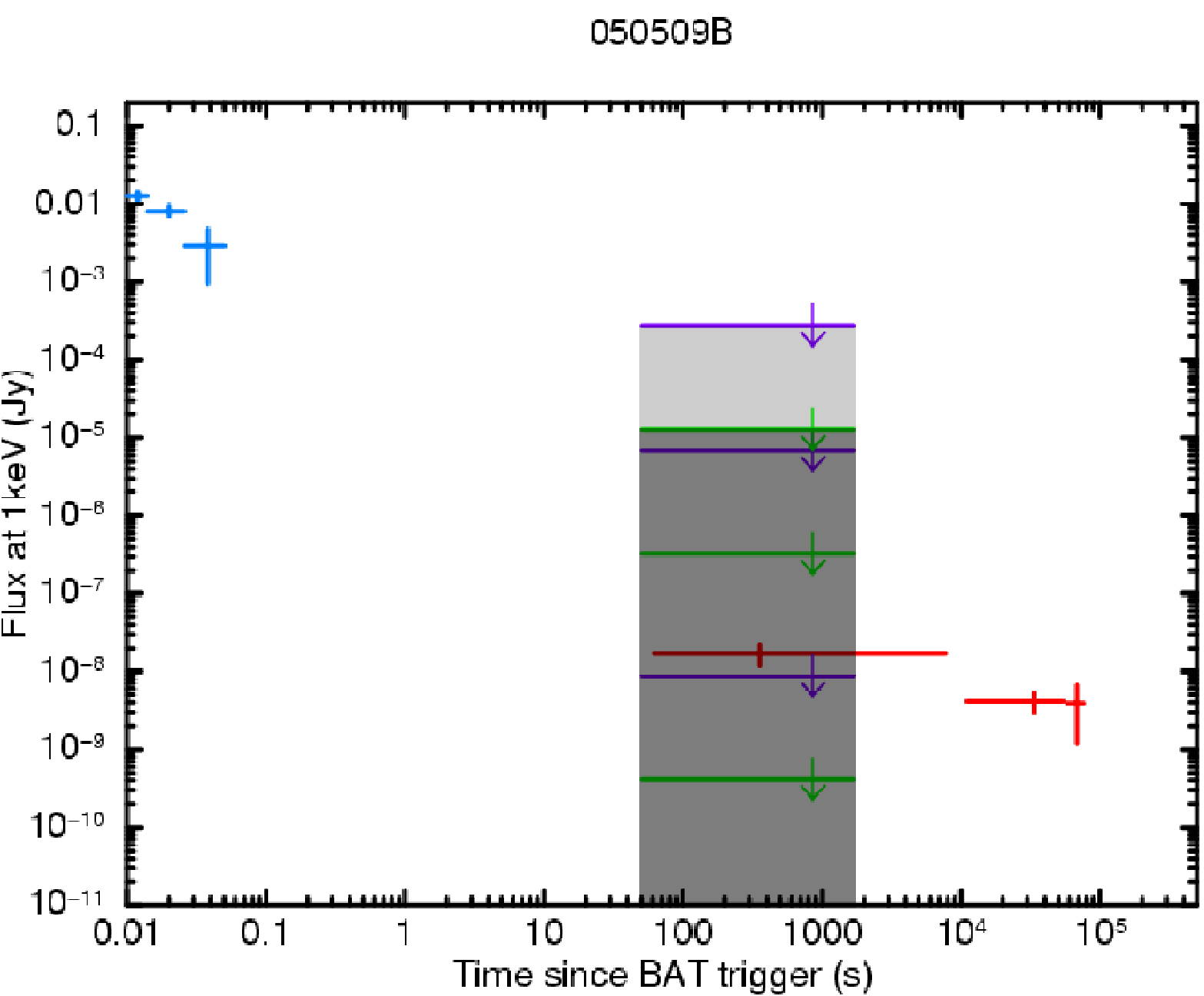}
\includegraphics[width=8.5cm]{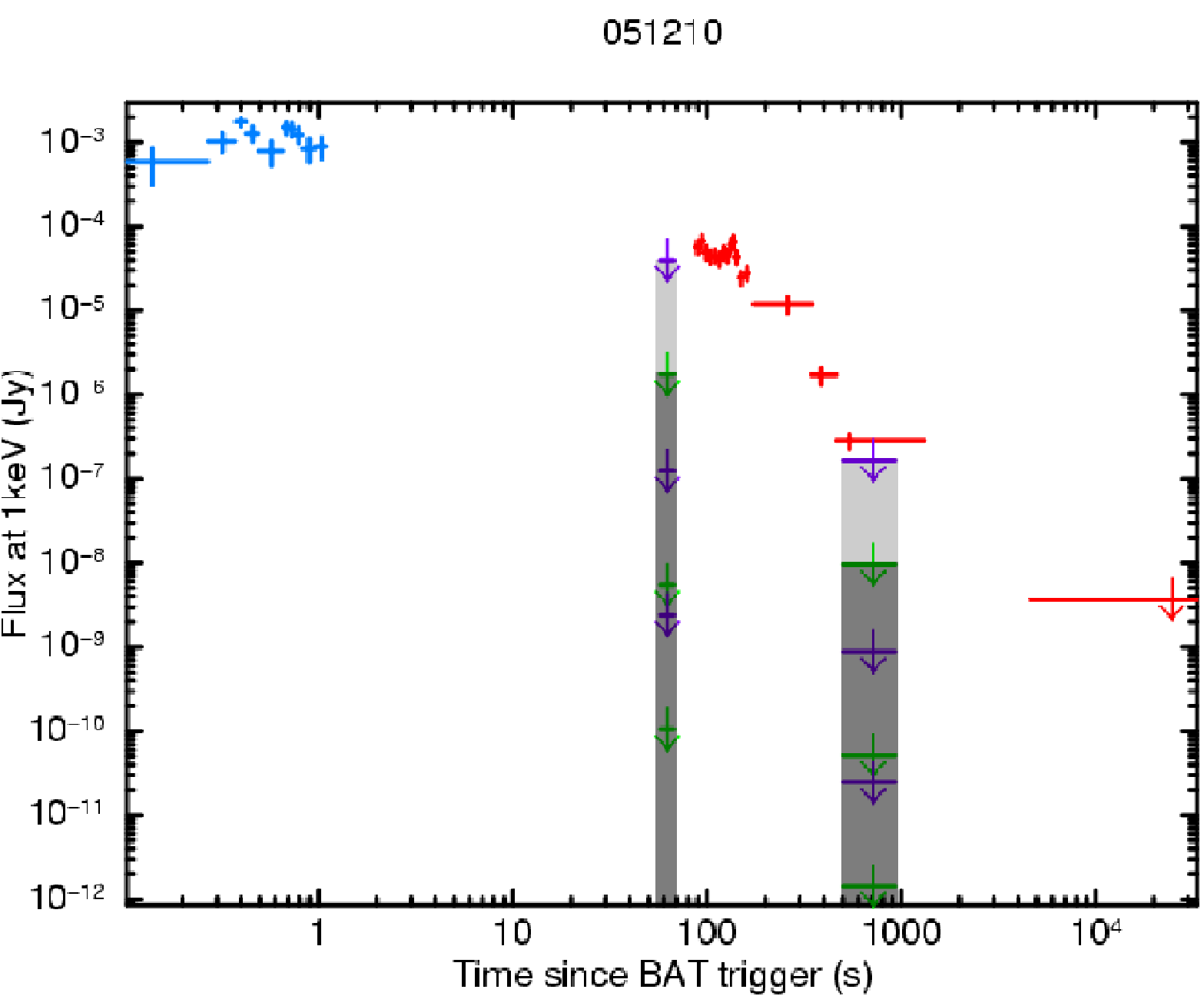}
\includegraphics[width=8.5cm]{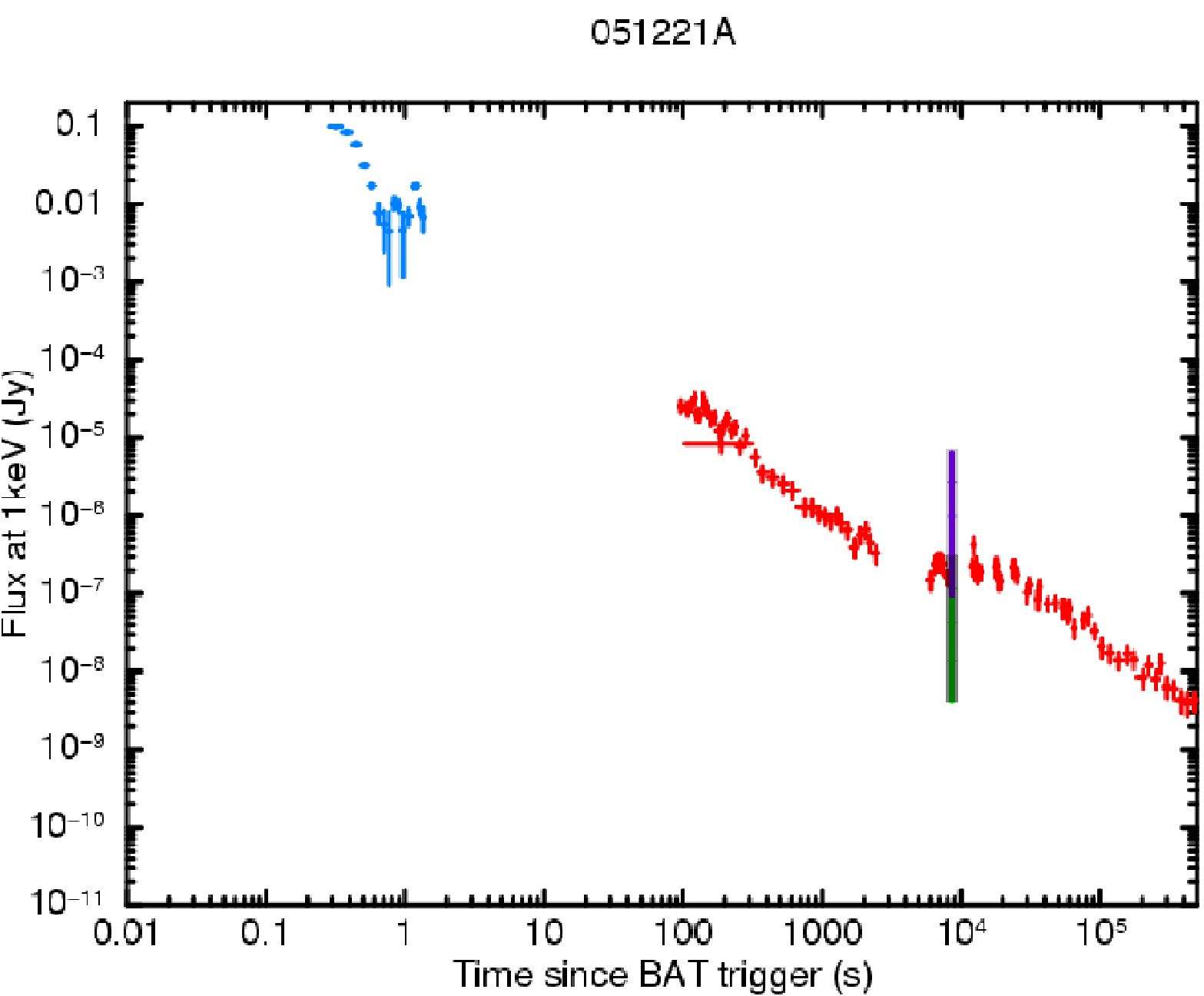}
\includegraphics[width=8.5cm]{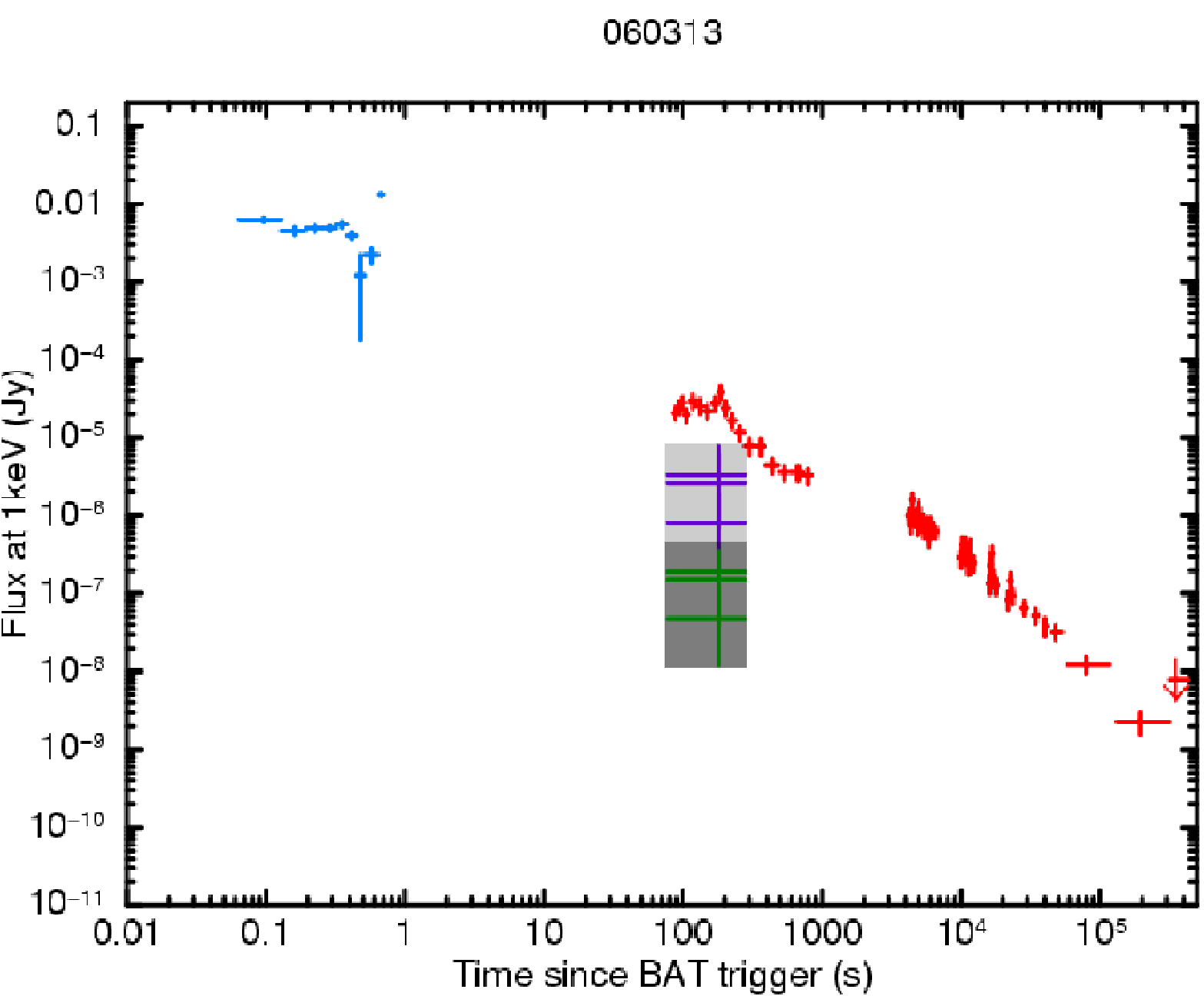}
\includegraphics[width=8.5cm]{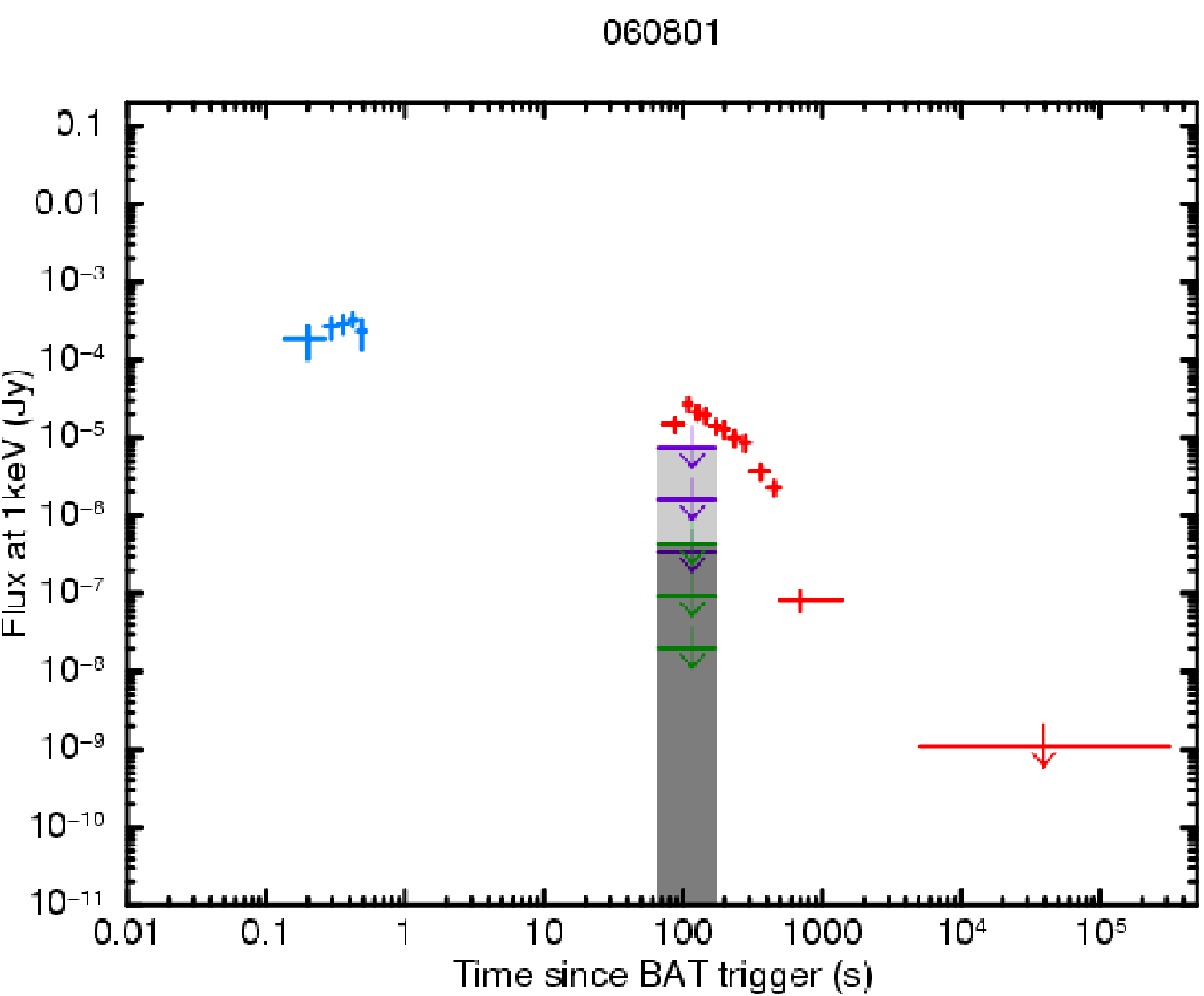}
\includegraphics[width=8.5cm]{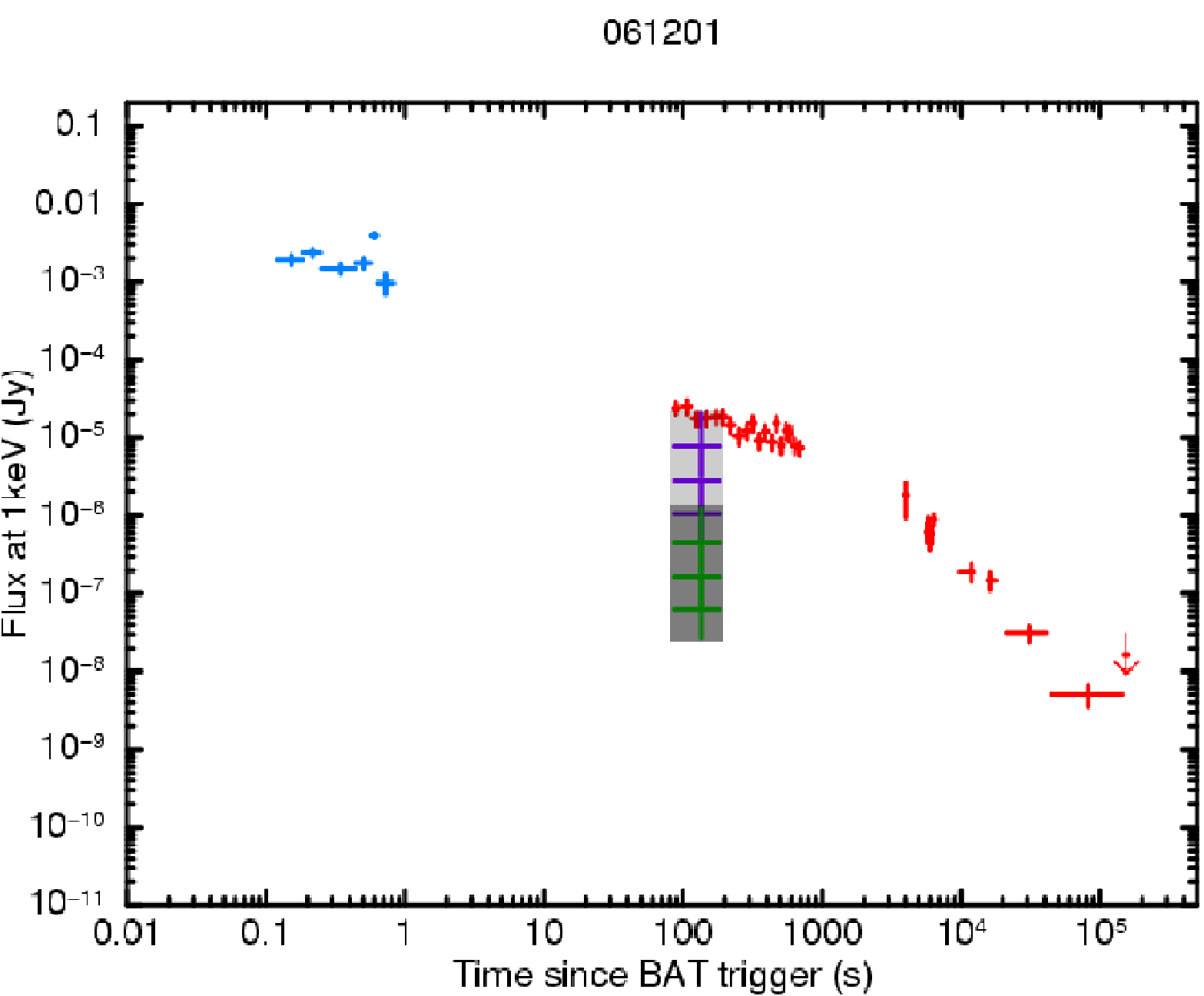}
\caption[Flux lightcurves comparing optical and X-ray data for the SGRB magnetar sample]{Comparison of the X-ray and optical data for the SGRBs fitted with the magnetar model. These are observed X-ray flux lightcurves at 1 keV with 1 extrapolated optical observation, light shaded region = optical observation assuming the most extreme cooling break between X-ray and optical and dark shaded region = optical observation assuming no cooling break. The references are for the optical observation used. If the X-ray and optical observations are consistent with originating from the same source, the X-ray data points should pass through the shaded regions. GRB 050509B - \citet{breeveld2005} - consistent, GRB 051210 - \citet{jelinek2005} - inconsistent, GRB 051221A - \citet{soderberg2006} - optical observations are consistent with X-ray observations, GRB 060313 - \citet{roming2006} - inconsistent, GRB 060801 - \citet{brown2006} - inconsistent and GRB 061201 - \citet{stratta2007} - only consistent for with most extreme cooling break and errors. }
\label{fig8d}
\end{figure*}

\begin{figure*}
\centering
\includegraphics[width=8.5cm]{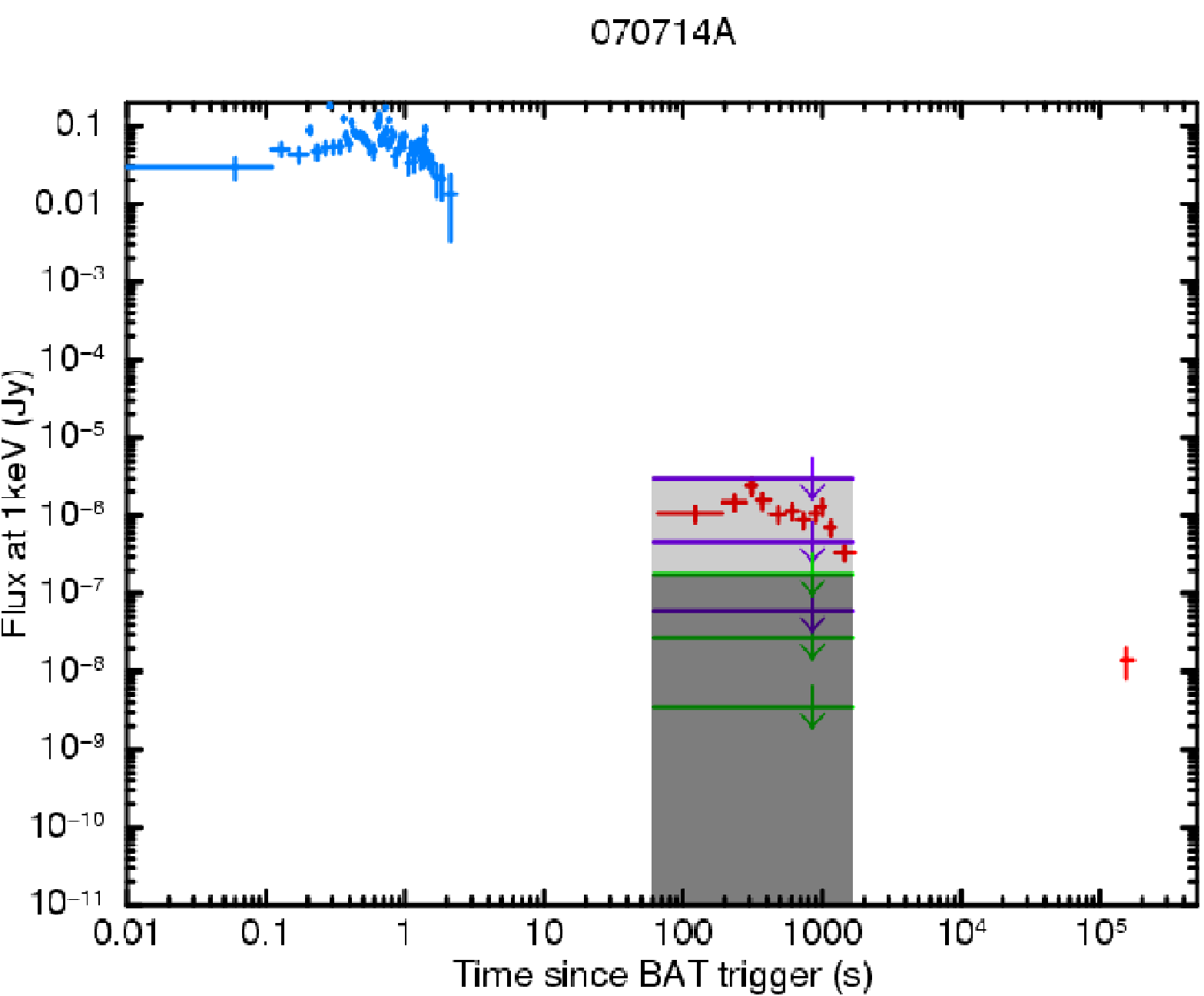}
\includegraphics[width=8.5cm]{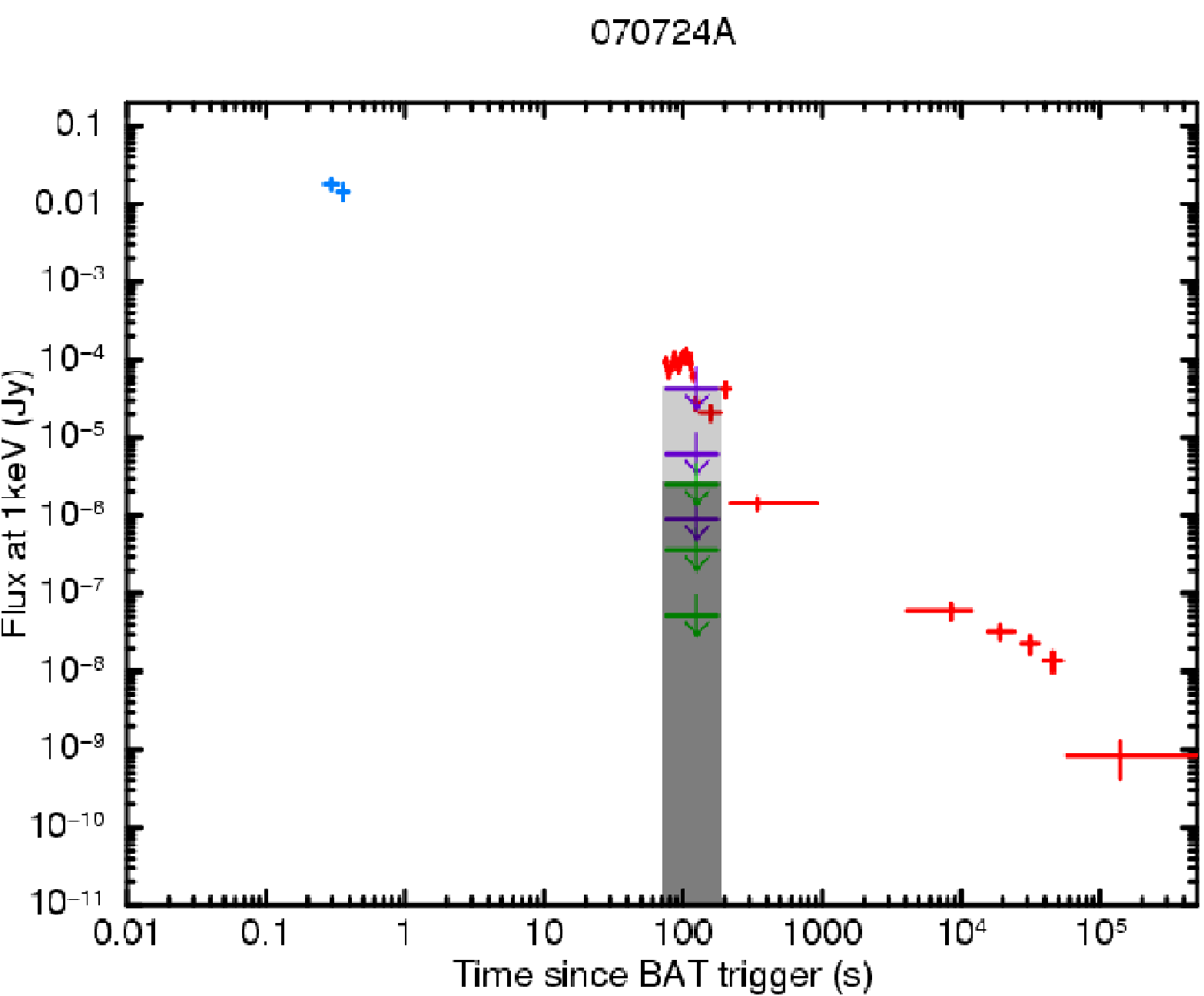}
\includegraphics[width=8.5cm]{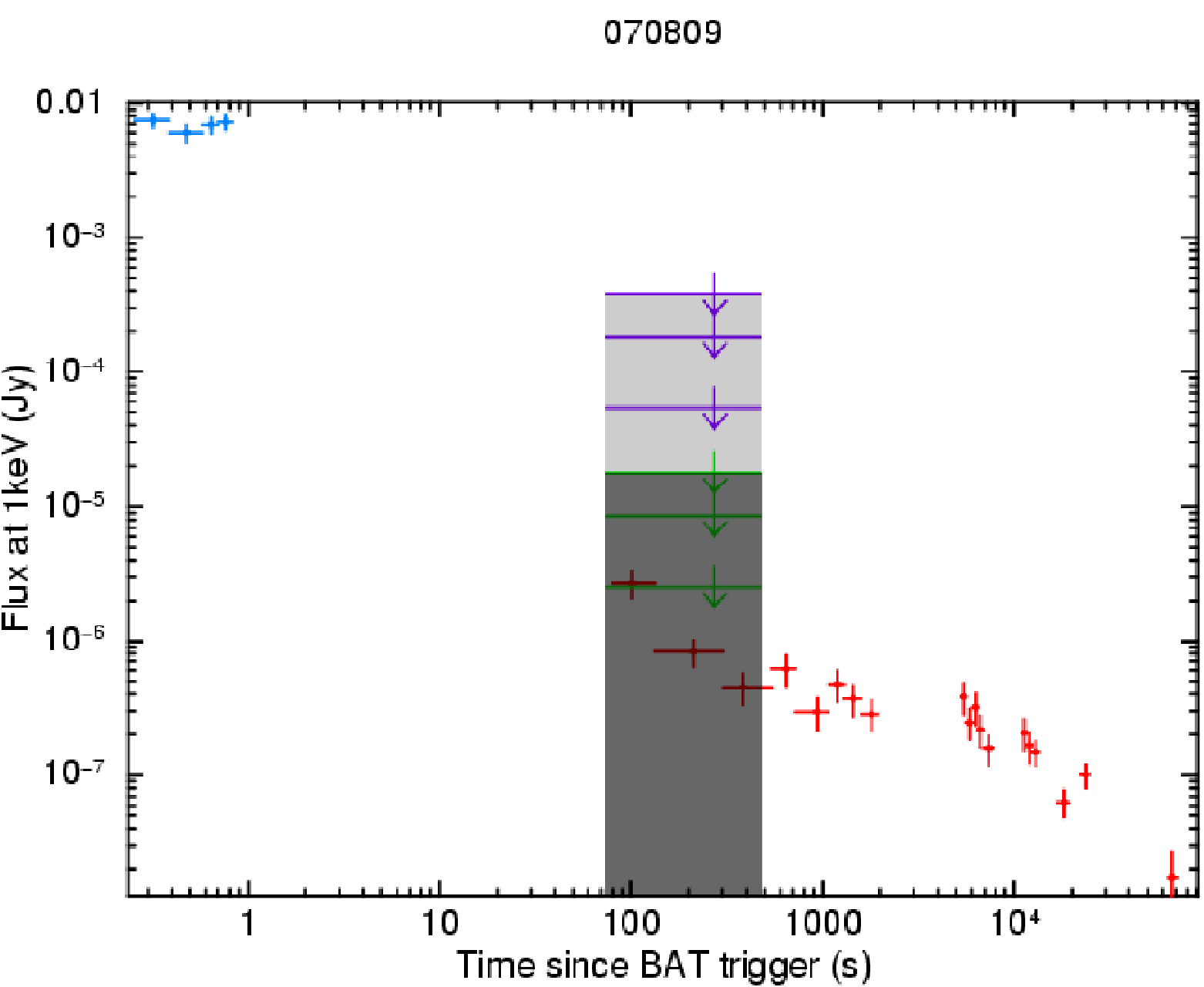}
\includegraphics[width=8.5cm]{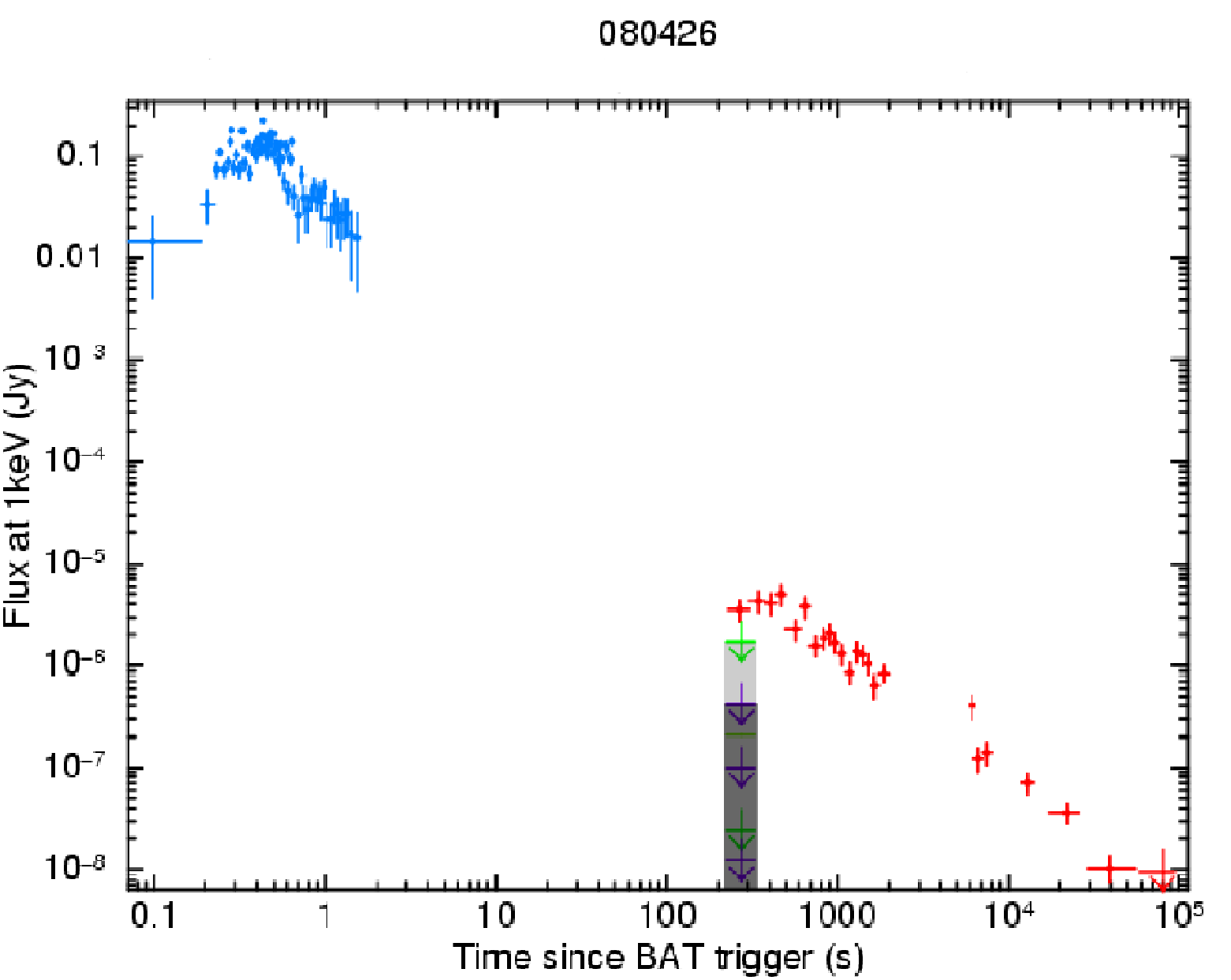}
\includegraphics[width=8.5cm]{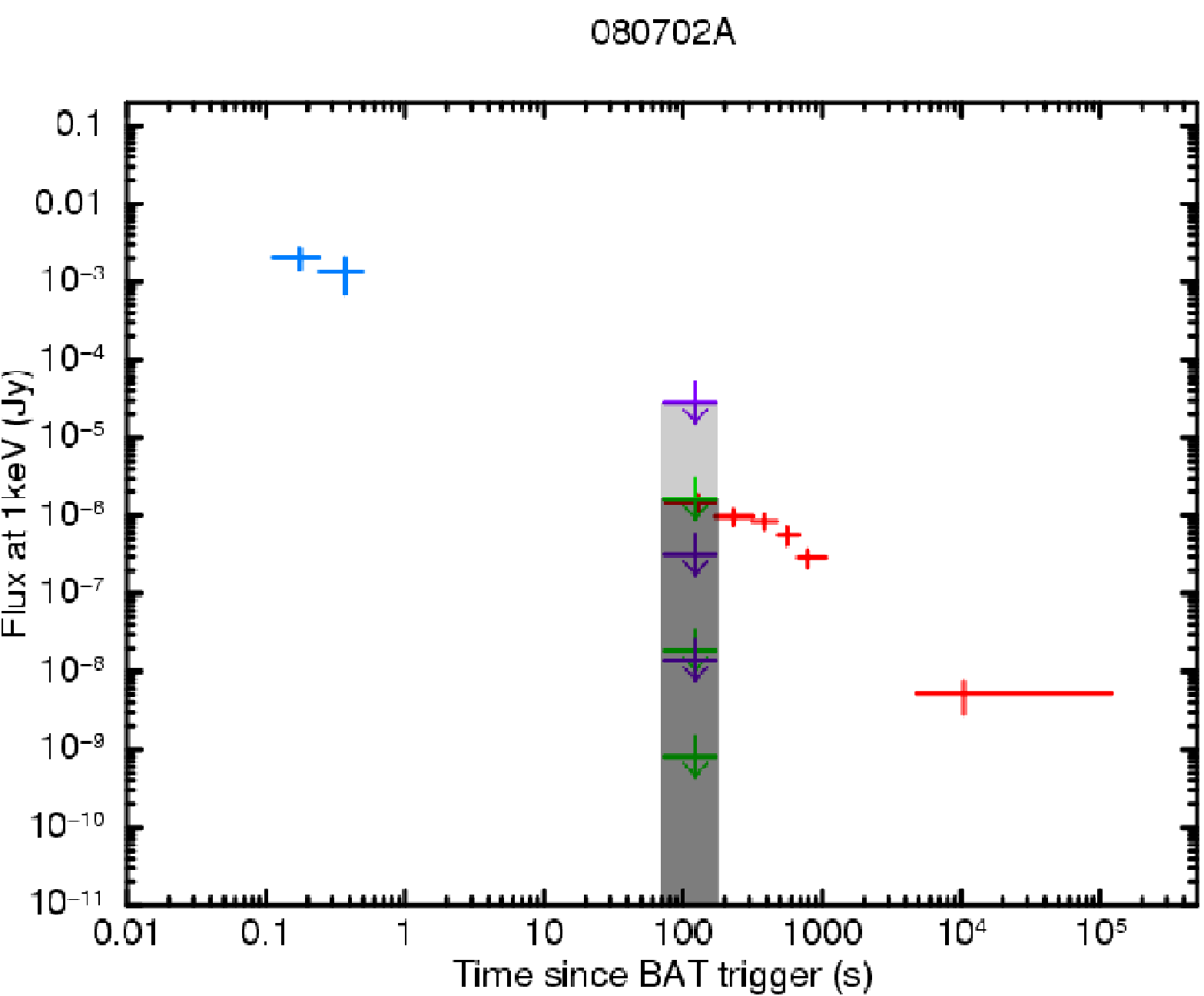}
\includegraphics[width=8.5cm]{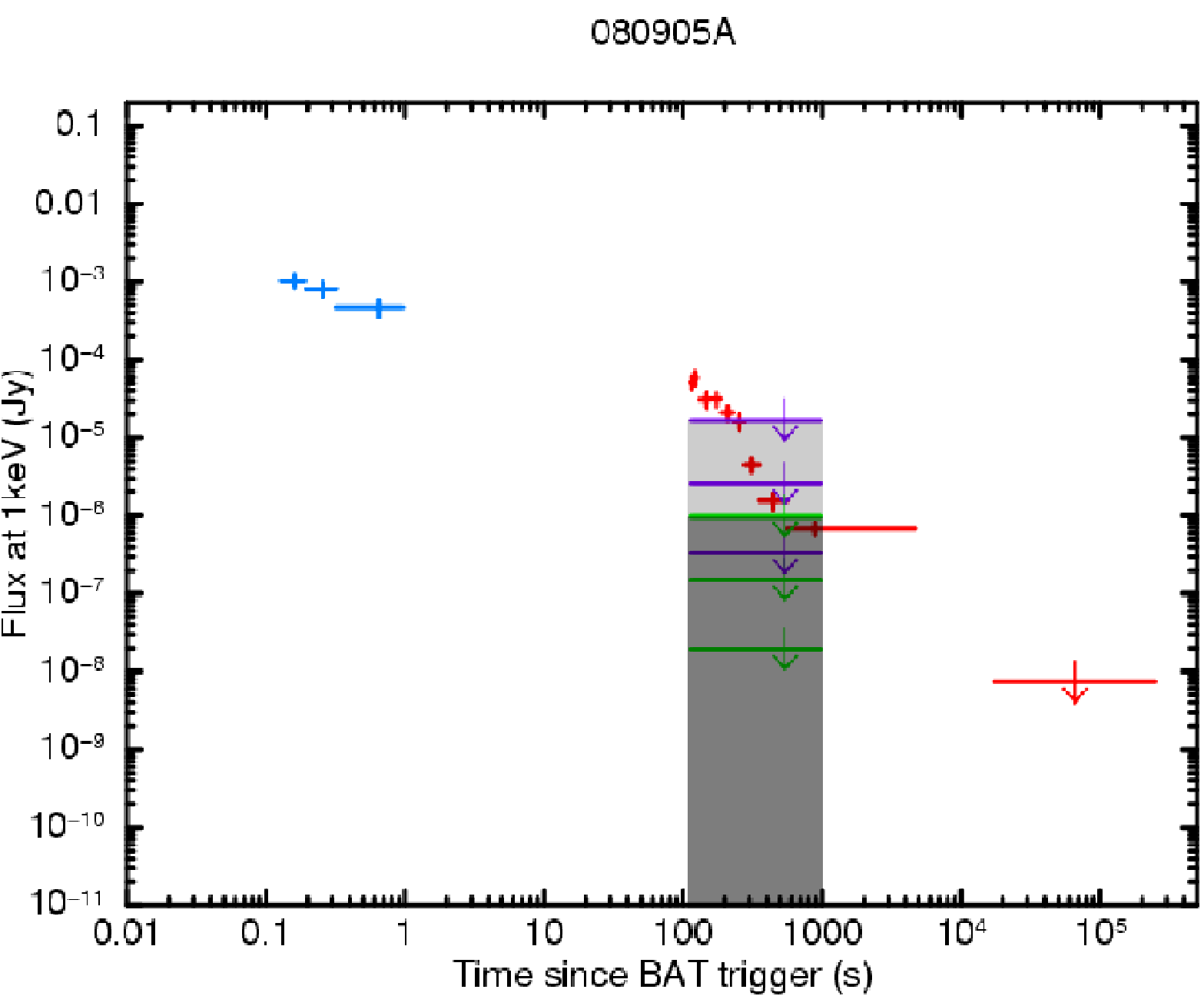}
\contcaption{GRB 070714A - \citet{chester2007} - upper limits inconclusive if there is an extreme cooling break, GRB 070724A - \citet{depasquale2007} - upper limits inconclusive if there is an extreme cooling break, GRB 070809 - \citet{chester2007b} - upper limits inconclusive, GRB 080426 - \citet{oates2008} - inconsistent, GRB 080702A - \citet{depasquale2008b} - upper limits inconclusive and GRB 080905A - \citet{brown2008} - upper limits inconclusive.}
\end{figure*}

\begin{figure*}
\centering
\includegraphics[width=8.5cm]{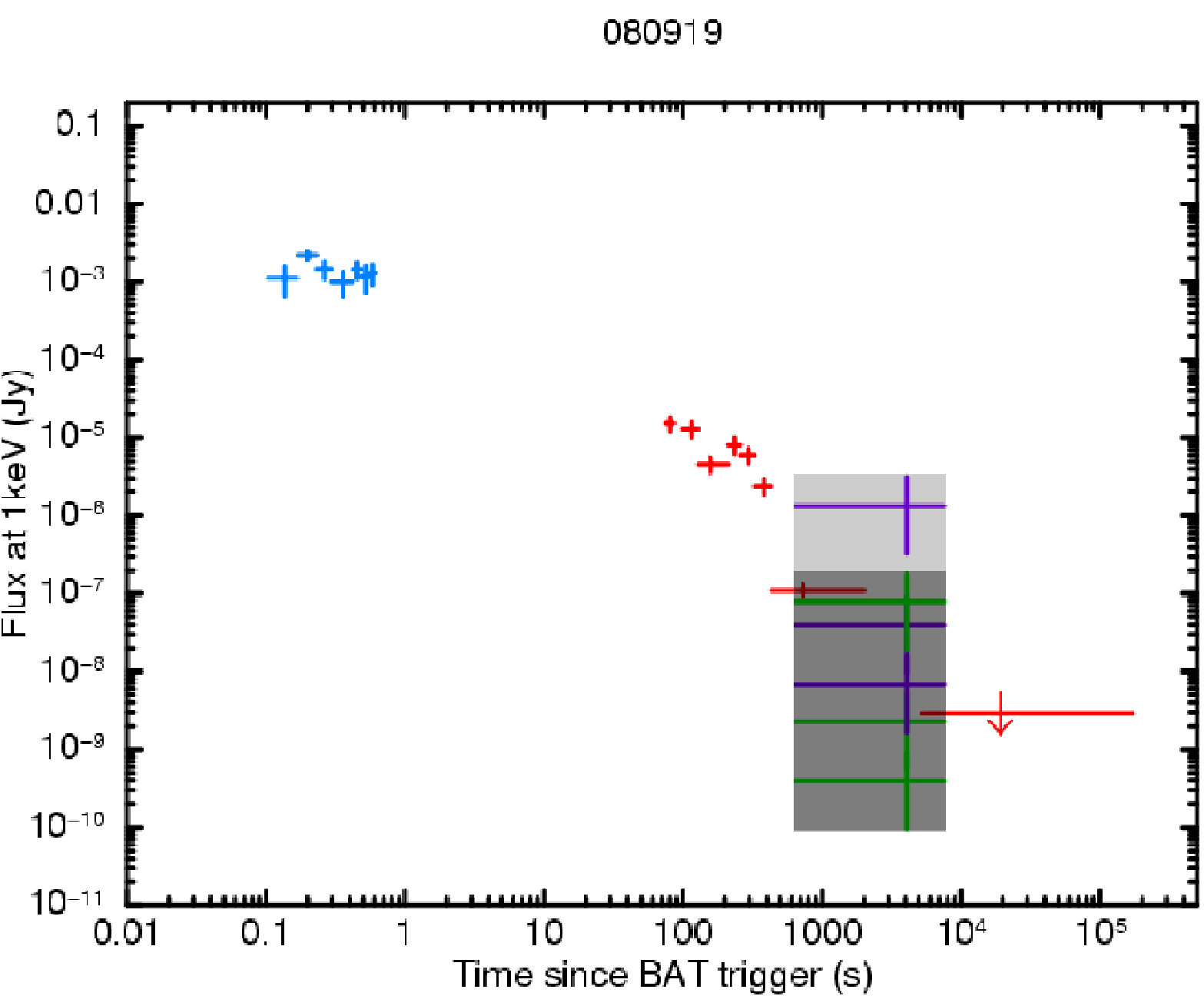}
\includegraphics[width=8.5cm]{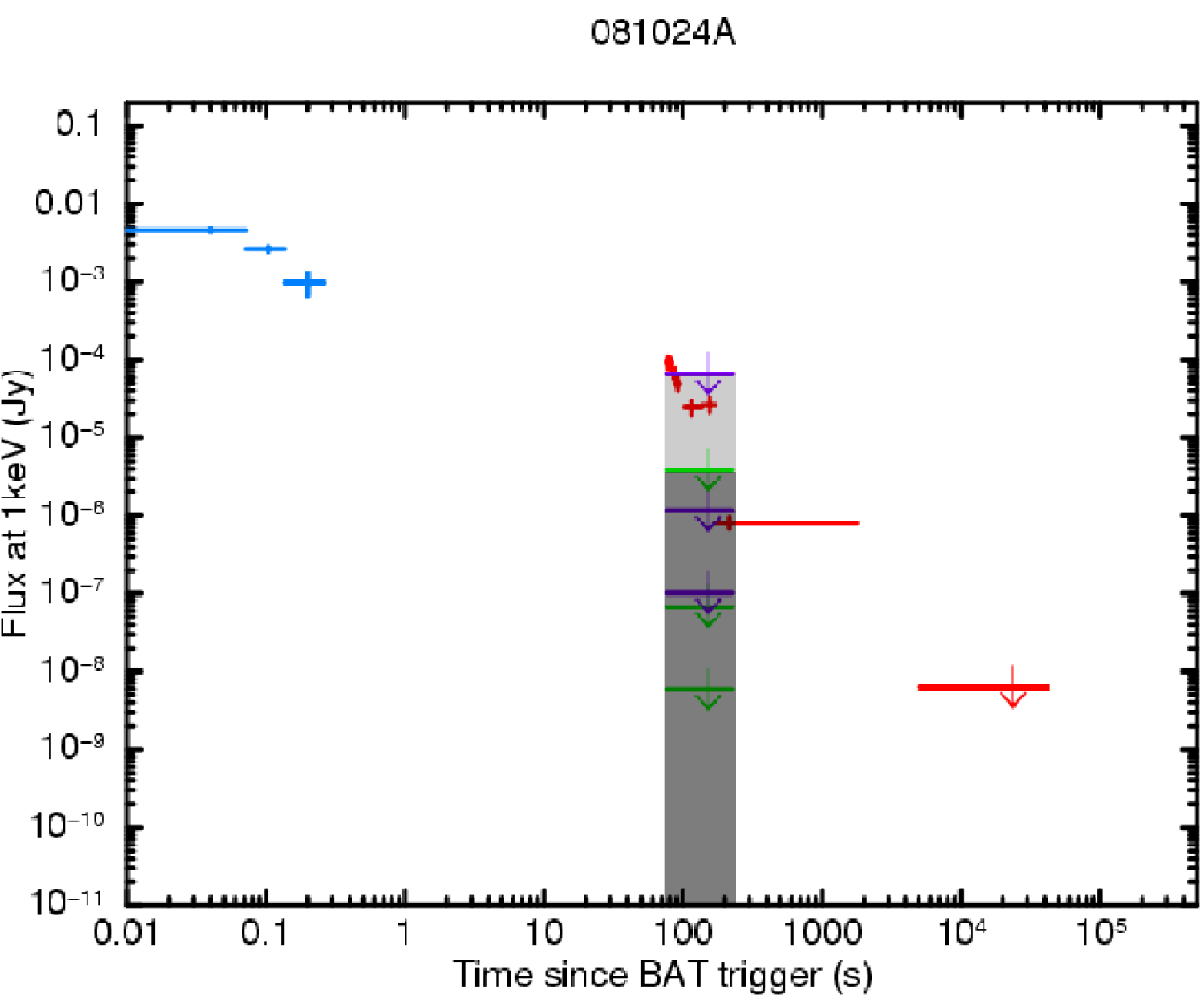}
\includegraphics[width=8.5cm]{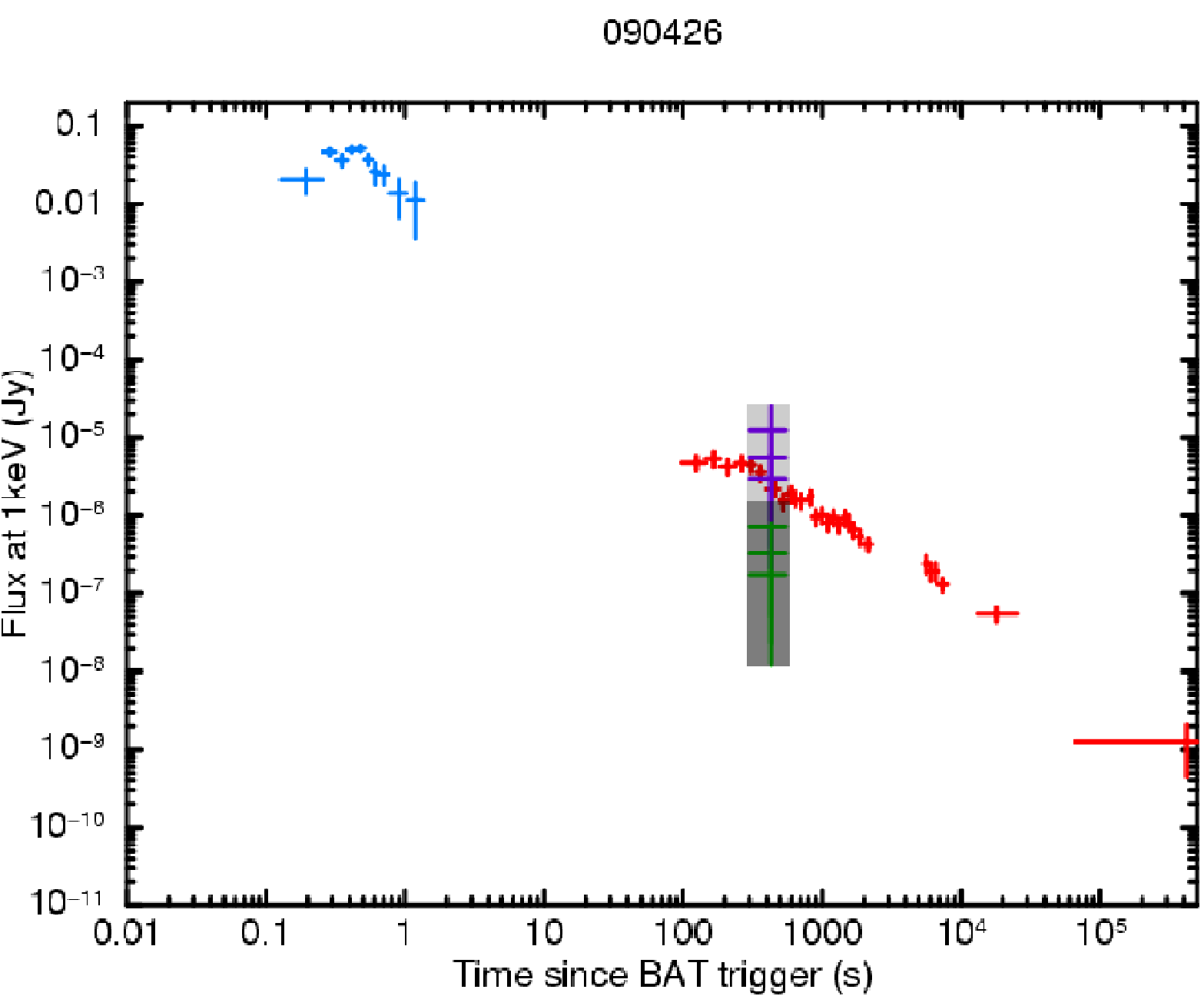}
\includegraphics[width=8.5cm]{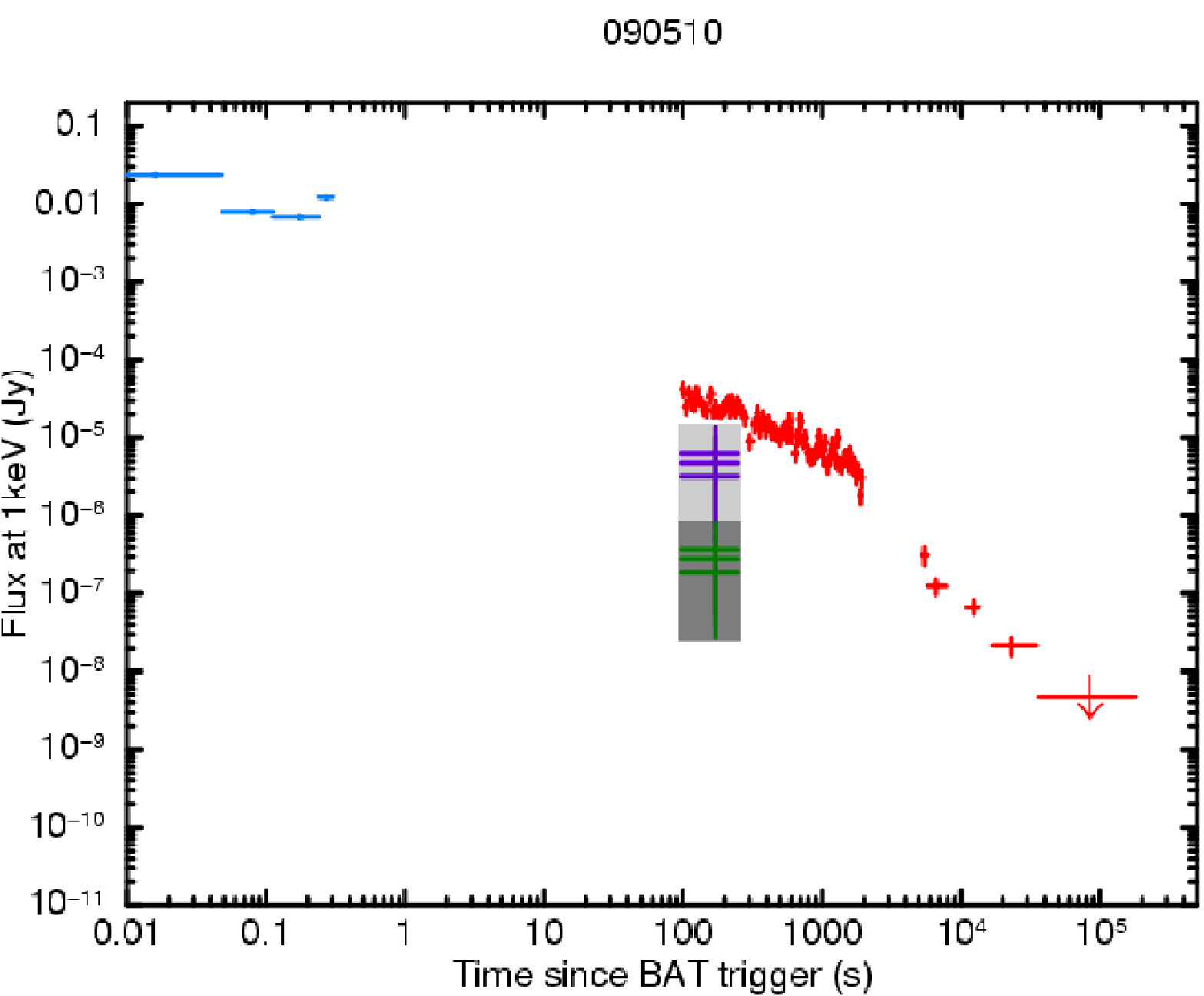}
\includegraphics[width=8.5cm]{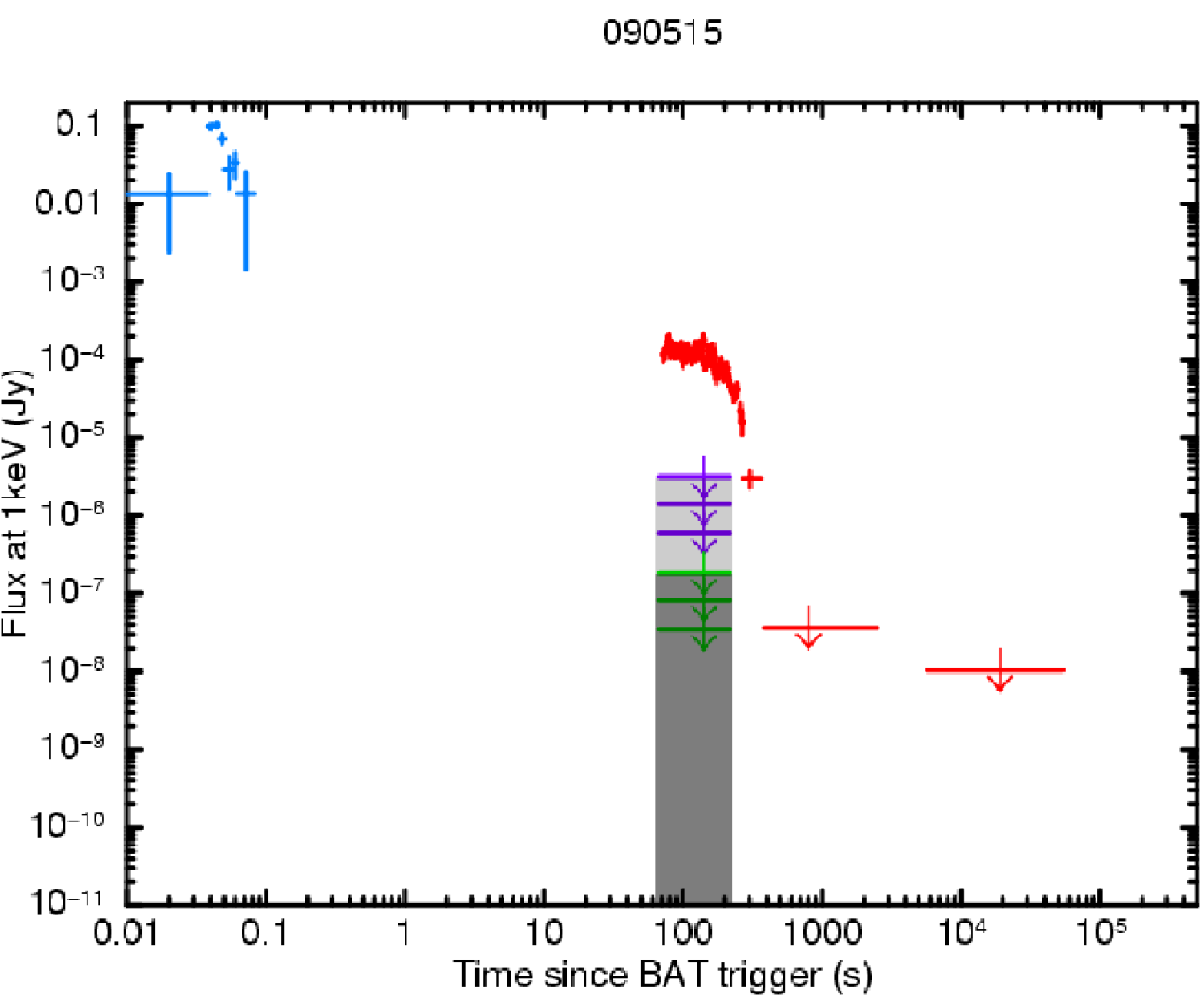}
\includegraphics[width=8.5cm]{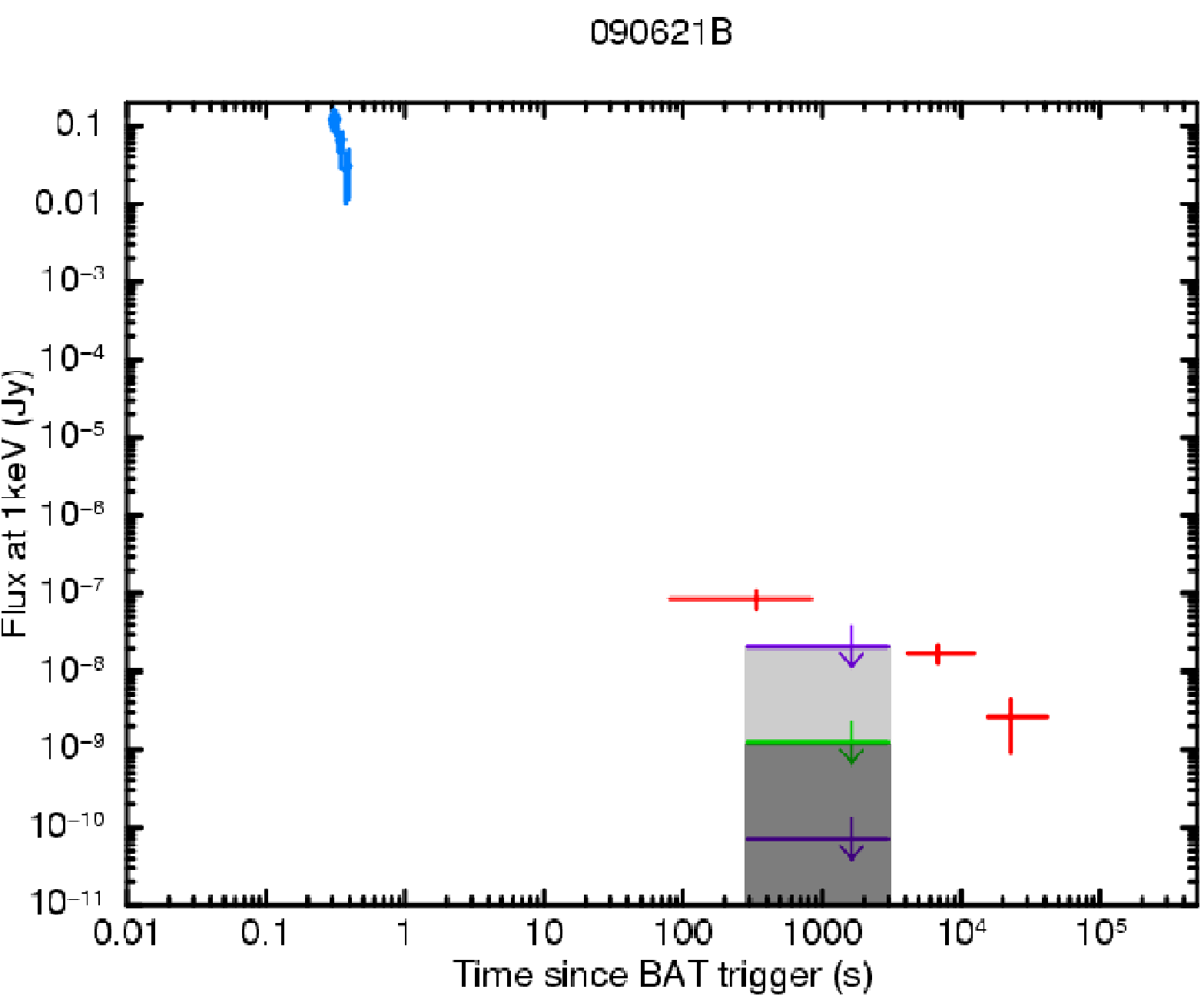}
\contcaption{GRB 080919 - \citet{immler2008} - likely consistent GRB 081024A - \citet{depasquale2008} - upper limits inconclusive and GRB 090426 - \citet{oates2009b} - optical observations are consistent with X-ray observations, GRB 090510 - \citet{kuin2009} - inconsistent, GRB 090515 - \citet{seigel2009} - inconsistent and GRB 090621B - \citet{curran2009} - inconsistent.}
\end{figure*}

\begin{figure*}
\centering
\includegraphics[width=8.5cm]{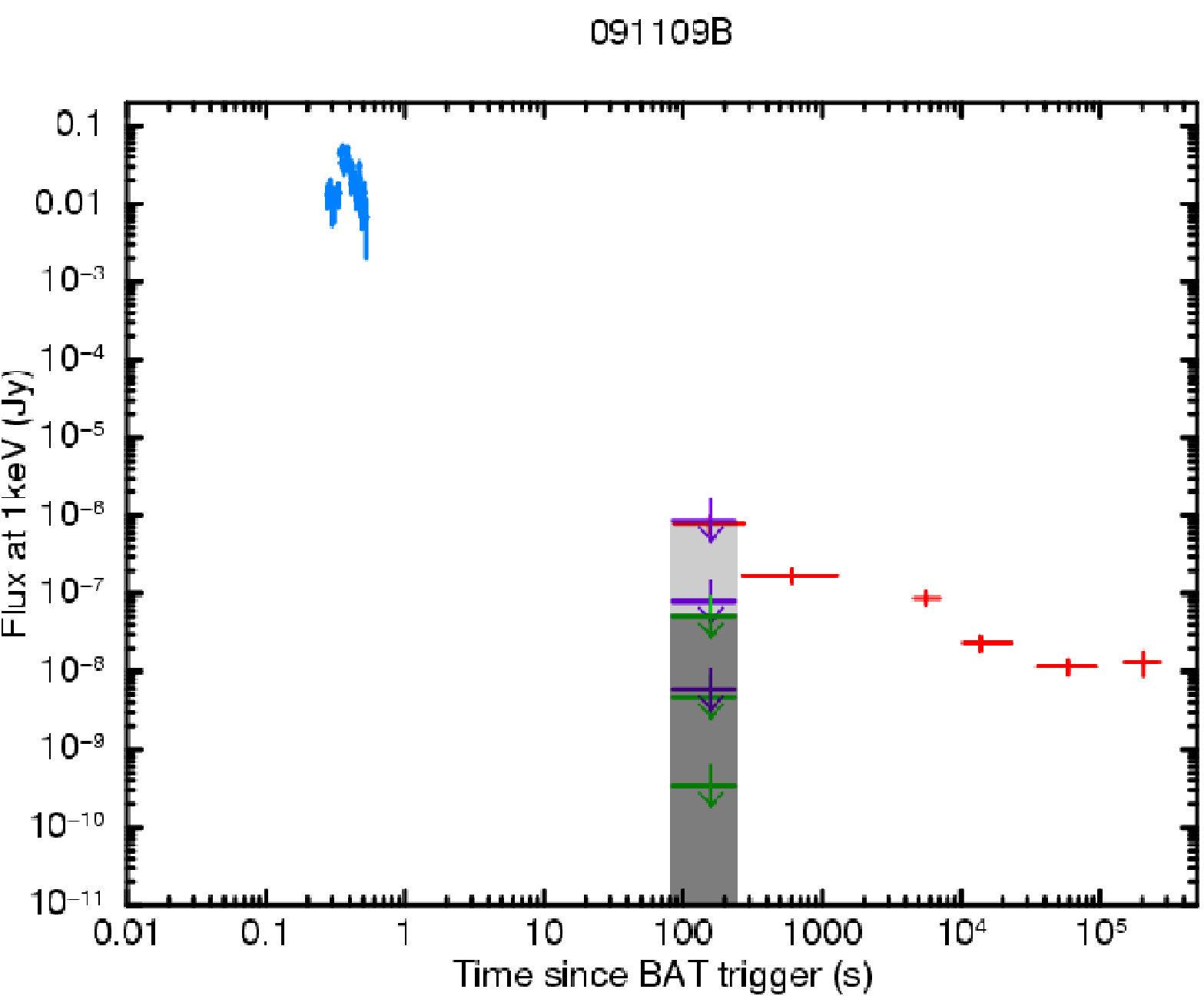}
\includegraphics[width=8.5cm]{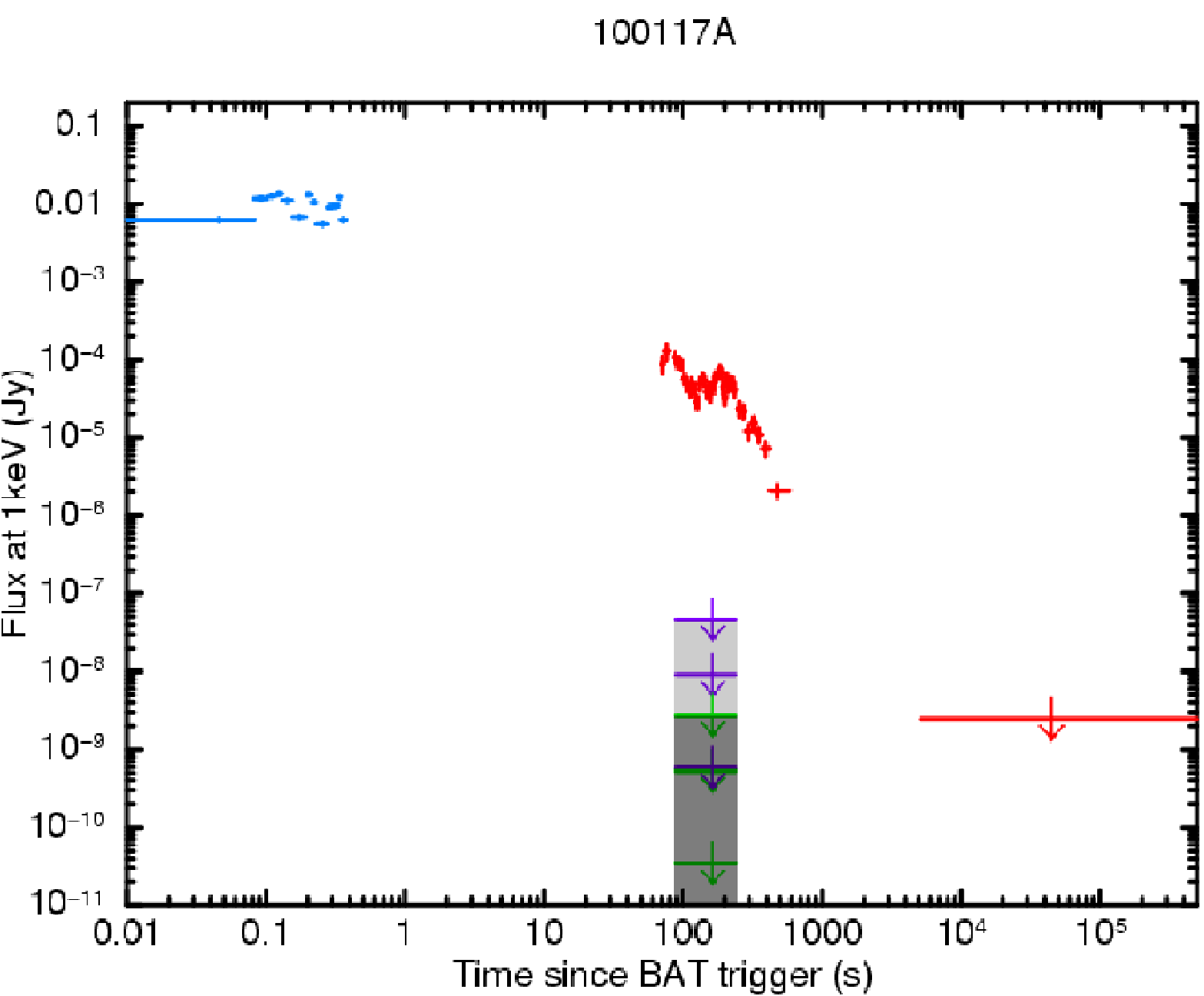}
\includegraphics[width=8.5cm]{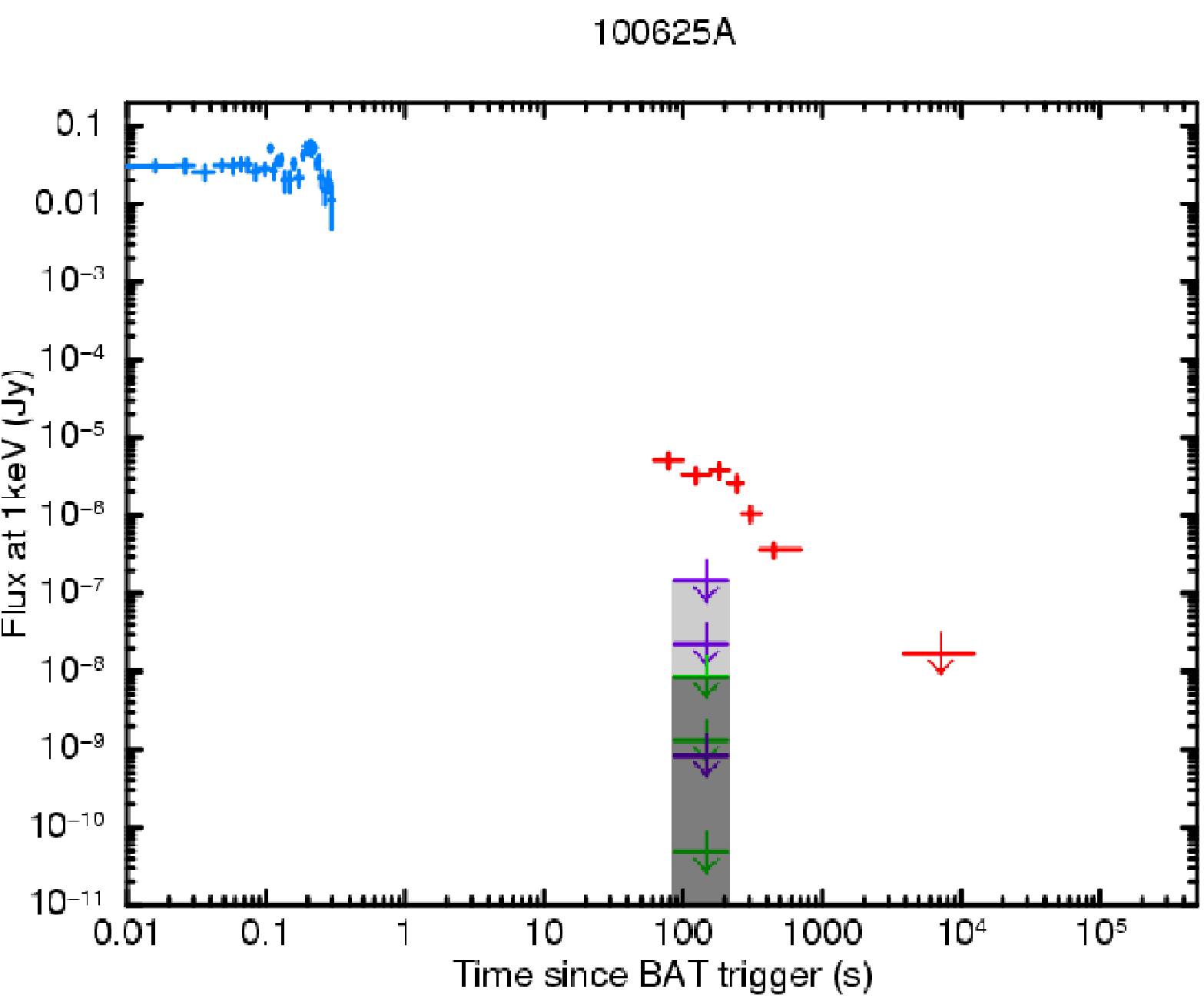}
\includegraphics[width=8.5cm]{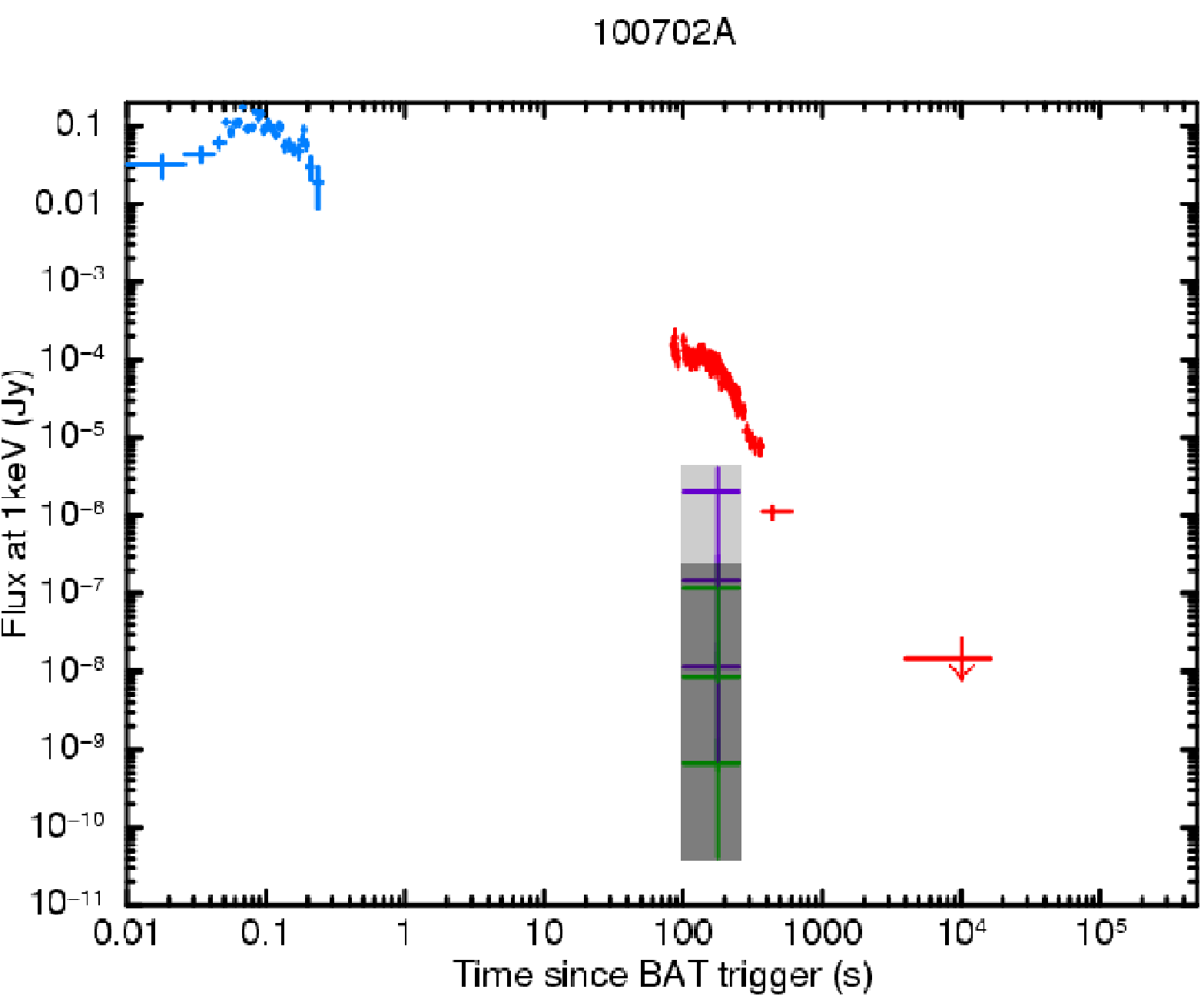}
\includegraphics[width=8.5cm]{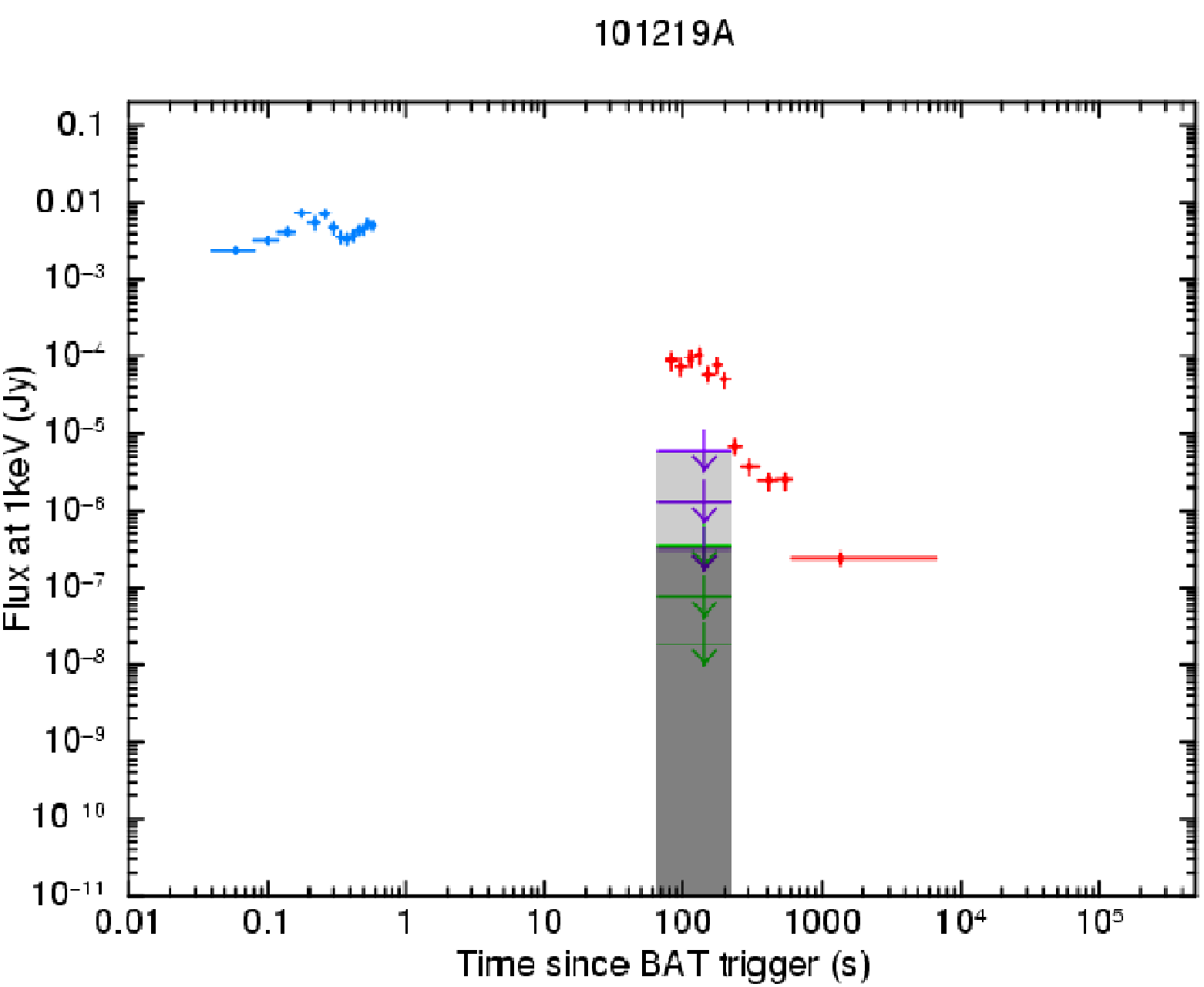}
\includegraphics[width=8.5cm]{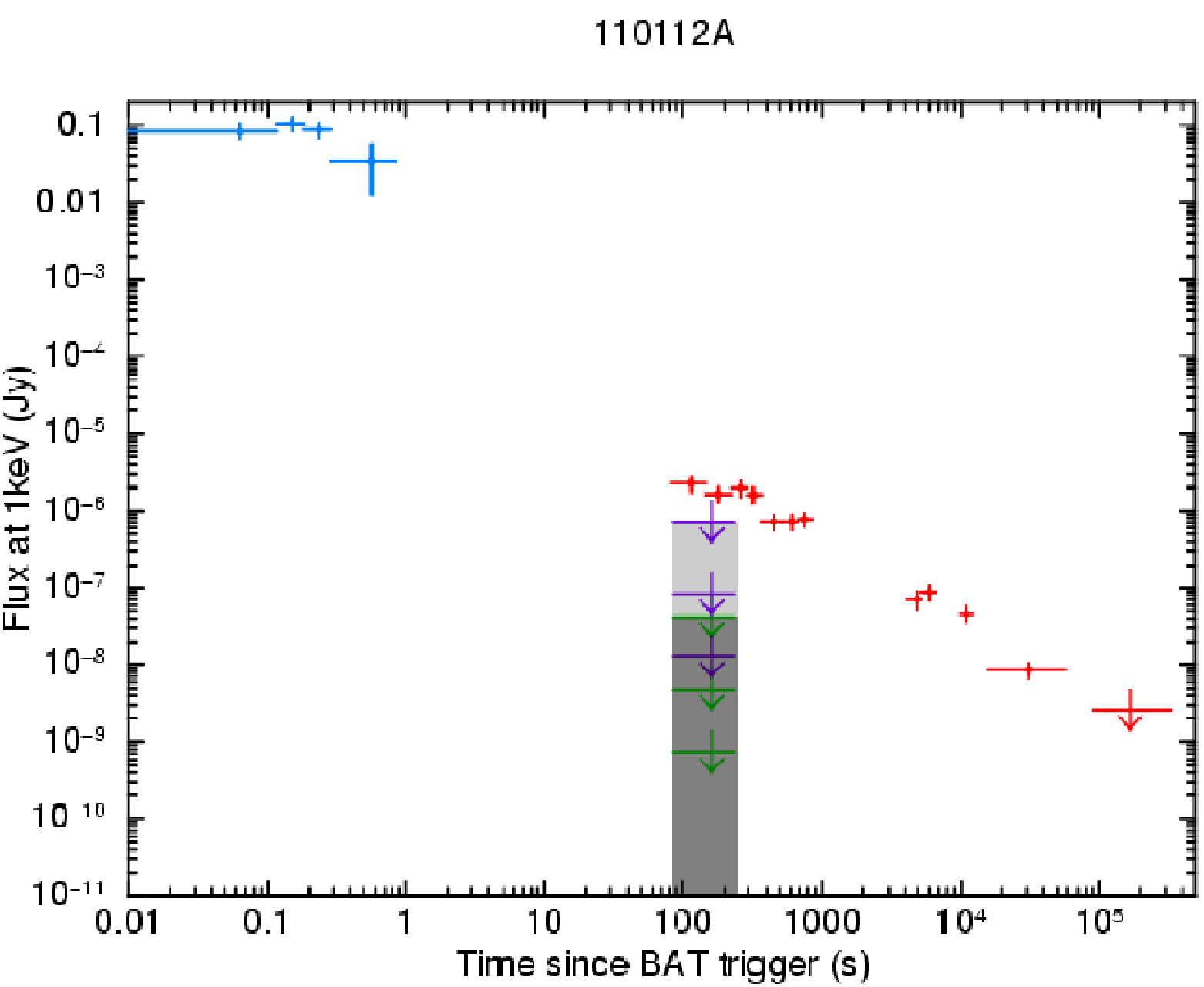}
\contcaption{GRB 091109B - \citet{oates2009} - upper limits inconclusive if there is an extreme cooling break, GRB 100117A - \citet{depasquale2010c} - extremely inconsistent, GRB 100625A - \citet{landsman2010} - inconsistent, GRB 100702A - \citet{depasquale2010b} - inconsistent, GRB 101219A - \citet{kuin2010} - inconsistent and GRB 110112A - \citet{breeveld2011} - inconsistent.}
\end{figure*}

\begin{figure*}
\centering
\includegraphics[width=8.5cm]{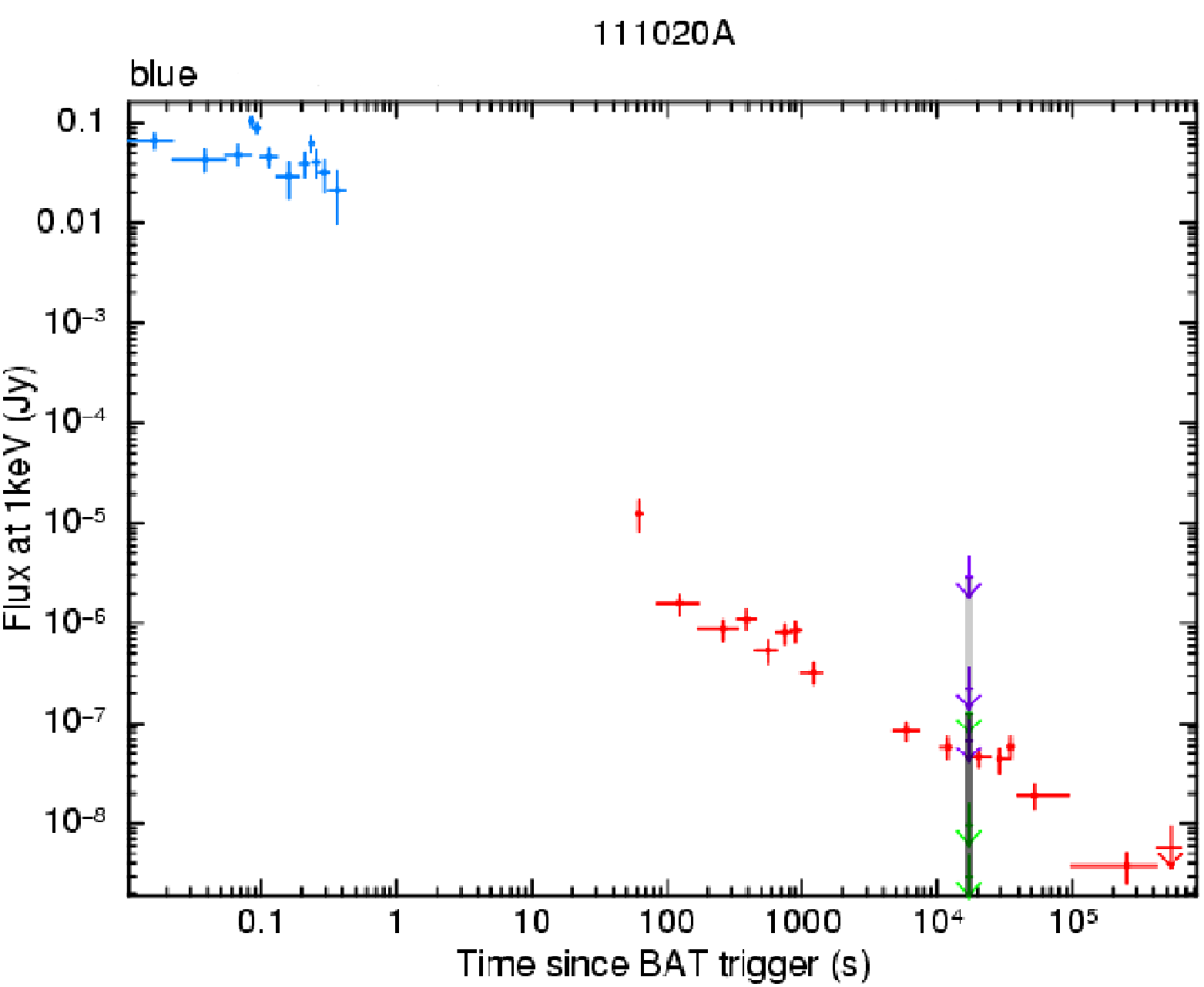}
\includegraphics[width=8.5cm]{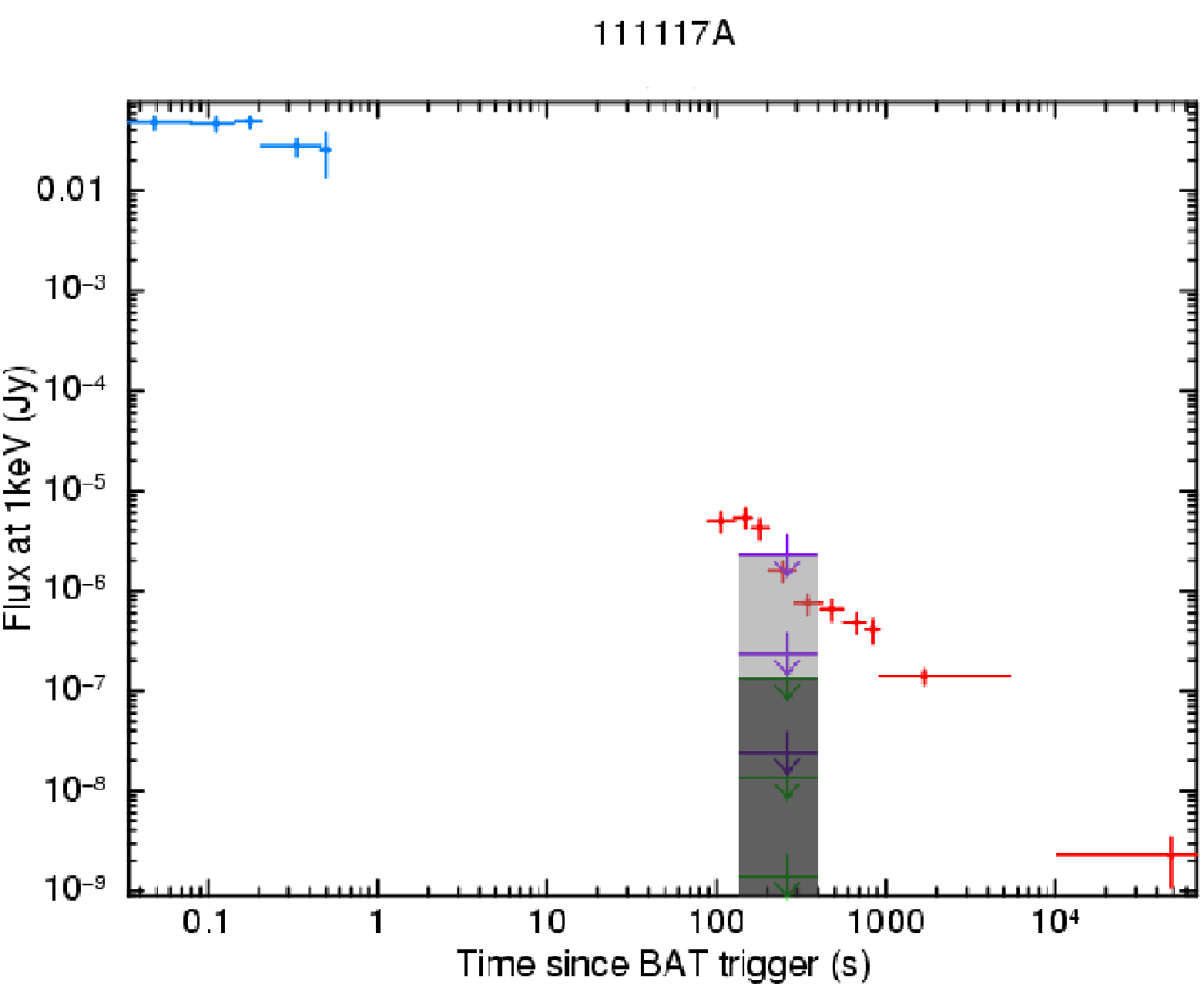}
\includegraphics[width=8.5cm]{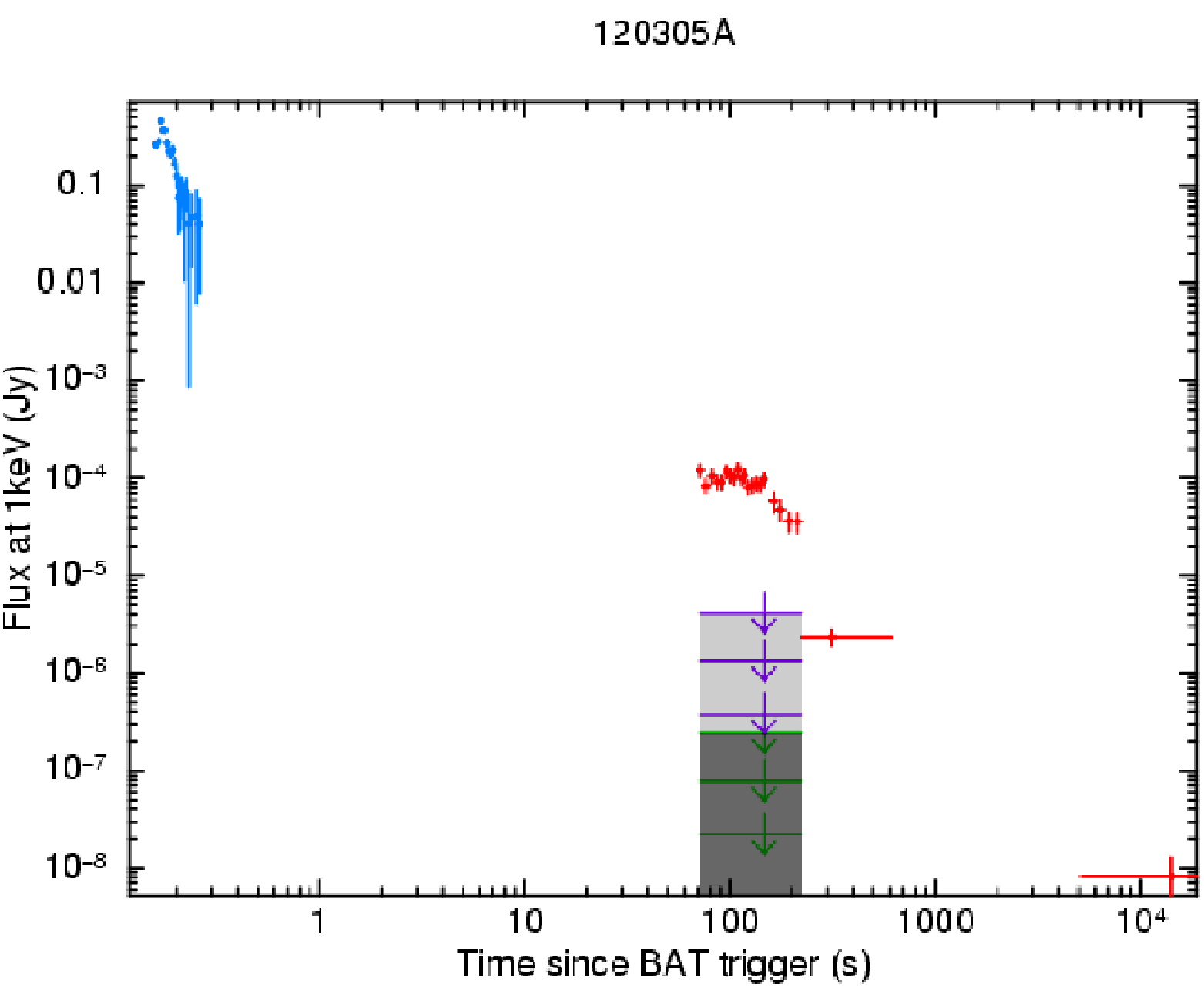}
\includegraphics[width=8.5cm]{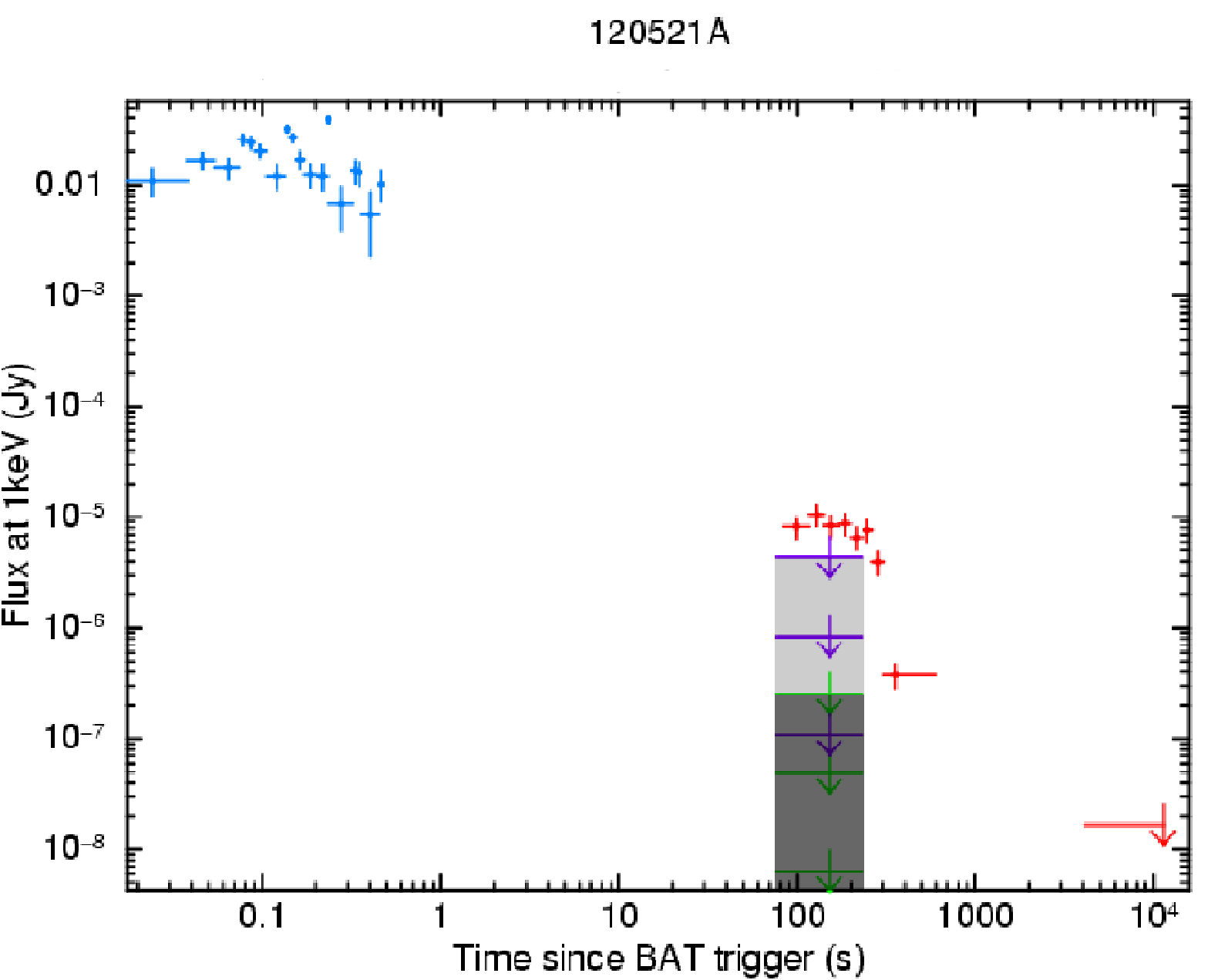}
\contcaption{GRB 111020A - \citet{guidorzi2011} - consistent GRB 111117A - \citet{oates2011} - upper limits inconclusive if there is an extreme cooling break, GRB 120305A - \citet{marshal2012} - inconsistent and GRB 120521A - \citet{oates2012} - inconsistent.}
\end{figure*}

A 1 keV observed flux lightcurve showing the prompt, X-ray and the most constraining optical observation during the plateau phase was created for each burst in the sample. These were produced using the simple relation given in equation \ref{flux} (assuming a simple power law spectrum and a spectral index $\beta_x=\Gamma_x-1$) to shift the observed fluxes at a measured energy to 1 keV.

\begin{figure*}
\centering
\includegraphics[width=7.5cm]{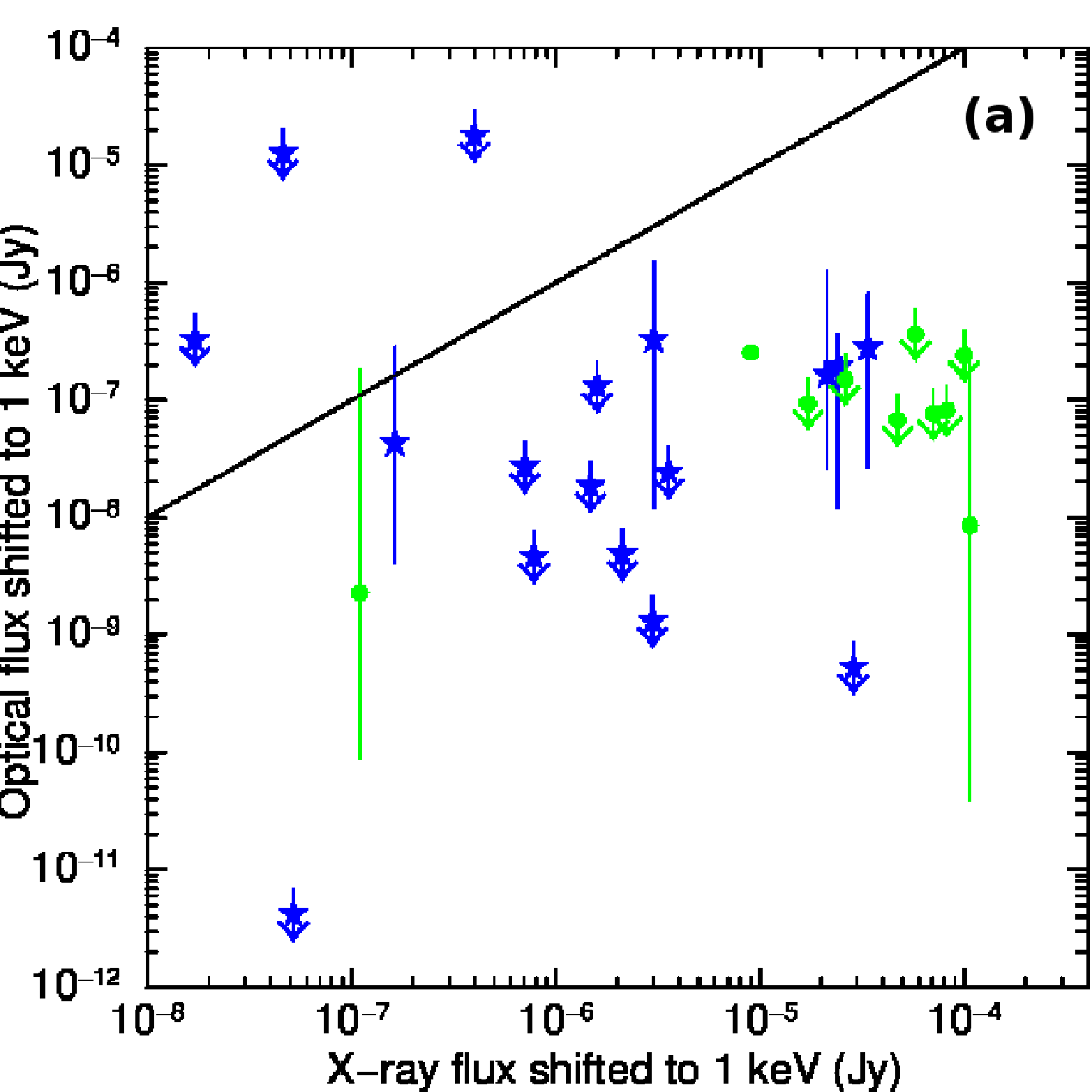}
\includegraphics[width=7.5cm]{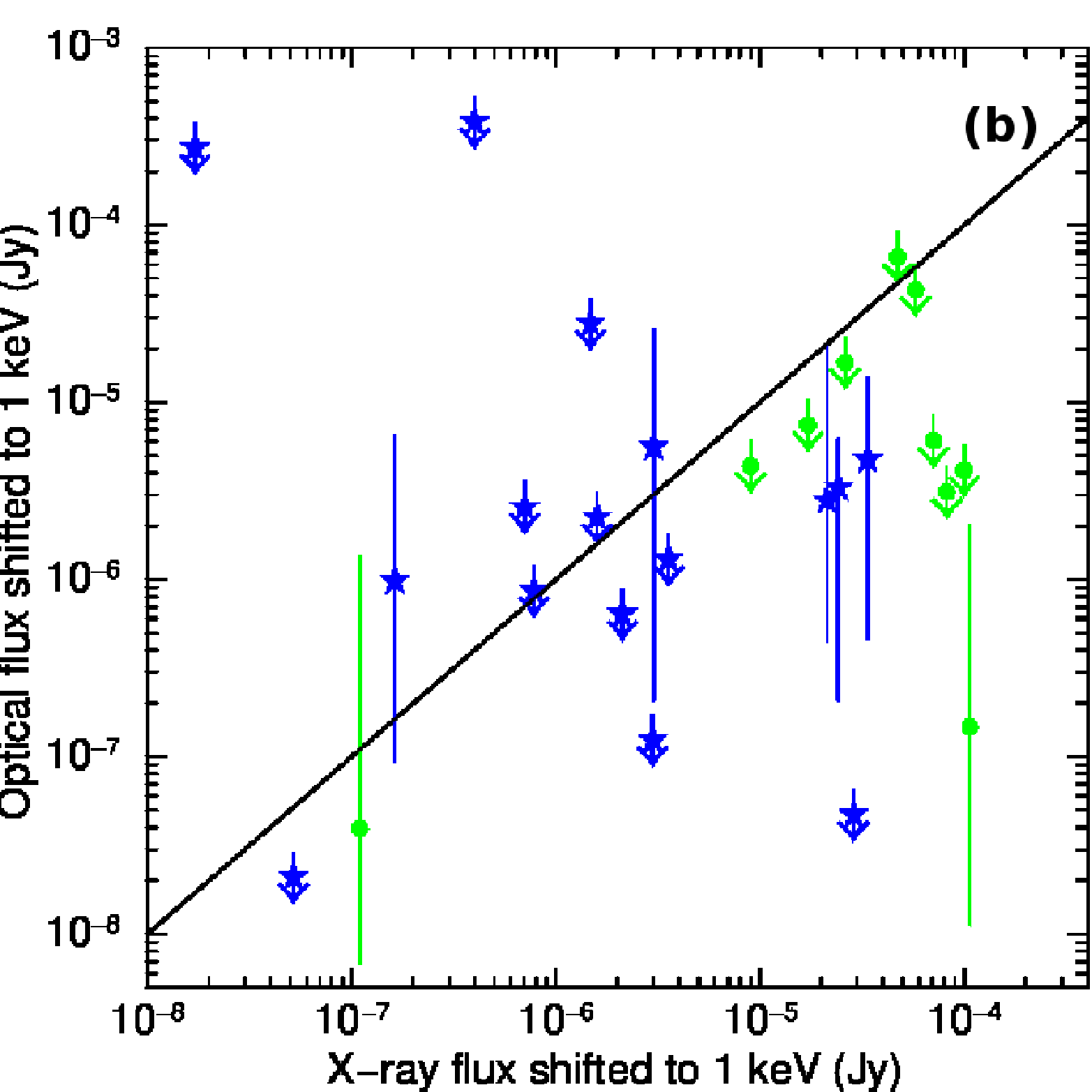}
\caption{The optical flux shifted to 1 keV is plotted against the average X-ray observed flux during the optical observation also shifted to 1 keV. The solid black line represents where these are equal, as expected if they are consistent with each other. In (a) we assume there is no cooling break between the optical and X-ray observations and in (b) we assume the most extreme cooling break. Symbols are as in Figure \ref{fig5}.}
\label{fig7b}
\end{figure*}

\begin{eqnarray}
F_{\nu(1keV)}=F_{\nu(measured)}\left(\frac{E_{(\rm measured)}}{1keV}\right)^{\beta_{x,o}} \label{flux}
\end{eqnarray}

$\Gamma_x$ was obtained from the time averaged PC mode spectra produced by the automated anaylsis on the UK {\it Swift} Data Centre website \citep{evans2007, evans2009}. The 0.3 -- 10 keV observed BAT-XRT lightcurves were extrapolated to flux at 1 keV using equation \ref{flux}. The optical magnitudes were converted into flux for the wavelength of the optical filter used and then shifted to 1 keV using equation \ref{flux}. As there may be a cooling break inbetween the optical and X-ray observations \citep{sari1998}, the two extreme cases are taken i.e. $\beta_o=\beta_x$ and $\beta_o=\beta_x-0.5$. The errors on the observed optical magnitudes and the errors on $\Gamma_x$ are used to define the region on the lightcurve that the optical data could reside in (dark grey - no cooling break, light grey - cooling break, note there is overlap between these two regimes). If the optical and X-ray data are consistent, then the X-ray data points should lie within the shaded regions for the optical data. 

The 1 keV flux lightcurves for SGRBs fitted with the magnetar model are shown in Figure \ref{fig8d} compared with the most constraining optical observation extrapolated to 1 keV. These compare the BAT-XRT lightcurve at 1 keV to the most constraining optical observation extrapolated to 1 keV. GRBs 051221A, 061201, 080905A, 080919 and 090426 have optical afterglows which are consistent with their X-ray afterglows, but many of these would require the most extreme errors on the spectral slope and cooling break. $\sim$55\% have optical afterglows that are underluminous with respect to their X-ray afterglows, signifying either significant optical absorption or an extra component in the X-ray afterglow. However, as shown in Section 3.3.1 using absorption in the X-ray spectra, the majority of the candidates are consistent with occuring in a low density environment. 

In Figure \ref{fig7b} we compare the average X-ray fluxes at 1 keV to the optical fluxes extrapolated to 1 keV with (b) and without (a) a cooling break in the spectrum. The average X-ray flux was calculated during the optical observation. There are several points which lie below the black line in both cases, showing there is less emission at optical wavelengths than expected. 

\begin{figure*}
\centering
\includegraphics[width=5.5cm]{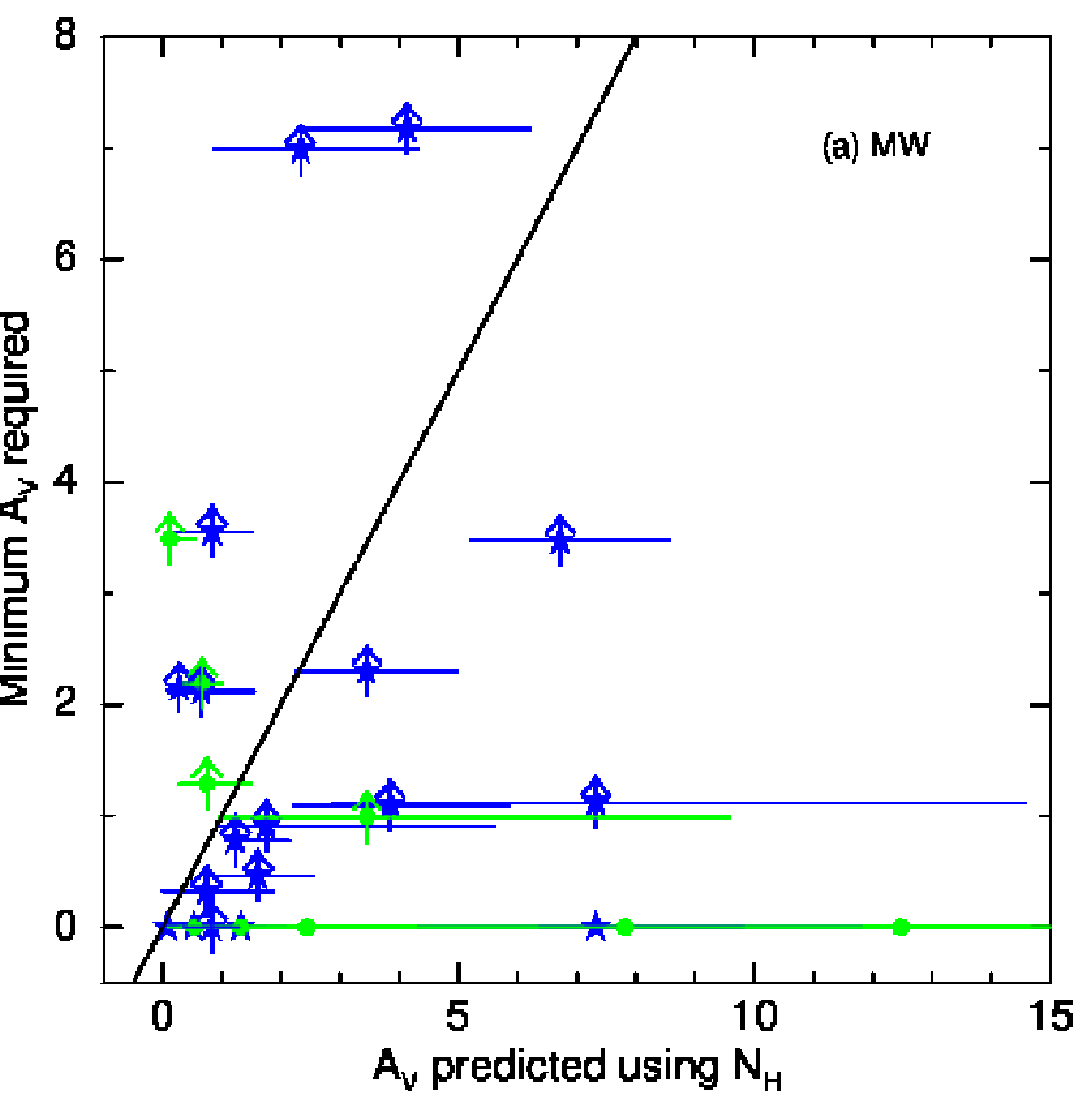}
\includegraphics[width=5.5cm]{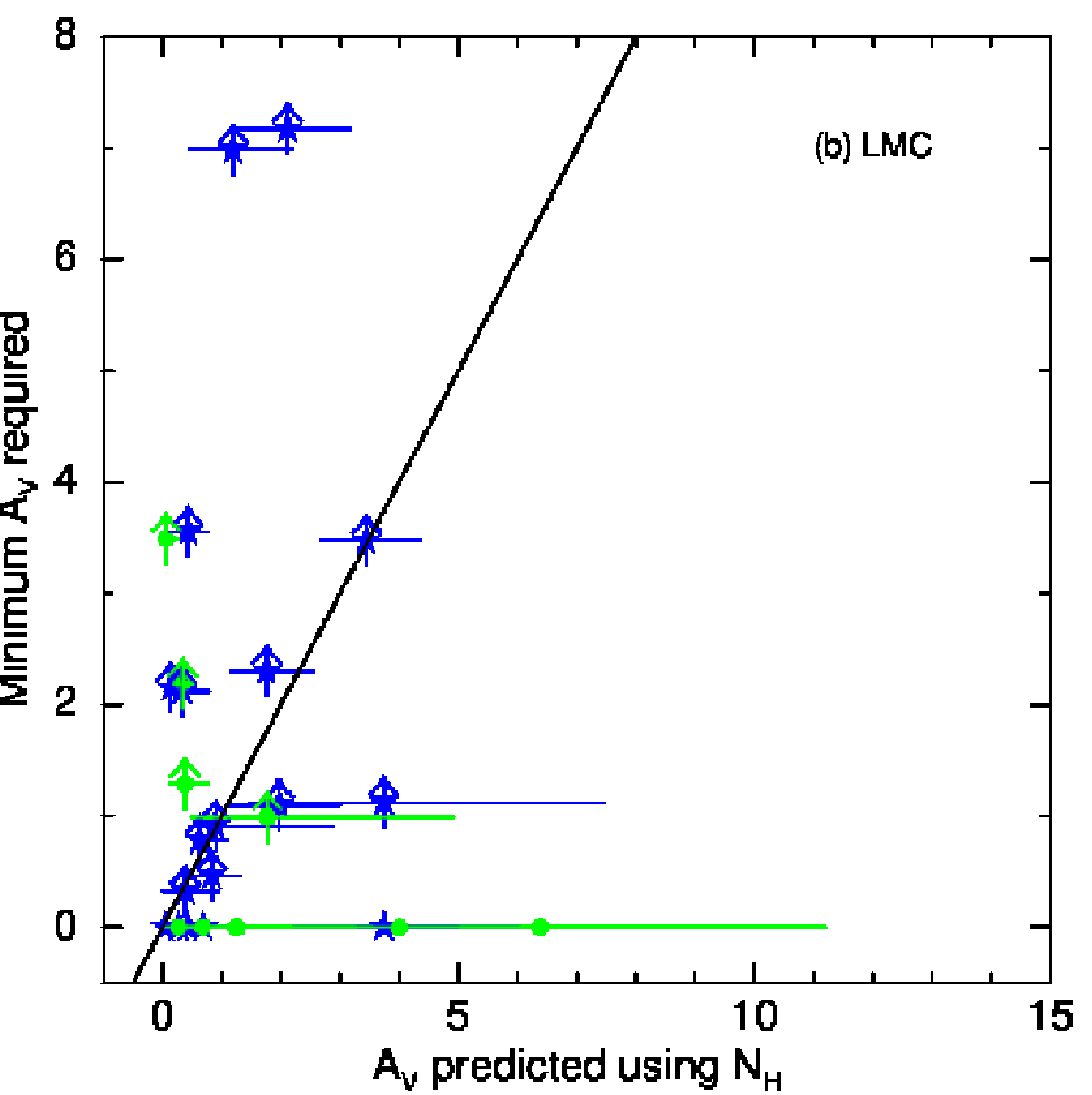}
\includegraphics[width=5.5cm]{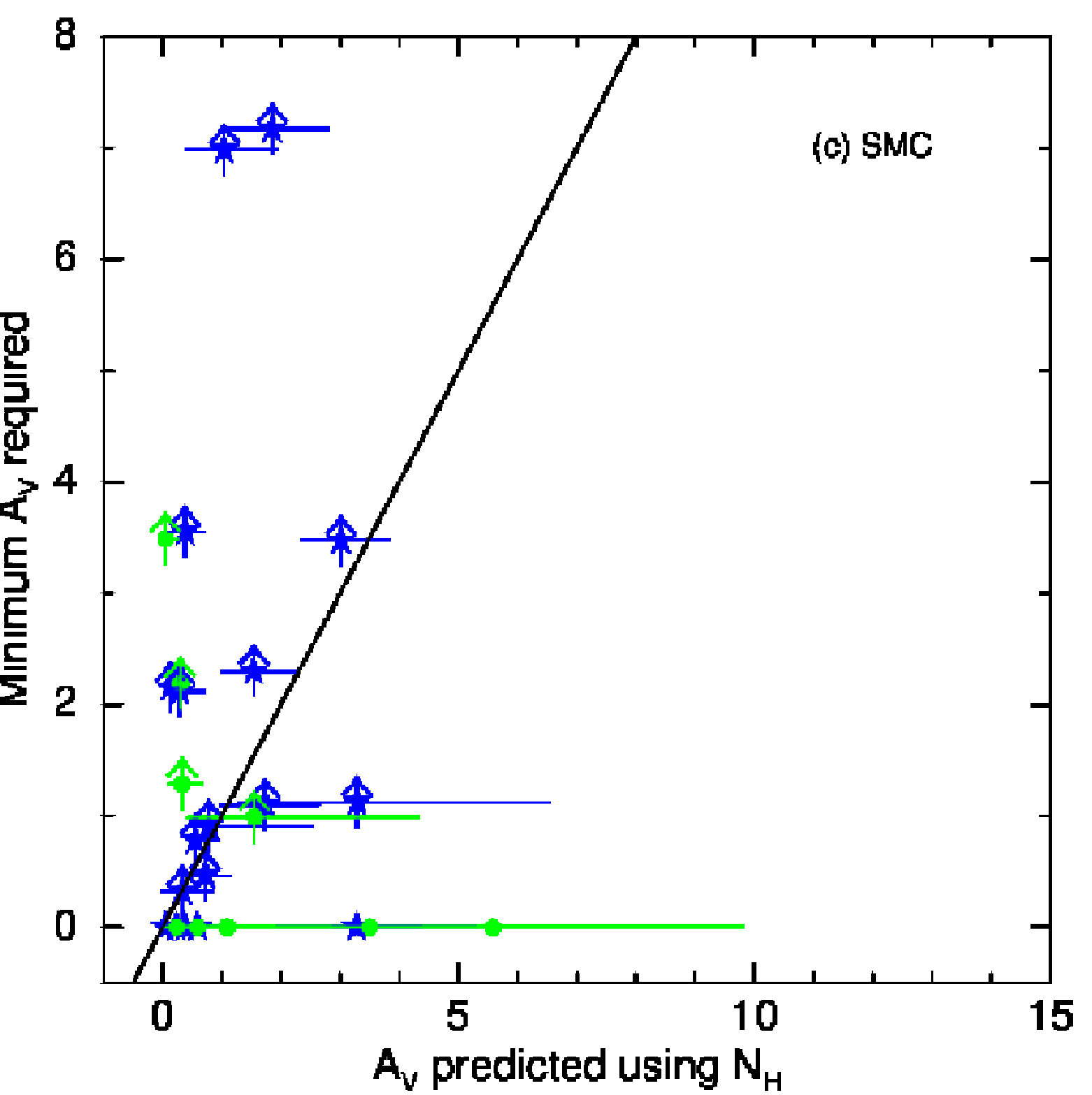}
\caption{A plot comparing the minimum optical absorption, A$_{V}$ required to explain the difference between the X-ray and optical absorptions to those predicted using the X-ray N$_{H}$. Unless the optical data are already consistent with the X-ray observations, all data points are lower limits given the assumptions made. These plots are for three different abundances: (a) Milky Way, (b) Large Magellanic Cloud and (c) Small Magellanic Cloud. Data points lying above the black line cannot be explained by simply using optical absorption. Symbols are as in Figure \ref{fig5}.}
\label{AV_fig}
\end{figure*}

\begin{table}
\begin{center}
\begin{tabular}{|c|c|c|c|c|}
\hline
GRB     & Minimum A$_{V}$ & MW A$_{V}$ & LMC A$_{V}$ & SMC A$_{V}$  \\
\hline
\multicolumn{4}{|l|}{Magnetar candidates}\\
\hline
051221A & 0.00                       & 1.32$^{+0.42}_{-0.39}$  & 0.68$^{+0.21}_{-0.20}$ & 0.59$^{+0.19}_{-0.17}$ \\
060313  & 2.15                       & 0.28$^{+0.39}_{-0.07}$  & 0.14$^{+0.20}_{-0.03}$ & 0.13$^{+0.18}_{-0.03}$ \\
060801  & 0.91                       & 1.75$^{+3.86}_{-0.85}$  & 0.89$^{+1.97}_{-0.43}$ & 0.78$^{+1.73}_{-0.38}$ \\
070724  & 0.32                       & 0.74$^{+1.12}_{-0.73}$  & 0.38$^{+0.57}_{-0.37}$ & 0.33$^{+0.50}_{-0.33}$ \\
070809  & 0.00                       & 0.52$^{+0.84}_{-0.17}$  & 0.27$^{+0.43}_{-0.09}$ & 0.23$^{+0.38}_{-0.08}$ \\
080426  & 1.10                       & 3.85$^{+2.00}_{-1.66}$  & 1.97$^{+1.02}_{-0.85}$ & 1.73$^{+0.89}_{-0.74}$ \\
080905A & 0.46                       & 1.61$^{+0.95}_{-0.73}$  & 0.83$^{+0.49}_{-0.37}$ & 0.72$^{+0.43}_{-0.33}$ \\
080919  & 1.12                       & 7.32$^{+7.25}_{-4.45}$  & 3.74$^{+3.71}_{-2.27}$ & 3.28$^{+3.24}_{-1.99}$ \\
081024  & 0.00                       & 7.32$^{+4.47}_{-3.02}$  & 3.74$^{+2.29}_{-1.54}$ & 3.28$^{+2.00}_{-1.35}$ \\
090426  & 0.00                       & 0.08$^{+2.02}_{-0.01}$  & 0.04$^{+1.03}_{-0.00}$ & 0.04$^{+0.90}_{-0.00}$ \\
090510  & 2.12                       & 0.65$^{+0.90}_{-0.56}$  & 0.33$^{+0.46}_{-0.29}$ & 0.29$^{+0.40}_{-0.25}$ \\
090515  & 3.55                       & 0.84$^{+0.66}_{-0.60}$  & 0.43$^{+0.34}_{-0.31}$ & 0.38$^{+0.30}_{-0.27}$ \\
100117A & 6.98                       & 2.33$^{+1.96}_{-1.45}$  & 1.19$^{+1.00}_{-0.74}$ & 1.04$^{+0.88}_{-0.65}$ \\
100702A & 7.17                       & 4.13$^{+2.07}_{-1.79}$  & 2.11$^{+1.06}_{-0.91}$ & 1.85$^{+0.93}_{-0.80}$ \\
101219A & 2.30                       & 3.45$^{+1.54}_{-1.19}$  & 1.76$^{+0.79}_{-0.61}$ & 1.54$^{+0.69}_{-0.53}$ \\
111020A & 0.00                       & 0.83$^{+0.47}_{-0.47}$  & 0.42$^{+0.24}_{-0.24}$ & 0.37$^{+0.21}_{-0.21}$ \\
120305A & 3.47                       & 6.72$^{+1.83}_{-1.49}$  & 3.44$^{+0.93}_{-0.76}$ & 3.01$^{+0.82}_{-0.67}$ \\
120521A & 0.78                       & 1.23$^{+0.89}_{-0.16}$  & 0.63$^{+0.45}_{-0.08}$ & 0.55$^{+0.40}_{-0.07}$ \\
\hline
\multicolumn{4}{|l|}{Possible candidates}\\
\hline
050509B & 0.00                       & 0.54$^{+0.45}_{-0.45}$  & 0.27$^{+0.23}_{-0.23}$ & 0.24$^{+0.20}_{-0.20}$ \\
051210  & 0.00                       & 2.45$^{+1.01}_{-1.01}$  & 1.25$^{+0.51}_{-0.51}$ & 1.10$^{+0.45}_{-0.45}$ \\
061201  & 2.19                       & 0.67$^{+0.33}_{-0.31}$  & 0.34$^{+0.17}_{-0.16}$ & 0.30$^{+0.15}_{-0.14}$ \\
070714A & 0.00                       & 12.47$^{+2.96}_{-2.62}$ & 6.38$^{+1.52}_{-1.34}$ & 5.58$^{+1.33}_{-1.17}$ \\
080702A & 0.00                       & 7.82$^{+14.11}_{-6.84}$ & 4.00$^{+7.21}_{-3.50}$ & 3.50$^{+6.31}_{-3.06}$ \\
090621B & 0.99                       & 3.45$^{+6.14}_{-2.50}$  & 1.77$^{+3.14}_{-1.28}$ & 1.55$^{+2.75}_{-1.12}$ \\
091109B & 0.00                       & 1.32$^{+1.61}_{-0.86}$  & 0.68$^{+0.82}_{-0.44}$ & 0.59$^{+0.72}_{-0.39}$ \\
100625A & 3.49                       & 0.12$^{+0.42}_{-0.00}$  & 0.06$^{+0.21}_{-0.00}$ & 0.05$^{+0.19}_{-0.00}$ \\
110112A & 1.29                       & 0.75$^{+0.73}_{-0.46}$  & 0.38$^{+0.37}_{-0.24}$ & 0.33$^{+0.33}_{-0.21}$ \\
111117A & 0.00                       & 2.43$^{+3.90}_{-1.76}$  & 1.24$^{+2.00}_{-0.90}$ & 1.09$^{+1.75}_{-0.79}$ \\
\hline
\end{tabular}
\caption{The minimum optical absorption, A$_{V}$, is the absorption required for the optical observations to just be consistent with the X-ray observations (0 means they are already consistent). The MW (Milky Way), LMC (Large Magellanic Cloud) and SMC (Small Magellanic Cloud) absorptions are the values predicted using the X-ray N$_{H}$.} 
\label{AV_table}
\end{center}
\end{table}

To determine if the observed X-ray excess could be caused by optical absorption, we compare the optical absorption (A$_{V}$) estimated using the observed X-ray N$_{H}$ to the minimum absorption that could explain the difference between the X-ray and optical fluxes. The observed spectra during the plateau regime (given in Table \ref{spectra}) are used when available and the other spectral fits are obtained from the automated data products from the UK {\it Swift} Science Data Centre \citep{evans2007,evans2009}. We convert the observed X-ray N$_{H}$ to optical absorptions using $\frac{N_{H}}{A_{V}}$ for Milky Way \citep[MW, $1.8\times10^{21}$;][]{predehl1995}, Large Magellanic Cloud \citep[LMC, $3.5\times10^{21}$;][]{koornneef1982, fitzpatrick1985} and Small Magellanic Cloud \citep[SMC, $4.0\times10^{21}$;][]{martin1989} abundances. Note that there are known to be significant scatter and uncertainties involved in this conversion \citep[e.g.][]{schady2010, campana2010}. To obtain the minimum A$_{V}$ which would be sufficient to explain the difference between the X-ray and optical fluxes, the maximum possible optical flux (including errors and assuming the most extreme cooling break) and the X-ray plateau flux are converted to V band magnitudes\footnote{using the webtool: http://www.stsci.edu/hst/nicmos/tools/conversion{\_}form.html}. The obtained optical absorptions are given in Table \ref{AV_table} and plotted in Figure \ref{AV_fig}. Many of the GRBs may be explicable via absorption however we note that $\sim$25\% of the sample are based on unconstraining optical upper limits while some rely on using the most extreme cooling breaks and uncertainties. In Figure \ref{AV_fig}, we also show that if some of the host galaxies are more consistent with LMC or SMC abundances then more of the GRBs cannot be explained via absorption. Results obtained by \cite{schady2010} for LGRBs, also suggest that $\frac{N_{H}}{A_{V}}$ may be an order of magnitude higher for GRB host galaxies, in which case even more GRBs in the sample would not be explicable via absorption. Despite all the uncertainties involved in this calculation, 8 GRBs in the sample clearly cannot have the difference between their X-ray and optical fluxes explained via absorption (GRBs 060313, 061201, 090510, 090515, 100117A, 100625A, 100702A and 110112A).

This analysis shows that at least some of the GRBs in this sample are consistent with there having an additional X-ray component. This may provide supporting evidence of energy injection, although energy injection is thought to cause an increase in flux at all wavelengths \citep[e.g.][however this also depends upon the electron energy distribution]{sari2000}.

Although there is some evidence that the magnetar candidates have additional X-ray emission, it is not known what spectrum is expected from a newly formed magnetar and hence we cannot completely discount those whose optical emission is consistent with their X-ray emission.

\begin{table*}
\begin{tabular}{ccccccc}
\hline
GRB     & Expected region & Extra component   & Predicted region  & Stable/Unstable \\
\hline
050509B & ?               & ?                 & No                & Stable \\
051210  & ?               & Yes               & ?                 & Unstable \\
051221A & Yes             & No                & No                & Stable \\
060313  & Yes             & Yes               & Yes               & Stable \\
060801  & Yes             & Yes               & Yes               & Unstable \\
061201  & ?               & ?                 & ?                 & Stable \\
070714A & ?               & ?                 & ?                 & Stable \\
070724A & Yes             & ?                 & No                & Unstable \\
070809  & Yes             & ?                 & Yes               & Stable \\
080426  & Yes             & Yes               & ?                 & Stable \\
080702A & ?               & ?                 & Yes               & Stable \\
080905A & Yes             & ?                 & No                & Unstable \\
080919  & Yes             & No                & No                & Unstable \\
081024  & Yes             & ?                 & Yes               & Unstable \\
090426  & Yes             & No                & Yes               & Stable \\
090510  & Yes             & Yes               & Yes               & Stable \\
090515  & Yes             & Yes               & Yes               & Unstable \\
090621B & ?               & Yes               & ?                 & Stable \\
091109B & ?               & ?                 & No                & Stable \\
100117A & Yes             & Yes               & No                & ? \\
100625A & ?               & Yes               & ?                 & ? \\
100702A & Yes             & Yes               & Yes               & Unstable \\
101219A & Yes             & Yes               & Yes               & Unstable \\
110112A & ?               & Yes               & ?                 & Stable \\
111020A & Yes             & No                & No                & Stable \\
111117A & ?               & ?                 & ?                 & Stable \\
120305A & Yes             & Yes               & Yes               & Unstable \\
120521A & Yes             & Yes               & Yes               & Unstable \\

\end{tabular}
\caption{A summary showing the main features studied. This gives best magnetar candidates found and the possible candidates. ``Expected region'' : fits within the required parameter space in Figure \ref{fig5} (? = could fit with various assumptions), ``Extra component'' : there is evidence of an extra component in the X-ray afterglow which is not observed in the optical note this could also be due to absorption (? = borderline case or optical upper limit not constraining), ``Predicted region'' : do the values for the plateau luminosity and the plateau duration, calculated using equations \ref{period} and \ref{luminosity}, lie within the predicted region in \citet{metzger2010}? (? = outside region but would fit with reasonable assumptions). ``Stable/Unstable'' : whether the magnetar is stable or if it collapses to form a BH (? = would be fitted well by either case.)}
\label{summary}
\end{table*}

\section{Discussion}

\subsection{The sample of SGRBs}

Here we discuss some particular SGRBs and then the sample as a whole.

GRB 070809 is one of the best fitting stable magnetar candidates and lies within the allowed regions. This GRB had a faint optical afterglow and is offset by 20 kpc from a galaxy at z = 0.219 \citep{perley2008a}, making it an ideal candidate for a magnetar formed via the merger of two NSs. However it is important to be cautious about this candidate host galaxy association as the likelihood that this is an unrelated field galaxy is 5 -- 10\% (Tunnicliffe et al. in prep).

GRB 061201, with a spin period of $\sim$16 ms, fits the magnetar model well but is spinning slower than expected. However the redshift used relies on the correct host galaxy identification which remains highly uncertain \citep[][; Tunnicliffe et al. in prep]{stratta2007}. If it actually occured at a higher redshift than used in this analysis it would lie within the expected region. Additionally, the approximate 10 ms limit imposed by \cite{usov1992} is dependent on the initial radius of the collapsing object and the radius of the final NS. This limit is also derived for the model involving AIC of a WD. Therefore there is some level of flexibility in this imposed limit. We still consider this, and other GRBs close to this boundary, to be potential candidate magnetars.

GRB 051221A is consistent with having energy injection in its lightcurve out to $\sim2\times10^4$ s \citep{burrows2006,soderberg2006}. \cite{fan2006} explained this as energy injection from a magnetar. Our model fits this GRB very well. \cite{jin2007} proposed an alternative two jet model to explain the lightcurves without requiring additional energy injection. 

GRB 060313 has been included in the magnetar sample by ignoring the first 50 -- 200 s of the lightcurve due to the flaring activity, this gives a good fit to the later data but this result should be treated with caution. Flares could be associated with on-going accretion onto the newly formed magnetar. Alternatively, \cite{dai2006} and \cite{gao2006} suggest that the X-ray flares originate from reconnection of twisted magnetic fields within the NS. \cite{margutti2011} have conducted a systematic study into SGRB flares, including the flares observed in GRB 060313, and concluded that the flares are consistent with a central engine origin.

Included in this sample are SGRBs whose progenitors are subject to significant debate, particularly GRB 090426 at z$\sim$2.6 which could have orginated from a collapsar instead of a binary merger \citep{antonelli2009,levesque2010, thone2011, xin2011}. GRB 090426 fits the model well, irrespective of the progenitor, but the progenitor debate is important to note as we are specifically studying possible NS binary merger progenitors.

Interestingly, 12 out of the 28 magnetar candidates require collapse to a BH. This implies that, if these SGRBs are making magnetars, they only collapse to a BH in a small number of cases. Comparing the derived plateau durations and the collapse times provided in Table \ref{table:log}, the magnetar typically (but not always) collapses to a BH after the plateau phase, i.e. when the magnetar has spun down significantly. The only exception to this is GRB 101219A where collapse occurs prior to the end of the plateau phase, however the collapse time and end of the plateau are consistent within errors. The collapse time is related to the mass of the magnetar and the spin period at which the differential rotation can no longer support gravitational collapse. The discrepancy between collapse time and plateau duration are hence likely to be reliant upon the mass of the magnetar. Additionally, there may be ongoing accretion on to the magnetar (remnants of the merger) which may raise the mass of the magnetar above the critical point prior to significant spin down. Interestingly, those candidates which collapse to form a BH and are within the allowed (unshaded) region of Figure \ref{fig5} have a higher magnetic field for a given spin period than the candidates which do not collapse to a BH.

Many of the magnetar candidates lie within, or near to, the predicted plateau luminosity and duration regions for newly formed magnetars given in \cite{metzger2010} when considering uncertainties due to redshift, efficiency and beaming. However, there are candidates whose plateaus are significantly shorter than predicted or at a lower luminosity. Our analysis and that of \cite{metzger2010} assumes a NS mass of 1.4M$_{\odot}$ and this is likely to be significantly higher for a NS merger progenitor (e.g. 2.1M$_{\odot}$). This has a small affect on the values of the magnetic field strength and the spin period calculated in our model \citep[as shown in][]{rowlinson2010b} but does not significantly affect the predicted regions for plateau luminosity and duration from \cite{metzger2010}. 

A summary of the properties of the whole magnetar sample are shown in Table \ref{summary}.

\subsection{Accretion Effects}

\begin{figure}
\centering
\includegraphics[width=8cm]{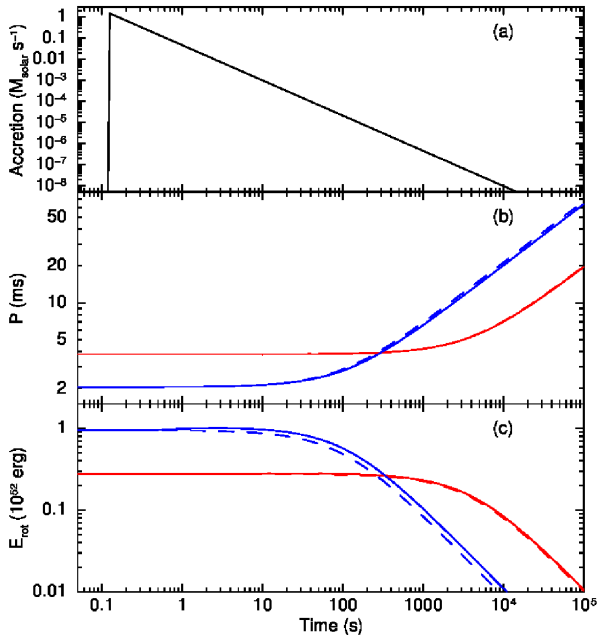}
\caption{(a) The accretion rate as a function of time assuming the accretion rate for a compact binary merger \citet{metzger2010b} starting at 0.16 s after the trigger time giving a total accretion disk mass of $\sim$0.3 M$_{\odot}$. (b) The evolution of the spin period of the magnetar for the two accretion rates, red - the magnetar predicted for GRB 060313 and blue - GRB 090515. Solid lines include accretion and dashed lines have no accretion. In these plots, accretion has a very small or negligible effect. (c) The amount of rotational energy available in the magnetar for each case.}
\label{fig10}
\end{figure}

\begin{figure}
\centering
\includegraphics[width=7.8cm]{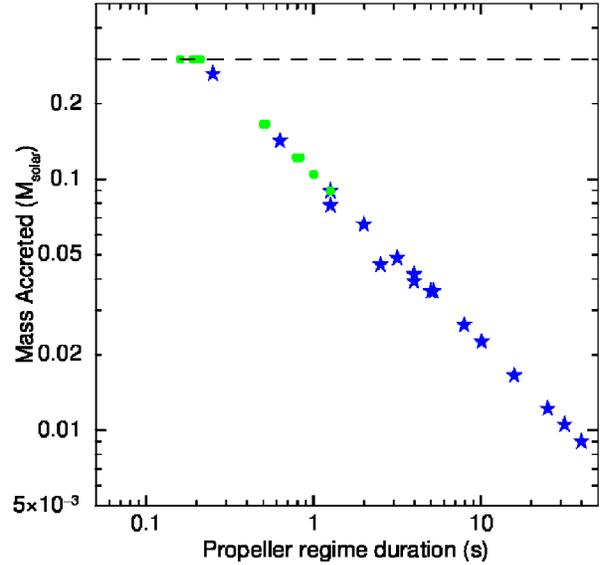}
\caption{The amount of mass accreted by the magnetar against the duration of the propeller regime. The dashed line represents the maximum mass available in the accretion disk and is 0.3 M$_{\odot}$ an upper limit for the amount of mass which can be accreted. Symbols are as in Figure \ref{fig5}.}
\label{fig10b}
\end{figure}

\begin{figure}
\centering
\includegraphics[width=7.8cm]{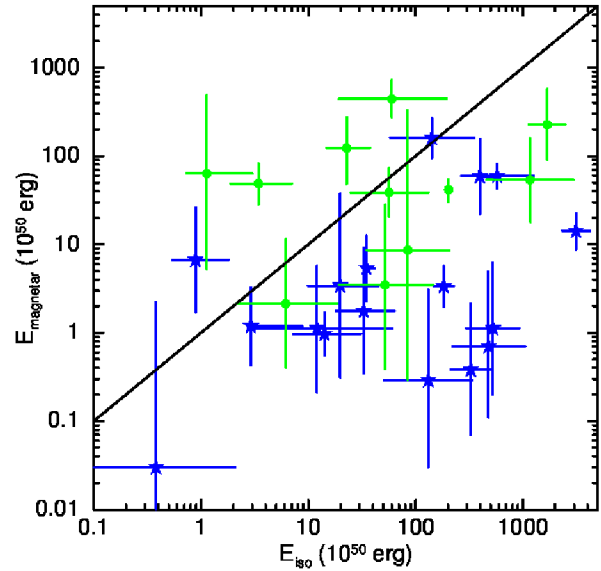}
\caption{The energy emitted during the plateau phase, calculated using the fits in Table \ref{table:log}, compared to the isotropic energy emitted during the prompt phase (1--10000 keV). Symbols are as in Figure \ref{fig5}.}
\label{fig7c}
\end{figure}

In our analysis we have not accounted for any ongoing accretion onto the magnetar from the surrounding torus of material formed during the merger. This could significantly effect the results obtained, especially if accretion increases the NS mass to more than can be supported as this results in collapse to a BH. Additionally, accretion could explain flares observed overlaying the plateau model. Flares may also be associated with ongoing magnetar activity as described in \cite{dai2006}. 

\cite{piro2011} studied the affect of accretion onto magnetars formed during SNe, however their results are also applicable to magnetars produced from neutron star binary mergers. The main difference for mergers is the significantly reduced reservoir of material available for accretion and the different accretion rate. In this section, we assume the simplest accretion rate published by \cite{metzger2010b} assuming that accretion starts at 0.16 s after the trigger time, giving a total accretion disk mass of $\sim$ 0.3 M$_{\odot}$. Accretion onto the magnetar occurs when the the propeller regime ends, given by equation \ref{propeller} from \cite{piro2011} where $\mu_{33} = B_{15}R_{6}^{3}$.

\begin{eqnarray}
\dot M < 6.0 \times 10^{-3} \mu_{33}^{2} M_{1.4}^{-5/3} P_{0,-3}^{-7/3} M_{\odot} s^{-1} \label{propeller}
\end{eqnarray}

As before, we assume an initial NS mass of 1.4 M$_{\odot}$ and radius of $10^6$ cm. In Figure \ref{fig10}a we show the accretion rate as a function of time after formation. In Figure \ref{fig10}b we show the evolution of the spin period of two different magnetars (using the parameters for GRBs 060313 and 090515 as these have contrasting magnetar properties) assuming there is either accretion onto the magnetar or no accretion. When there is significant accretion (e.g. GRB 090515) it can marginally prevent spin down and increase the rotational energy (Figure \ref{fig10}c) available, although these are negligible effects for the low accretion rates considered.

It is worth noting that accretion would potentially have a very large effect on the results obtained for LGRB magnetar candidates \citep[e.g. the sample in][]{lyons2009} as these are thought to have a significantly higher mass accretion disk and an accretion rate similar to that proposed by \cite{piro2011}. In that case, the energy reservoir could reach values in excess of 10$^{53}$ ergs for particular combinations of the initial conditions. This additional energy source could be a potential explanation for large flares observed in some of the LGRB candidate lightcurves \citep[e.g.][]{margutti2011}.

In Figure \ref{fig10b} we show the total mass accreted after the propeller regime has ended. The linear correlation between the duration of the propeller regime and the mass accreted is caused by the relationship: $\dot M \propto t^{-5/3}$ (i.e. the sooner the propeller regime ends, the greater the mass that can be accreted). The candidates which accrete the most mass are those which also collapse to form a BH within a few hundred seconds, leading to the suggestion that accretion is an alternative to drive this collapse. Typically, the magnetar is thought to collapse when the fast rotation can no longer support the mass of the magnetar. The stable magnetar outliers are GRBs 100625A and 100117A which were also well fit by the unstable magnetar model but we chose the stable model to reduce the number of free parameters. Additionally GRB 090426 is again a clearly stable magnetar candidate which is separate from the other stable candidates.

\subsection{Energy Constraints}

Including all of the possible candidates, the SGRBs in our sample can be fitted with the magnetar model. In Table \ref{table:log} we show the isotropic energy released during the prompt emission phase of the GRB. These values tend to be consistent with the maximum expected energy output from the magnetar central engine model, $E_{iso} < 3\times 10^{52}$ erg \citep{metzger2010}. Within the uncertainties many of the magnetar candidates are consistent with this limit while some others exceed it. However, we have not corrected for beaming and had to assume redshifts in many cases. Not correcting for beaming will undoubtedly affect these results by increasing the spin period and the magnetic field strengths as shown in \cite{rowlinson2010b}. Beaming, with a half-opening angle of 30$^{\circ}$, has been shown to form via the formation of an ordered magentic field during the merger of two 1.5 M$_{\odot}$ NSs which collapse to form a BH \citep{rezzolla2011}. However, the beaming angles of SGRBs and associated magnetars remain unconstrained \citep[see recent work on SGRB jets by ][]{fong2012}. With a reasonable beaming correction, all of the GRBs which exceed the energy constraint would lie well below the maximum expected energy output.

Another consideration is that $E_{iso} \propto M_{1.4} P_{0,-3}^{-2}$, so if magnetars can have masses up to 2.1M$_{\odot}$ then the maximum energy output could be as high as E$_{iso} \sim 1 \times 10^{53}$ erg. 

In Figure \ref{fig7c}, we show the energy emitted during the magnetar plateau phase (the plateau luminosity multiplied by the duration from Table \ref{table:log}, these values were calculated from the fitted $B_{15}$ and $P_{-3}$ using Equations \ref{luminosity} and \ref{period}) against the isotropic energy emitted during the prompt emission. Only five GRBs which fit the magnetar model emit more energy during the plateau phase, GRBs 051210, 070724A, 070809, 090515 and 100702A.

We have also assumed 100\% efficiency in the conversion of rotational energy into EM radiation. This will not be the case and assuming a lower efficiency would act counter to the effect of any beaming, in the sense of reducing the inferred spin period and the magnetic field strengths. For example, GRB 090515 has $B \sim 1.4 \times 10^{16}$ G and $P \sim 2.3$ ms assuming 100\% efficiency, at 10\% efficiency these drop to $B \sim 4.4 \times 10^{15}$ G and $P \sim 0.73$ ms. Given the uncertainties in both beaming and efficiency, we note that the real values of the magnetic field strength and the spin period may be uncertain by at least a factor of 3. 

\subsection{Gravitational Wave Signals}

\begin{table*}
\begin{tabular}{ccccccc}
\hline
Phase	& Citation & Predicted Amplitude & Distance used & AdLIGO/LCGT limit & ET limit  & Amplitude at z$\sim$0.1\\
        &          & (h)                 &         (Mpc) & (Mpc)        & (Mpc)     & (h) \\
\hline
Inspiral       & \cite{abadie2010} & $4 \times 10^{-24}$  & 445 & 445       & 5900   & $4.6\times10^{-24}$ \\
Magnetar Spindown & \cite{corsi2009}  & $7\times10^{-24}$ & 100  & 175  & 2300 & $1.8\times10^{-24}$\\
Collapse to BH & \cite{novak1998}  & $4\times10^{-23}$    & 10  & 100       & 1300   & $1\times10^{-24}$
\end{tabular}
\caption{Gravitational wave predictions for the three different regimes in this magnetar model and applied to future observatories. The distances quoted are luminosity distances. The magnetar spindown values are calculated using Equation 14 in \citet{corsi2009}.}
\label{grav1}
\end{table*}

Systems of the kind we have considered represent interesting sources of gravitational waves as there are predicted signals for all of the stages this system can go through: inspiral, magnetar spindown and final collapse to BH. In Table \ref{grav1}, we show the distances out to which each phase would be visible, assuming the amplitude ($h$) of the gravitational waves is inversely proportional to distance for Advanced LIGO (AdLIGO, with a sensitivity of $h \sim 4 \times 10^{-24}$), the Large Cryogenic Gravitational Telescope \citep[LCGT, comparable sensitivity to AdLIGO;][]{kuroda2010} and the Einstein Telescope \citep[ET, $h \sim 3 \times10^{-25}$;][]{hild2011}. The gravitational wave amplitude is quoted for a distance of $z\sim0.1$ or 390 Mpc. The magnetar phase prediction is an upper limit assuming a spin period of 1 ms, $I_{45}=1.5$ for a binary merger progenitor, and an ellipticity $\epsilon=1$. AdLIGO predictions by \cite{abadie2010} are for NS-NS mergers.

Using the lowest and maximum possible rates for NS-NS mergers per Milky Way Equivalent Galaxy from \cite{abadie2010}, it is possible to predict the number of unstable magnetars (i.e. one source giving 2 distinct gravitational wave signals) we might expect to detect with AdLIGO and ET. To detect all the stages for the formation and collapse of a magnetar, AdLIGO would require it to be at a distance $\sim$100 Mpc and ET would require $\sim$1300 Mpc. Within these volumes there is predicted to be a NS-NS merger rate of $2\times10^{-5}$ -- $0.08$ yr$^{-1}$ for AdLIGO and $10$ -- $4\times10^{5}$ yr$^{-1}$ for ET. However, the rates need modification as not all NS-NS mergers will lead to an unstable magnetar which will give both signals. From the analysis in this paper, only 11 SGRBs in the total sample of 28 magnetar candidates (39\%, assuming NS-NS mergers always produce a magnetar) are thought to form unstable magnetars, giving rates of $8\times10^{-6}$ -- $0.03$ yr$^{-1}$ for AdLIGO and $4$ -- $2\times10^{5}$ yr$^{-1}$ for ET. Therefore, it is unlikely that AdLIGO or LCGT will observe both the formation and collapse of an unstable magnetar but ET should detect many cases. On a more optimistic note, \cite{bausswein2012} estimate that AdLIGO will be able to detect a post-merger signal associated with a newly formed massive NS with a rate of 0.015 -- 1.2 yr$^{-1}$.

\cite{shibata2006} also study different masses relative to the maximum mass of a NS. They determined that if $M<M_{\rm max}$ then the NS will emit gravitational waves during the magnetar spindown phase until it is a stable sphere and collapse to a black hole is dependant on the gravitational wave emission (possibly collapsing within 50 ms) or on forces such as magnetic breaking. In this case, they predict that advanced gravitational wave detectors will be able to observe these events out to 50 Mpc using detectors such as AdLIGO. Alternatively if $M\sim M_{\rm max}$, then it collapses rapidly to spherical shape and hence is more likely to create a stable NS which may collapse at late times due to magnetic breaking. The gravitational waves from the more massive NS would be detectable to 10 Mpc. 

In both \cite{baiotti2008} and \cite{shibata2006}, instabilities in the NS formed by a compact merger produce detectable gravitational waves in contrast to the spherical collapse model of \cite{piro2011}. However \cite{piro2011} showed that accretion may have an important affect on the gravitational wave signal. Therefore, these objects are potentially important sources of gravitational waves and further analysis combining all these factors and the new limits on maximum NS masses is required.

The predictions by \cite{metzger2010} do not take into account the loss of energy via gravitational waves and this may play a significant role for the formation of a magnetar via the merger of two NSs. Some of our candidates have shorter plateau durations than predicted by \cite{metzger2010} however if the energy losses via gravitational waves are more significant then the magnetar will spin down more rapidly.

\section{Conclusions}

We have analysed the BAT-XRT lightcurves of all the {\it Swift} GRBs with prompt durations $T_{90} \le2$ s detected until May 2012. About half of these SGRBs require fitting with a broken powerlaw model showing a plateau phase. Although the plateau phases show many similarities with those observed in LGRB lightcurves, they are typically orders of magnitude earlier. The initial temporal indicies ($\alpha_{1}$ and $\alpha_{2}$) are comparable to those found for the ``canonical'' LGRBs but there is much more variation in the final decay ($\alpha_{3}$). The correlation between luminosity and duration of the plateau phase is found to be consistent with the identified correlation for ``canonical'' LGRB lightcurves identified by \cite{dainotti2010}. 

Following on from the study of GRB 090515, this work has shown that the X-ray lightcurves of some SGRBs considered could be explained with energy injection from a magnetar which can collapse to form a BH. 18 firm candidates ($64\%$) and 10 possible candidates were found. Of the 18 firm candidates, 10 are thought to collapse to form a BH and when including possible candidates, 11 out of 28 magnetar candidates may collapse to form a BH. This implies that $29$--$56\%$ of events forming magnetars would collapse to a BH within the first few hundred seconds. In some cases the magnetar plateau phase is not directly observed as it occurs prior to the XRT observations. This predicts plateau emission that may be observable with future missions that are able to slew faster than {\it Swift}.

The X-ray fluxes at 1000 s and 10000 s are typically higher for the stable magnetar candidates. The late time fluxes are significantly lower for the unstable magnetar cases. There is excess emission in the X-ray afterglows not observed in the optical afterglows for many of the magnetar sample. Many of the magnetar candidates lie within or close to the predicted regions for plateau luminosity and duration for newly formed magnetars given in \cite{metzger2010}.

Accretion onto the newly formed magnetar formed by a NS-NS binary merger has a negligible affect on the spin periods and hence the rotational energy budget of the magnetar. However, it can be shown that accretion can have a significant affect for collapsar progenitors. This may explain late time flares for collapsar progenitors and our calculations suggest the rotational energy budget could exceed $10^{53}$ erg for some combinations of initial spin periods and magnetic fields. The unstable magnetar candidates, those which collapse to form a BH, are potentially accreting more material than the stable candidates. We suggest this is an additional solution for why they collapse at late times which would work alongside the theory that the magnetar spins down to a critical point where it can no longer support its mass using rotation.

These objects are highly interesting targets for future gravitational wave observatories as they are predicted to emit gravitational waves during merger, the magnetar phase (likely to be increased via accretion and bar mode instabilities) and, in some cases, the final collapse to form a BH. In this paper, we have focused on NS-NS merger progenitors, however the accretion induced collapse of a WD could also produce a SGRB and leave behind a rapidly rotating magnetar with similar X-ray emission properties. Among other observational signatures, the very different gravitational wave signals between these events may someday allow these progenitors to be distinguished however the inspiral remains the most luminous phase of gravitational wave emission.

For the candidates which form a stable magnetar, \cite{duncan1992} showed that the amount of energy available for an SGR giant flare is $E~\propto~3\times10^{47}~B_{15}^{2}$ erg. Hence a young magnetar, with a magnetic field of $B_{15}\sim 10$, could produce a giant flare with an energy of $3\times10^{49}$ erg. This value is comparable to the isotropic energy of some SGRBs \citep[e.g. GRB 080905A at $z\sim0.12$,][]{rowlinson2010a} so would be observable in the local universe. Both of the merger and giant flare events are very rare, however considering these models it is possible (although very unlikely) that in the future we may have two spatially co-incident SGRBs. This has also been proposed for LGRBs by\cite{giannios2010} and they suggest that these magnetar candidates could be identified by discovering an old spatially coincident radio GRB afterglow in nearby galaxies.

We have shown that a model of SGRB production from binary NS mergers that result in the formation of a magnetar can explain the plateaus seen in many SGRB X-ray lightcurves. Although this is not conclusive proof of such a model, it would tie in to the evidence for late time central engine activity in SGRBs and may have important observational consequences.

\section{Acknowledgements}
AR acknowledges funding from the Science and Technology Funding Council. This work makes use of data supplied by the UK {\it Swift} Science Data Centre at the University of Leicester and the {\it Swift} satellite. {\it Swift}, launched in November 2004, is a NASA mission in partnership with the Italian Space Agency and the UK Space Agency. Swift is managed by NASA Goddard. Penn State University controls science and flight operations from the Mission Operations Center in University Park, Pennsylvania. Los Alamos National Laboratory provides gamma-ray imaging analysis. BDM is supported by NASA through Einstein Postdoctoral Fellowship grant number PF9-00065 awarded by the Chandra X-ray Center, which is operated by the Smithsonian Astrophysical Observatory for NASA under contract NAS8-03060.

\end{document}